\documentclass[12pt]{article}
\pdfoutput=1

\usepackage{pdflscape}

\usepackage{diagbox}
\usepackage{epsfig}
\usepackage{psfrag}
\usepackage{latexsym}
\usepackage{indentfirst}
\usepackage{fancyhdr}
\usepackage{dsfont}
\usepackage{amsmath}
\usepackage{amssymb}
\usepackage{amsfonts}
\usepackage{mathrsfs}
\usepackage{amsthm}
\usepackage{pifont}
\usepackage{dsfont}
\usepackage{multirow}
\usepackage{array}
\usepackage{chngpage}
\usepackage{longtable}
\usepackage{cite}
\usepackage{bbold}
\usepackage{color}
\usepackage{braket}
\usepackage{colordvi}
\usepackage{fancybox}
\usepackage[footnotesize]{caption2}
\usepackage{graphicx}
\usepackage[center,footnotesize,hang]{subfigure}
\usepackage{bbm}
\usepackage[colorlinks, linkcolor=red, anchorcolor=black, citecolor=green]{hyperref}
\usepackage{multirow}
\usepackage{array}
\usepackage{textcomp}  
\usepackage{enumitem}  
\usepackage{mathrsfs}  
\newcommand{\PreserveBackslash}[1]{\let\temp=\\#1\let\\=\temp}
\newcolumntype{C}[1]{>{\PreserveBackslash\centering}p{#1}}
\newcolumntype{R}[1]{>{\PreserveBackslash\raggedleft}p{#1}}
\newcolumntype{L}[1]{>{\PreserveBackslash\raggedright}p{#1}}
\allowdisplaybreaks

\newcommand{\bq}{\begin{eqnarray}}
\newcommand{\nq}{\end{eqnarray}}
\newcommand{\ignore}[1]{}

\newcommand*{\Scale}[2][4]{\scalebox{#1}{$#2$}}%

\usepackage{tikz}
\newcommand*{\circled}[1]{\lower.7ex\hbox{\tikz\draw (0pt, 0pt)%
    circle (.5em) node {\makebox[1em][c]{\small #1}};}}

\begin{document}

\title{\hfill ~\\[-15mm] \textbf{\large Texture-zero patterns of lepton mass matrices from modular symmetry}}

\author{
Gui-Jun Ding$^{1}$\footnote{E-mail: {\tt
dinggj@ustc.edu.cn}},  \
F. R. Joaquim$^{2}$\footnote{E-mail: {\tt filipe.joaquim@tecnico.ulisboa.pt}}, \
Jun-Nan Lu$^{1}$\footnote{E-mail: {\tt
hitman@mail.ustc.edu.cn}}
\\*[20pt]
\centerline{
\begin{minipage}{\linewidth}
\begin{center}
$^1${\it \normalsize Department of Modern Physics, University of Science and Technology of China, \\ Hefei, Anhui 230026, China}\\[2mm]
$^2${\it \normalsize
Departamento de F\'{\i}sica and CFTP, Instituto Superior T\'ecnico, Universidade de Lisboa, Lisboa, Portugal}
\end{center}
\end{minipage}}
\date{}
\\[10mm]}
\maketitle
\thispagestyle{empty}

\begin{abstract}

Texture zeros in fermion mass matrices have been widely considered in tackling the Standard Model flavour puzzle. In this work, we perform a systematic analysis of texture zeros in lepton mass matrices in the framework of $\Gamma_{3}'\cong T'$ modular symmetry. Assuming that the lepton fields transform as irreducible representations of $T'$, we obtain all possible texture-zero patterns for both charged-lepton and neutrino mass matrices which can be achieved from $T'$ modular symmetry. We provide representative models for the phenomenologically-viable textures which can accommodate the experimental data. The predictions for lepton mixing angles, CP-violating phases, light neutrino masses and effective neutrino mass relevant for neutrinoless doble beta decay, are discussed. We find that the minimal viable lepton model depends on only $7$ real free parameters including the modulus $\tau$ (the corresponding charged-lepton mass matrix contains $4$ vanishing entries, and the neutrino mass matrix has $1$ texture zero). Finally, we study in detail three benchmark models, one for each neutrino mass generation mechanism considered (Dirac, Majorana via Weinberg operator and Majorana via minimal type-I seesaw mechanism).

\end{abstract}
\newpage

\section{Introduction}

Understanding the fermion flavour pattern is still one of the most challenging and outstanding problems in particle physics. In the last decades, the quark mass and mixing parameters have been precisely measured, while neutrino oscillation experiments have firmly established the existence of neutrino masses and lepton mixing. Among all aspects related to the {\em flavour problem}, perhaps the most intriguing one is the huge hierarchy among the quark and lepton masses. Namely, the lightest neutrino mass is (at most) of $\mathcal{O({\rm 1\, eV})}$, while the heaviest (top) quark mass is over eleven orders of magnitude heavier. Moreover, the quark and lepton mixing patterns turn out to be drastically different: the Cabibbo-Kobayashi-Maskawa (CKM) mixing matrix in quark sector is close to the identity with small mixing angles, while the leptonic Pontecorvo-Maki-Nakagawa-Sakata (PMNS) matrix features two large (atmospheric and solar) mixing angles, and the small (reactor) one. In the Standard Model (SM), the fermion mass matrices are determined by the Yukawa couplings, which are arbitrary complex numbers, unconstrained by the SM gauge symmetry. The fact that the number of free parameters is much larger than the number of physical observables makes the SM unpredictive in the flavour sector. 

With the purpose of finding a solution for the flavour puzzle, several approaches have been developed. One of the ways of reducing the number of free parameters in fermion mass matrices, is to assume some of their entries to be vanishing~\cite{Fritzsch:1977za,Weinberg:1977hb,Wilczek:1977uh}, i.e. to assume what are commonly known as texture-zero {\em ansätze}. A typical example is the Fritzsch-type quark mass matrices which can relate the Cabibbo angle $\theta_{\rm C}$ to the ratio between the down and strange quark masses~\cite{Fritzsch:1977vd,Fritzsch:1979zq}. The phenomenology of texture zeros in both quark and lepton sectors have been widely studied in literature -- see Refs.~\cite{Fritzsch:1999ee,Gupta:2012fsl} for reviews. Systematical and complete analyses of all possible texture zeros have been carried out for both quark~\cite{Ludl:2015lta} and lepton mass matrices~\cite{Ludl:2014axa}. It is found that the predictivity of pure texture zero models is quite weak and the predictive mass matrices need relations among the non-zero matrix elements.  The most straightforward way to impose vanishing Yukawa couplings is by enforcing them with Abelian flavour symmetries~\cite{Grimus:2004hf,GonzalezFelipe:2014zjk,Camara:2020efq}. In such case, the non-zero entries of the fermion mass matrices are uncorrelated since the three generations of matter fields are not linked by the Abelian symmetry group. 

Recently, modular symmetries have been  proposed to address the flavour puzzle~\cite{Feruglio:2017spp}. In this approach, the Yukawa couplings are level-$N$ modular forms, which are holomorphic functions of a single complex scalar field -- the modulus $\tau$ -- and transform nontrivially under the action of the modular group. The matter fields are usually assumed to be in irreducible representations of the modular group. Models with modular flavour symmetries can be quite predictive, in the sense that fermion masses and mixing parameters depend on few inputs. The phenomenology of modular invariance has been widely studied, and a plethora of modular-invariant models for lepton masses and mixing have been constructed by using the inhomogeneous finite modular group $\Gamma_N$ for $\Gamma_{2}\cong S_{3}$~\cite{Kobayashi:2018vbk,Kobayashi:2018wkl,Kobayashi:2019rzp,Okada:2019xqk}, $\Gamma_{3}\cong A_{4}$~\cite{Feruglio:2017spp,Criado:2018thu,Kobayashi:2018vbk,Kobayashi:2018scp,deAnda:2018ecu,Okada:2018yrn,Kobayashi:2018wkl,Novichkov:2018yse,Nomura:2019jxj,Okada:2019uoy,Ding:2019zxk,Kobayashi:2019xvz,Asaka:2019vev,Gui-JunDing:2019wap,Zhang:2019ngf,King:2020qaj,Ding:2020yen,Okada:2020rjb,Asaka:2020tmo,Okada:2020ukr,
Okada:2020brs,Yao:2020qyy,Feruglio:2021dte,Chen:2021zty,Ding:2021eva,Ding:2022bzs,Gunji:2022xig}, $\Gamma_{4}\cong S_{4}$~\cite{Penedo:2018nmg,Novichkov:2018ovf,deMedeirosVarzielas:2019cyj,Kobayashi:2019mna,King:2019vhv,Criado:2019tzk,Wang:2019ovr,Gui-JunDing:2019wap,Qu:2021jdy}, $\Gamma_{5}\cong A_{5}$~\cite{Novichkov:2018nkm,Ding:2019xna,Criado:2019tzk} and $\Gamma_{7}\cong PSL(2,Z_{7})$~\cite{Ding:2020msi}, the homogeneous finite modular group $\Gamma'_N$ for $\Gamma'_3\cong T'$~\cite{Liu:2019khw,Lu:2019vgm,Okada:2022kee}, $\Gamma'_4\cong S'_4$~\cite{Novichkov:2020eep,Liu:2020akv,Ding:2022nzn}, $\Gamma'_5\cong A'_5$~\cite{Wang:2020lxk,Yao:2020zml} and $\Gamma'_6$~\cite{Li:2021buv}, and the finite metaplectic group $\widetilde{\Gamma}_N$~\cite{Liu:2020msy,Yao:2020zml}. The most general finite modular groups beyond $\Gamma_N$ and $\Gamma'_N$ are discussed in~\cite{Liu:2021gwa} from the view of vector-valued modular form.

Texture-zero patterns for fermion mass matrices can naturally arise from modular symmetries. In comparison to the models based on Abelian flavour symmetries, the texture-zero models relying on modular invariance are more predictive since the nonzero entries of the mass matrices are related by the modular symmetry. In Ref.~\cite{Lu:2019vgm}, texture zeros in quark mass matrices from $T'$ modular group symmetry have been investigated . In the same token, the specific case of texture zeros of quark mass matrices with nearest neighbor interaction has been discussed in~\cite{Kikuchi:2022svo}. Regarding the lepton sector, two-zero textures of the Majorana neutrino mass matrix with diagonal charged-lepton mass matrix have been studied in the framework of the $A_{4}$ modular group~\cite{Zhang:2019ngf}. However, all matter fields were assigned to $A_4$ singlets. As a consequence, the modular forms appearing in the lepton mass matrices can be absorbed into the Yukawa coupling constants. Since the triplet representation of $A_4$ is not used, the effect of $A_{4}$ modular symmetry is equivalent to an Abelian flavour symmetry and, thus, the correlations among nonzero entries of lepton mass matrices are lost. Thus, a more general and systematic study of lepton texture-zero in the context of modular symmetries is needed.

In this work, we will extend the analysis of~\cite{Lu:2019vgm} to the lepton sector, following a systematic approach in seeking the modular-symmetry realisations of lepton texture zeros and studying their phenomenology. We shall use the modular group $\Gamma'_3\cong T'$ as cornerstone. In contrast with Ref.~\cite{Zhang:2019ngf}, we will not impose any specific basis for charged leptons, and all possible ways of assigning the lepton fields to irreducible representations of the $T'$ modular group will be explored. Since the particle nature of neutrinos is still unclear, we will consider both Dirac and Majorana neutrino masses. For the latter, we analyse two kinds of mass generation mechanisms: the effective Weinberg operator and the type-I seesaw mechanism. To the combination of a given pattern for the charged-lepton and neutrino mass matrix realised with a specific representation and modular weight assignment we call a {\em texture-zero lepton model} (TZLM).

This paper is organised as follows: in Section~\ref{sec:modular_symmetry_N3}, we briefly review some key aspects about modular symmetry and modular forms of level 3. In Section~\ref{sec:charged_lepton_sector}, we systematically analyse the texture-zero patterns of the charged-lepton mass matrix stemming from $T'$ modular symmetry
 by considering all possible representation assignments of matter fields. This procedure is repeated for the neutrino mass matrices in Section~\ref{sec:neutrino_sector}. The combination of charged-lepton and neutrino patterns is carried out in Section~\ref{sec:lepton_models}, and the phenomenologically viable pairs are identified in Section~\ref{sec:numerical_analysis} by means of a complete numerical analysis in view of latest neutrino experimental data. Among all viable cases, we focus on three benchmark TZLMs which will be studied in Section~\ref{sec:benchmark_models}. Finally, we draw ourconclusions in Section~\ref{sec:conclusion}. Group-theoretical aspects of the modular group $T'$ are covered in Appendix~\ref{app:Tp_group}. In particular, we provide the level 3 modular forms of weight 2, 3, 4, 5 and 6 in Appendix~\ref{subsec:modular_forms_higher}. The details of the representative models for the viable textures of lepton mass matrices and the corresponding predictions for lepton observables are summarized in Appendix~\ref{sec:appendix_C}.

\section{Modular symmetry and modular forms of level 3}
\label{sec:modular_symmetry_N3}

The modular group $\Gamma=SL(2, \mathbb{Z})$ is the special linear group of two-dimensional matrices with integer entries defined as:
\begin{equation}
\Gamma=SL(2, \mathbb{Z})=\left\{\begin{pmatrix}
a ~&~ b\\
c ~&~ d
\end{pmatrix}\bigg| a,b, c,d \in \mathbb{Z}, ad-bc=1
\right\}\,.
\end{equation}
The group $\Gamma$ has infinite elements and it can be generated by two generators $S$ and $T$, namely
\begin{equation}
S=\begin{pmatrix}
0  ~&~ 1 \\
-1 ~&~ 0
\end{pmatrix},~~~~T=\begin{pmatrix}
1 ~&~ 1 \\
0  ~&~ 1
\end{pmatrix}\,,
\end{equation}
which satisfy the relations
\begin{equation}
\label{eq:ST-gen}S^4=(ST)^3=\mathbb{1}_2,~~~S^2T=TS^2\,,
\end{equation}
where $\mathbb{1}_2$ denotes the $2\times2$ unit matrix and $S^2=-\mathbb{1}_2$. Notice that Eq.~\eqref{eq:ST-gen} implies $(TS)^3=\mathbb{1}_2$. The modular group $\Gamma$ acts on the complex modulus $\tau$ with fractional linear transformations:
\begin{equation}
\tau\rightarrow\gamma\tau=\frac{a\tau+b}{c\tau+d},~~\text{with}~~\mathrm{Im}\tau>0,~~~\gamma=\begin{pmatrix}
a ~&~ b\\
c ~&~d
\end{pmatrix}\in\Gamma\,,
\end{equation}
being the action of $S$ and $T$ on $\tau$
\begin{equation}
S:\tau\rightarrow-\frac{1}{\tau},~~~~~~T:\tau\rightarrow\tau+1\,.
\end{equation}
We see that the modulus $\tau$ transforms in the same way under the action of $\gamma$ and $S^2\gamma=-\gamma$. Hence, the group of fractional linear transformations is isomorphic to the projective matrix group $PSL(2, \mathbb{Z})\cong SL(2, \mathbb{Z})/\{\mathbb{1}_2, -\mathbb{1}_2\}$. Moreover, the modular transformation of a set of chiral supermultiplets $\Phi_{i}$ under the action of $\gamma$ is given by
\begin{equation}
\Phi_{i}\stackrel{\gamma}{\longmapsto} (c\tau+d)^{-k_{\Phi}}\rho_{ij}(\gamma)\Phi_{j} \,,
\end{equation}
where $k_{\Phi}\in \mathbb{Z}$ is the modular weight of the superfield multiplet $\Phi_i$, and $\rho$ is the unitary representation of $\Gamma$ with finite image. For a modular flavour symmetry~\cite{Feruglio:2017spp}, it is assumed that the representation matrix $\rho(\gamma)$ is a unit matrix when $\gamma$ belongs to the principal congruence subgroup of level $N$\footnote{In fact, one can start from any irreducible representation $\rho$ of $SL(2, \mathbb{Z})$ with finite image~\cite{Liu:2021gwa}, and here $\rho(\Gamma(N))=1$ is a particular choice.}
\begin{equation}
\Gamma(N)=\left\{\left(\begin{array}{cc}
a  ~&~ b \\
c  ~&~ d
\end{array}\right)\in SL(2, \mathbb{Z}),~~ \left(\begin{array}{cc}
a  ~&~ b \\
c  ~&~ d
\end{array}\right)=\left(
\begin{array}{cc}
1 ~&~0 \\
0  ~&~ 1
\end{array}
\right)\,(\mathrm{mod}~N)
\right\}\,,
\end{equation}
which is an infinite normal subgroup of $SL(2, \mathbb{Z})$. We see $\Gamma(1)= SL(2, \mathbb{Z})$ and $T^{N}\in \Gamma(N)$. The fundamental theorem of homomorphisms implies that $\rho$ is effectively a representation of the quotient group $\Gamma'_N\equiv SL(2, \mathbb{Z})/\Gamma(N)\cong SL(2, Z_N)$ which is called homogeneous finite modular group. $\Gamma'_N$ can be obtained by further imposing the condition $T^N=1$ besides those in Eq.~\eqref{eq:ST-gen}:
\begin{equation}
\Gamma'_N=\langle S,T\big|S^4=(ST)^3=T^N=1,~S^2T=TS^2\rangle\,.
\end{equation}
Additional relations are necessary in order to render the group $\Gamma'_{N}$ finite for $N\geq6$~\cite{deAdelhartToorop:2011re}. The $\Gamma'_N$ group can also be expressed in terms of three generators $S$, $T$ and $R$ as follow
\begin{equation}
\Gamma'_N=\langle S,T, R\big|S^{2}=R, (ST)^3=T^N=R^{2}=1,~RT=TR\rangle\,.
\end{equation}
If $\rho$ cannot distinguish between $\gamma$ and $-\gamma$ with $\rho(S^2)=1$, $\rho$ would be the representation of the inhomogeneous finite modular group $\Gamma_N\equiv SL(2, \mathbb{Z})/\pm \Gamma(N)\cong \Gamma'_N/Z^{S^2}_2$, where $Z^{S^2}_2$ is the order-two cyclic group generated by $S^2$. For $N\leq 5$, the defining relations of $\Gamma_N$ are
\begin{equation}
\Gamma_N=\langle S,T\big|S^2=(ST)^3=T^N=1\rangle\,.
\end{equation}
Note that $\Gamma'_2=\Gamma_2$ since $S^2\in\Gamma(2)$, while $\Gamma'_N$ is the double cover of $\Gamma_N$ for $N\geq 3$ due to $S^2\notin \Gamma(N)$. It is notable that the groups $\Gamma_N$ for $N=2, 3, 4, 5$ are isomorphic to the permutation groups $S_3$, $A_4$, $S_4$ and $A_5$ respectively.

Implementation of modular flavour symmetries requires modular forms $Y(\tau)$ of weight $k$ and level $N$. $Y(\tau)$ is a holomorphic function of $\tau$ with a well-defined transformation property under $\Gamma(N)$:
\begin{equation}
Y\left(\gamma\tau\right)=(c\tau+d)^kY(\tau),~~~~\gamma=\left(\begin{array}{cc}
a  ~&~  b \\
c  ~&~  d
\end{array}
\right)\in\Gamma(N)\,,
\end{equation}
where $k$ is a generic non-negative integer\footnote{The rational weight modular forms have been studied in~\cite{Liu:2020msy,Yao:2020zml}}. The modular forms of weight $k$ and level $N$ span a linear space $\mathcal{M}_{k}(\Gamma(N))$ of finite dimension, they are invariant under $\Gamma(N)$ but transform under the quotient group $\Gamma'_{N}$. It is always possible to choose a basis in the linear space $\mathcal{M}_{k}(\Gamma(N))$ such that the modular forms can
be arranged into some modular multiplets $Y^{(k)}_{\mathbf{r}}=\left(Y_1(\tau), Y_2(\tau), \ldots\right)^T$ which transform as irreducible representation $\mathbf{r}$ of the finite modular group $\Gamma'_N$ or $\Gamma_N$~\cite{Feruglio:2017spp,Liu:2019khw}:
\begin{equation}
\label{eq:Mod-def}Y^{(k)}_{\mathbf{r}}(\gamma\tau)=(c\tau+d)^k\rho_{\mathbf{r}}(\gamma)Y^{(k)}_{\mathbf{r}}(\tau)~~~\mathrm{for}~~\forall~\gamma\in SL(2,\mathbb{Z})\,,
\end{equation}
where $\gamma$ is the representative element of the coset
$\gamma\Gamma(N)$ in $\Gamma'_N$, and $\rho_{\mathbf{r}}(\gamma)$ is the representation matrix of the element $\gamma$ in the irreducible representation $\mathbf{r}$.

The superpotential $\mathcal{W}(\Phi_I, \tau)$ can be expanded in power series of the supermultiplets $\Phi_I$,
\begin{equation}
\mathcal{W}(\Phi_I, \tau)=\sum_{n}Y_{I_1\ldots I_n}(\tau)\Phi_{I_1}\ldots\Phi_{I_n}\,,
\end{equation}
where the sum is taken over all possible combinations of the fields $\{I_1, \ldots, I_n \}$. The supermultiplet $\Phi_{I_i}$ is assumed to transform in the representation $\rho_{I_i}$ of $\Gamma'_N$, being $-k_{I_i}$ its modular weight. Modular invariance requires $\mathcal{W}(\Phi_I, \tau)$ to be invariant under the finite modular group $\Gamma'_N$ and, thus, the total weight of each of its terms must vanish. As a consequence, $Y_{I_1\ldots I_n}(\tau)$ should be a modular multiplet of weight $k_Y$ transforming in the representation $\rho_{Y}$ of $\Gamma'_N$, i.e.
\begin{equation}
Y_{I_1\ldots I_n}(\tau)\rightarrow Y_{I_1\ldots I_n}(\gamma\tau)=(c\tau+d)^{k_Y}\rho_{Y}(\gamma)Y_{I_1\ldots I_n}(\tau)\,,
\end{equation}
with $k_Y$ and $\rho_{Y}$ satisfying
\begin{equation}
k_Y=k_{I_1}+\ldots+k_{I_n},~~~~\rho_Y\otimes \rho_{I_1}\otimes\ldots\otimes \rho_{I_n}\ni \mathbf{1}\,,
\end{equation}
where $\mathbf{1}$ denotes the trivial singlet representation of $\Gamma'_N$.

\subsection{\label{subsec:modular_forms_1}Weight-1 modular forms of level $N=3$ }

The linear space $\mathcal{M}_{k}(\Gamma(3))$ spanned by level-3 and weight-$k$ modular forms has dimension $k+1$, and can be explicitly constructed by using the Dedekind eta function $\eta(\tau)$:
\begin{equation}
\mathcal{M}_{k}(\Gamma(3))=\sum_{m+n=k;\\ m,n\ge0} c_{mn} \frac{\eta^{3m}(3\tau)\eta^{3n}(\tau /3 )}{\eta^{m+n}(\tau)}=\sum_{m+n=k;\, m,n\ge0}c_{mn} \left[\frac{\eta^3(3\tau)}{\eta(\tau)}\right]^m\left[\frac{\eta^{3}(\tau/3)}{\eta(\tau)}\right]^n\,,
\end{equation}
where $c_{mn}$ are general complex coefficients and
\begin{equation}
\eta(\tau)=q^{1/24}\prod^{\infty}_{n=1}(1-q^{n}),~~~q=e^{2\pi i\tau}\,.
\end{equation}
Hence, $\mathcal{M}_{k}(\Gamma(3))$ can be generated by polynomials of degree $k$ in the two functions $u_1(\tau)$ and $u_2(\tau)$:
\begin{equation}
u_1(\tau)=\frac{\eta^{3}(3\tau)}{\eta(\tau)},~~~ u_{2}(\tau)=\frac{\eta^{3}(\tau / 3)}{\eta(\tau)}\,,
\end{equation}
which are the two linearly independent weight-1 modular forms at level 3. The functions $u_1(\tau)$ and $u_2(\tau)$ transform under the modular generators $S$ and $T$ as
\begin{eqnarray}
\nonumber&&u_1(\tau)\stackrel{T}{\longmapsto} e^{2\pi i /3}u_1(\tau),\qquad u_{2}(\tau)\stackrel{T}{\longmapsto} 3\sqrt{3} e^{-i\pi/6}u_1 + u_2 \,,\\
&&u_1(\tau) \stackrel{S}{\longmapsto} 3^{-3/2}(-i\tau)u_2(\tau),\qquad u_{2}(\tau)\stackrel{S}{\longmapsto} 3^{3/2}(-i\tau)u_1(\tau) \,.
\end{eqnarray}
and can be arranged in a $T'$ doublet $Y^{(1)}_{\mathbf{2}}$ up to the automorphy factor $c\tau+d$~\cite{Liu:2019khw}. In the group representation basis given by Eq.~\eqref{eq:irr-Tp}, we find that the modular form doublet $Y^{(1)}_{\mathbf{2}}(\tau)$ is defined as
\begin{equation}
\label{eq:Y1-Y2}
Y^{(1)}_{\mathbf{2}}(\tau)=\begin{pmatrix}
Y_1(\tau) \\
Y_2(\tau)
\end{pmatrix}\,,
\end{equation}
with
\begin{equation}
Y_1(\tau)=\sqrt{2}\,u_1(\tau),~~~~ Y_2(\tau)=-u_1(\tau)-\frac{1}{3}u_2(\tau)\,.
\end{equation}
One can check that $Y^{(1)}_{\mathbf{2}}(\tau)$ transforms under the modular generator $S$ and $T$ as expected, namely
\begin{equation}
Y^{(1)}_{\mathbf{2}}(-1/\tau)=-\tau\rho_\mathbf{2}(S)Y^{(1)}_{\mathbf{2}}(\tau),\qquad Y^{(1)}_{\mathbf{2}}(\tau+1)=\rho_\mathbf{2}(T)Y^{(1)}_{\mathbf{2}}(\tau)\,,
\end{equation}
where the representation matrices $\rho_{\mathbf{2}}(S)$ and $\rho_{\mathbf{2}}(T)$ are given in Appendix~\ref{app:Tp_group}. The $q$-expansion of the weight-1 modular forms $Y_{1,2}(\tau)$ reads
\begin{eqnarray}
\nonumber Y_1(\tau)&=& \sqrt{2}\,  q^{1/3} \left(1 + q + 2 q^2 + 2 q^4 + q^5 + 2q^6+q^8+2q^9+2q^{10}+2q^{12} +\ldots\right) ,\\
\label{eq:q_exp_weigh1}Y_2(\tau)&=& -1/3 -2 q - 2 q^3 - 2 q^4- 4 q^7 - 2 q^9 -2q^{12}-4q^{13}+\ldots\,.
\label{q-expansion}
\end{eqnarray}
Notice that $Y_1(\tau)$ and $Y_2(\tau)$ are algebraically independent. The level-3 modular form of weight $k\geq2$ can be expressed as polynomials of degree $k$ in $Y_1(\tau)$ and $Y_2(\tau)$, as shown explicitly in Appendix~\ref{subsec:modular_forms_higher}.

\section{\label{sec:charged_lepton_sector}Texture zeros in the charged-lepton mass matrix}

As antecipated in the Introduction, in this work we will carry out a systematic analysis of texture zeros in the lepton mass matrices stemming from a $\Gamma_{4}\cong T'$ modular symmetry. The matter fields are assumed to transform as irreducible representations of $T'$ which has three one-dimensional irreducible representations denoted by $\mathbf{1}$, $\mathbf{1}'$, $\mathbf{1}''$, three two-dimensional irreducible representations denoted by $\mathbf{2}$, $\mathbf{2}'$, $\mathbf{2}''$ and a three-dimensional irreducible representation denoted by $\mathbf{3}$. Thus, the lepton fields can be assigned to any of these representations. 

The transformation assignments of the three generations of left-haned (LH) leptons and the right-handed (RH) charged leptons can be classified into the following cases, i.e.,
\begin{eqnarray} \label{eq:assign_L}
\hskip-0.3in &&L\equiv\begin{pmatrix}
L_1 \\
L_2 \\
L_3 \\
\end{pmatrix}\sim\mathbf{3}\,,~~\text{or}~~L_D\equiv\begin{pmatrix}
L_1 \\
L_2
\end{pmatrix}\sim\mathbf{2}^{i},\quad L_3\sim \mathbf{1}^{j}\,,~~\text{or}~~
L_a\sim\mathbf{1}^{j_{a}},~~\textrm{with}~~a=1, 2, 3\,,\\
\hskip-0.3in &&E^{c}\equiv\begin{pmatrix}
e^{c} \\
\mu^{c} \\
\tau^{c} \\
\end{pmatrix}\sim\mathbf{3}\,,~~\text{or}~~E^{c}_D\equiv\begin{pmatrix}
e^{c} \\
\mu^{c}
\end{pmatrix}\sim\mathbf{2}^{k},\quad \tau^{c}\sim \mathbf{1}^{l}\,,~~\text{or}~~
E^{c}_a\sim\mathbf{1}^{l_{\alpha}},~~\textrm{with}~~\alpha=1, 2, 3\,.
\end{eqnarray}
where $i,j,k,l,j_{1,2,3},l_{1,2,3}=0, 1, 2$ with $\mathbf{1}\equiv \mathbf{1}^{0}$, $\mathbf{1}^{'}\equiv \mathbf{1}^{1}$, $\mathbf{1}^{''}\equiv \mathbf{1}^{2}$ ($\mathbf{2}\equiv \mathbf{2}^{0}$, $\mathbf{2}^{'}\equiv \mathbf{2}^{1}$, $\mathbf{2}^{''}\equiv \mathbf{2}^{2}$) for the singlet (doublet) representations. $E^{c}_{a}$ stands for $e^{c},\mu^{c},\tau^{c}$ when $a=1,2,3$, respectively.
The above assignment for the three lepton generations is not unique in the sense that permutations among them can the be considered. This amounts to multiplying the charged-lepton mass matrix on the left and/or right by permutation matrices. However, this does not change lepton mixing and, at the end, we can identify nine classes of charged-lepton mass matrices according to how the lepton fields transform under $T'$. 

In Table~\ref{tab:classes_charged_lepton_mass_matrix}, we summarize all possible texture-zero patterns for the charged-lepton mass matrix that can be obtained with $T'$ modular group. In the following, we discuss how those structures can be achieved by properly assigning the representations and weights of the lepton fields under the $T'$ modular group. We should emphasize that the matrix entries marked with $\times$ in the mass matrices considered in this work are not arbitrary as in usual texture-zero frameworks based on Abelian symmetries. As already mentioned in the Introduction, this is due to the fact that extra relations among the non-zero entries appear as a result of the $T'$ modular symmetry. Therefore, compatibility of a given texture with arbitrary non-vanishing entries does not guarantee its viability in the present framework.

We will now discuss the 9 distinct assignments of lepton fields, and require the charged-lepton mass matrix to have rank three in order to accommodate the three nonzero masses of the electron, muon and tau.

\begin{table}[t!]
 \centering
 \begin{tabular}{|l|} \hline\hline
$\mathcal{C}_{1}^{(1)}:\left( \begin{array}{ccc} 0 & \times & \times \\
                 \times & \times & \times \\
                                \times & \times & \times \end{array} \right)$ \\ \hline
$\mathcal{C}_{2}^{(1)}:\left( \begin{array}{ccc} 0 & \times & \times \\
                 \times & 0 & \times \\
                 \times & \times & \times \end{array} \right)\,,~~
\mathcal{C}_{2}^{(2)}:\left( \begin{array}{ccc} \times & \times & \times \\
                 \times & \times & \times \\
                 0 & 0 & \times \end{array} \right)\,,~~
\mathcal{C}_{2}^{(3)}:\left( \begin{array}{ccc} \times & \times & 0 \\
                 \times & \times & 0 \\
                               \times & \times & \times \end{array} \right)$ \\ \hline
   $\mathcal{C}_{3}^{(1)}: \left( \begin{array}{ccc} \times & \times & 0 \\
                 \times & \times & \times \\
                 0 & 0 & \times \end{array} \right)\,,~~
\mathcal{C}_{3}^{(2)}:\left( \begin{array}{ccc} \times & 0 & \times \\
                 \times & 0 & \times \\
                 0 & \times & \times \end{array} \right)\,,~~
\mathcal{C}_{3}^{(3)}:\left( \begin{array}{ccc} 0 & \times & \times \\
                 \times & 0 & \times \\
                               \times & \times & 0 \end{array} \right)$ \\ \hline
 $\mathcal{C}_{4}^{(1)}:\left( \begin{array}{ccc} \times & \times & 0 \\
                 \times & \times & 0 \\
                 0 & 0 & \times \end{array} \right)\,,~~
\mathcal{C}_{4}^{(2)}:\left( \begin{array}{ccc} 0 & \times & 0 \\
                 \times & 0 & 0 \\
                 \times & \times & \times \end{array} \right)\,,~~
\mathcal{C}_{4}^{(3)}:\left( \begin{array}{ccc} 0 & \times & \times \\
                 \times & 0 & \times \\
                               0 & 0 & \times \end{array} \right)$  \\ \hline
 $\mathcal{C}_{6}^{(1)}:\left( \begin{array}{ccc} 0 & \times & 0 \\
                 \times & 0 & 0 \\
                 0 & 0 & \times \end{array} \right)$ \\
   \hline \hline
\end{tabular}
\caption{\label{tab:classes_charged_lepton_mass_matrix} Texture-zero classification for the (rank-3) charged-lepton mass matrices which can be realised from $T'$ modular symmetry, up to row and column permutations.}
\end{table}

\begin{itemize}
\item{$L\equiv (L_{1},L_{2},L_{3})^{T}\sim \mathbf{3},~~E^{c}\equiv (e^{c},\mu^{c},\tau^{c})^{T}\sim \mathbf{3}$}

In this case, the most general effective Yukawa terms for the charged leptons in the superpotential are given by
\begin{equation}
\label{eq:W_E_1}
\mathcal{W}_{E}=\sum_{\mathbf{r}, a}g^{E}_{\mathbf{r}, a}\left[E^{c}LY^{(k_{L}+k_{E^{c}})}_{\mathbf{r}{a}}\right]_{\mathbf{1}}H_{d}\,,
\end{equation}
where $g^{E}_{\mathbf{r}, a}$ are coupling constants, and $Y^{(k)}_{\mathbf{r} a}$ stands for the modular form of weight $k$ and representation $\mathbf{r}$, with $a$ possibly labelling linearly independent multiplets of the same type. Notice that all possible contractions into the $T'$ singlet should be considered. As shown in Table~\ref{Tab:Level3_MM}, the allowed representation $\mathbf{r}$ of modular forms in Eq.~\eqref{eq:W_E_1} depends on the modular weights $k_{L}+k_{E^{c}}$. From Eq.~\eqref{eq:W_E_1}, we can read out the expressions of the elements of charged lepton mass matrix,
\begin{eqnarray} \nonumber
\hskip-0.3in (M_{E})_{\alpha\beta}&=&v_{d}\sum_{a,b,c,d} \left\{\left[\, g^{E}_{\mathbf{3}_{S},a}(3\delta_{\alpha\beta}-1)+ g^{E}_{\mathbf{3}_{A},a}(1-\delta_{\alpha\beta})(-1)^{(\alpha-\beta)|3}\,\right]Y^{(k_{L}+k_{E^{c}})}_{\mathbf{3}a,3-(\alpha+\beta)|3}\right.\\
&~&\hskip-0.3in\left.+g^{E}_{\mathbf{1},b}\delta_{2,(\alpha+\beta)|3}Y^{(k_{L}+k_{E^{c}})}_{\mathbf{1}b}+g^{E}_{\mathbf{1}',c}\delta_{1,(\alpha+\beta)|3}Y^{(k_{L}+k_{E^{c}})}_{\mathbf{1}'c}+ g^{E}_{\mathbf{1}'',d}\delta_{0,(\alpha+\beta)|3}Y^{(k_{L}+k_{E^{c}})}_{\mathbf{1}''d}\right\}\,,
\end{eqnarray}
where $v_{d}=\langle H^{0}_d\rangle$ is the vacuum expectation value (VEV) of the Higgs field $H_{d}$ and $a|3\equiv (a~\text{mod}~3)$. From now on, $Y^{(k_{L}+k_{E^{c}})}_{\mathbf{r}_{a},j}$ denotes the $j^{\rm th}$ component of the modular form $Y^{(k_{L}+k_{E^{c}})}_{\mathbf{r}_{a}}$ in the vector notation of Eq.~\eqref{eq:Y1-Y2} and
Eqs.~\eqref{eq:modfvec1}-\eqref{eq:modfvec2}. The charged-lepton mass matrix $M_E$ is defined in the right-left basis with $E^{c}_\alpha(M_E)_{\alpha\beta}L_\beta$. We find that only texture $\mathcal{C}_{6}^{(1)}$ can be reproduced, requiring total weight $k_{L}+k_{E^{c}}=0$. None of the other cases can be obtained for a generic value of the modulus $\tau$ when the three lepton doublets and singlets are in triplet representations of $T^\prime$. It turns out that the modular realisation of $\mathcal{C}_{6}^{(1)}$ is not viable since, in this case, modular symmetry imposes $(M_E)_{12}=(M_E)_{21}=(M_E)_{33}$, leading to a fully degenerate charged-lepton mass spectrum.

\item{$L\equiv (L_{1},L_{2},L_{3})^{T}\sim \mathbf{3},~~E^{c}_{D}\equiv (e^{c},\mu^{c})^{T}\sim \mathbf{2}^{k},~~E_{3}^{c}\equiv \tau^{c}\sim \mathbf{1}^{l}$} 

In this case, the superpotential terms relevant for charged-lepton mass generation are:
\begin{equation}\label{eq:W_E_2}
\mathcal{W}_{E}=\sum_{\mathbf{r},a,b}g^{E}_{\mathbf{r},a}\left[E_{D}^{c}LY^{(k_{L}+k_{E_{D}^{c}})}_{\mathbf{r}_{a}}\right]_{\mathbf{1}}H_{d}+g^{E}_{\mathbf{3},b}\left[E_{3}^{c}LY^{(k_{L}+k_{E_{3}^{c}})}_{\mathbf{3}b}\right]_{\mathbf{1}}H_{d}\,.
\end{equation}
where $g^{E}_{\mathbf{r},a}$, $g^{E}_{\mathbf{3},b}$ are coupling constants. Modular invariance requires the representation $\mathbf{r}$ to be $\mathbf{2},\mathbf{2'},\mathbf{2''}$, for which the corresponding modular forms of weight $k=k_{L}+k_{E_{D}^{c}}$ are given in Table~\ref{Tab:Level3_MM}. The elements in the first two rows of $M_{E}$ read
\begin{eqnarray}\nonumber
\Scale[0.8]{(M_{E})_{\alpha\beta}}&=&\Scale[0.8]{v_{d}\sum_{a,b,c} \Bigg\{ g^{E}_{\mathbf{2}^{2-k},a}\left(1-\delta_{1,\beta-\alpha}\right) \left[ \delta_{1,(\alpha-\beta)|3}Y^{(k_{E^{c}_{D}}+k_{L})}_{\mathbf{2}^{2-k}a,1}(\delta_{\alpha,2}-\sqrt{2}\delta_{\alpha,1}) + \delta_{\alpha\beta}Y^{(k_{E^{c}_{D}}+k_{L})}_{\mathbf{2}^{2-k}a,2}(\delta_{\alpha,1}+\sqrt{2}\delta_{\alpha,2})\right]} \\ \nonumber
&~&\hskip-0.1in \Scale[0.8]{+ g^{E}_{\mathbf{2}^{(1-k)|3},b}\left(1-\delta_{1,(\alpha-\beta)|3}\right) \left[ \delta_{\alpha\beta}Y^{(k_{E^{c}_{D}}+k_{L})}_{\mathbf{2}^{(1-k)|3}b,1}(\delta_{\alpha,2}-\sqrt{2}\delta_{\alpha,1}) + \delta_{1,\beta-\alpha}Y^{(k_{E^{c}_{D}}+k_{L})}_{\mathbf{2}^{(1-k)|3}b,2}(\delta_{\alpha,1}+\sqrt{2}\delta_{\alpha,2})\right]}\\ \label{eq:exp_U} &~&\hskip-0.1in \Scale[0.8]{+g^{E}_{\mathbf{2}^{(-k)|3},c}\left(1-\delta_{\alpha\beta}\right) \left[ \delta_{1,\beta-\alpha}Y^{(k_{E^{c}_{D}}+k_{L})}_{\mathbf{2}^{(-k)|3}c,1}(\delta_{\alpha,2}-\sqrt{2}\delta_{\alpha,1}) +\delta_{1,(\alpha-\beta)|3}Y^{(k_{E^{c}_{D}}+k_{L})}_{\mathbf{2}^{(-k)|3}c,2}(\delta_{\alpha,1}+\sqrt{2}\delta_{\alpha,2})\right] \Bigg\}\,.}
\end{eqnarray}
with $\alpha=1,2$ and $\beta=1,2,3$, while the elements in the third row are given by
\begin{equation}\label{eq:exp_R0}
  (M_{E})_{3\beta}=\sum_{b}g^{E}_{\mathbf{3}, b}v_{d}Y^{(k_{L}+k_{E_{3}^{c}})}_{\mathbf{3}b,3-(l+\beta+1)|3}\;,\;\beta=1,2,3.
\end{equation}
In this case, only the $\mathcal{C}_{2}^{(1)}$ texture-zero pattern can be realised by the modular symmetry with $k_{E^{c}_{D}}+k_{L}=1$, and there are no zero elements in $M_E$ for the remaining weights.

\item{$L\equiv (L_{1},L_{2},L_{3})^{T}\sim \mathbf{3},~~E^{c}_{\alpha}\sim \mathbf{1}^{l_{\alpha}}$}

With the LH leptons transforming as a triplet of $T'$, and the RH charged leptons transforming as singlets of $T'$, the Yukawa terms for charged leptons are
\begin{equation}
\label{eq:W_E_3}
  \mathcal{W}_{E}=\sum_{\alpha=1}^{3}\sum_{a}g^{E}_{\alpha \mathbf{3},a}\left[E_{\alpha}^{c}LY_{\mathbf{3}a}^{(k_{L}+k_{E^{c}_{\alpha}})}\right]_{\mathbf{1}}H_{d}\,,
\end{equation}
where $g^{E}_{\alpha \mathbf{3},a}$ are coupling constants. The matrix elements of $M_{E}$ are in this case:
\begin{equation}
\label{eq:general-ele-lepton} (M_E)_{\alpha\beta}=\sum_{a}g^{E}_{\alpha \mathbf{3},a}v_{d}Y_{\mathbf{3}a,3-(l_{\alpha}+\beta+1)|3}^{(k_{L}+k_{E^{c}_{\alpha}})}\,.
\end{equation}
Demanding that $M_{E}$ is of rank 3, no texture zeros can be obtained in this case.

\item{$L_{D}\equiv (L_{1},L_{2})^{T}\sim \mathbf{2}^{i},~~L_{3}\sim \mathbf{1}^{j},~~E^{c}\equiv (e^{c},\mu^{c},\tau^{c})^{T}\sim \mathbf{3}$}

This case can be related to the second case described above by exchanging the assignments of LH and RH lepton fields. Hence, one can obtain the corresponding $M_E$ transposing the mass matrix given in Eqs.~(\ref{eq:exp_U}) and \eqref{eq:exp_R0}). Consequently, only texture $\mathcal{C}_{2}^{(1)}$ can be realised for $k_{L_{D}}+k_{E^{c}}=1$ in this case.

\item{$L_{D}\equiv (L_{1},L_{2})^{T}\sim \mathbf{2}^{i},~~L_{3}\sim \mathbf{1}^{j},~~E_{D}^{c}\equiv (e^{c},\mu^{c})^{T}\sim \mathbf{2}^{k},~~E^{c}_{3}\sim \mathbf{1}^{l}$} 

In the case that both the left- and RH charged leptons transform as direct sums of one- and two-dimensional representations of $T'$, the superpotential $\mathcal{W}_{E}$ is given as 
 \begin{eqnarray}\nonumber\mathcal{W}_{E}&=&\sum_{\mathbf{r},a,b,c,d}g^{E}_{1\mathbf{r},a}\left[E_{D}^{c}L_{D}Y^{(k_{L_{D}}+k_{E_{D}^{c}})}_{\mathbf{r}a}\right]_{\mathbf{1}}H_{d}+g^{E}_{2\mathbf{r},b}\left[E_{D}^{c}L_{3}Y^{(k_{L_{3}}+k_{E_{D}^{c}})}_{\mathbf{r}b}\right]_{\mathbf{1}}H_{d}\\ \label{eq:W_E_5}
&~&~~+g^{E}_{3\mathbf{r},c}\left[E_{3}^{c}L_{D}Y^{(k_{L_{D}}+k_{E_{3}^{c}})}_{\mathbf{r}c}\right]_{\mathbf{1}}H_{d}+g^{E}_{4\mathbf{r},d}\left[E_{3}^{c}L_{3}Y^{(k_{L_{3}}+k_{E_{3}^{c}})}_{\mathbf{r}d}\right]_{\mathbf{1}}H_{d}\,.
  \end{eqnarray}
The corresponding charged-lepton mass matrix can be divided into four blocks which correspond to $2\times 2$, $2\times 1$, $1\times 2$ and $1\times 1$ sub-matrices. Using the Kronecker products and the CG coefficients of $T'$ given in Appendix~\ref{app:Tp_group}, we find the explicit forms of the four submatrices:
%
\begin{eqnarray}\nonumber
 \hskip-0.2in  (M_{E})_{\alpha\beta}&=&v_{d}\sum_{a,b,c,d}\Bigg\{ g^{E}_{1\mathbf{3},a} (\sqrt{2})^{\delta_{\alpha\beta}}(-1)^{\delta_{2,\alpha+\beta}}Y^{(k_{E^c_D}+k_{L_D})}_{\mathbf{3}a,3-(k+i-\alpha-\beta)|3} +(1-\delta_{\alpha\beta})(-1)^{(\alpha-\beta)|3}\delta_{3,\alpha+\beta}\\ \label{eq:exp_S}
    &~&\hskip-0.8in\left[ g^{E}_{1\mathbf{1},b}\delta_{2,(i+k)|3}Y^{(k_{E^c_D}+k_{L_D})}_{\mathbf{1}b}+g^{E}_{1\mathbf{1}',c}\delta_{1,(i+k)|3}Y^{(k_{E^c_D}+k_{L_D})}_{\mathbf{1}'c}+g^{E}_{1\mathbf{1}'',d}\delta_{0,(i+k)|3}Y^{(k_{E^c_D}+k_{L_D})}_{\mathbf{1}''d}\right]\Bigg\}\,,\\ \nonumber
  (M_{E})_{\alpha 3}&=&v_{d}\sum_{a,b,c}\Bigg[ g^{E}_{2\mathbf{2},a} \delta_{2,(k+j)|3}(-1)^{\alpha+1}Y^{(k_{E^{c}_{D}}+k_{L_{3}})}_{\mathbf{2}a,3-\alpha} + g^{E}_{2\mathbf{2'},b} \delta_{1,(k+j)|3}(-1)^{\alpha+1}Y^{(k_{E^{c}_{D}}+k_{L_{3}})}_{\mathbf{2}'b,3-\alpha}\\ \label{eq:exp_C}
                   &~&\qquad \quad +g^{E}_{2\mathbf{2}'',c} \delta_{0,(k+j)|3}(-1)^{\alpha+1}Y^{(k_{E^{c}_{D}}+k_{L_{3}})}_{\mathbf{2}''c,3-\alpha}\Bigg]\,,\\ \nonumber
   (M_{E})_{3\beta}&=&v_{d}\sum_{a,b,c}\Bigg[ g^{E}_{3\mathbf{2},a} \delta_{2,(i+l)|3}(-1)^{\beta+1}Y^{(k_{E^{c}_{3}}+k_{L_{D}})}_{\mathbf{2}a,3-\beta} + g^{E}_{3\mathbf{2}',b} \delta_{1,(i+l)|3}(-1)^{\beta+1}Y^{(k_{E^{c}_{3}}+k_{L_{D}})}_{\mathbf{2}'b,3-\beta}\\ \label{eq:exp_R}
            &~&\qquad \quad+g^{E}_{3\mathbf{2}'',c} \delta_{0,(i+l)|3}(-1)^{\beta+1}Y^{(k_{E^{c}_{3}}+k_{L_{D}})}_{\mathbf{2}''c,3-\beta}\Bigg]\,,\\ \nonumber
  \hskip-0.3in (M_{E})_{33}&=&v_{d}\sum_{a,b,c}\Bigg[ g^{E}_{4\mathbf{1},a} \delta_{0,(j+l)|3}Y^{(k_{E^{c}_{3}}+k_{L_{3}})}_{\mathbf{1}a}+g^{E}_{4\mathbf{1}',b} \delta_{2,(j+l)|3}Y^{(k_{E^{c}_{3}}+k_{L_{3}})}_{\mathbf{1}'b} \\
  &~&\qquad \quad+g^{E}_{4\mathbf{1}'',c} \delta_{1,(j+l)|3}Y^{(k_{E^{c}_{3}}+k_{L_{3}})}_{\mathbf{1}''c}\Bigg]\;,\;\alpha,\beta=1,2\,.
\end{eqnarray}
The possible forms of the sub-matrices for different representation and weight assignments are summarised in Table~\ref{Tab:Submatrix}. When combined, these submatrix configurations may lead to the following 9 texture-zero patterns for $M_{E}$:
\begin{equation}\mathcal{C}_{1}^{(1)}\,,~\mathcal{C}_{2}^{(1)}\,,~\mathcal{C}_{2}^{(2)}\,,~\mathcal{C}_{2}^{(3)}\,,~\mathcal{C}_{3}^{(3)}\,,~\mathcal{C}_{4}^{(1)}\,,~\mathcal{C}_{4}^{(2)}\,,~\mathcal{C}_{4}^{(3)}\,,~\mathcal{C}_{6}^{(1)}\,.
\end{equation}
Note that the texture $\mathcal{C}_{6}^{(1)}$ can be achieved only if $k_{L_{D}}+k_{E_{D}^{c}}=0$ and $k+i=2\,(\text{mod}~3)$, with $M_{E}$ satisfying $(M_{E})_{12}=-(M_{E})_{21}$ and  $(M_{E})_{33}\neq 0$, thus featuring two degenerate charged-lepton masses.

\item{$L_{D}\equiv (L_{1},L_{2})^{T}\sim \mathbf{2}^{i},~~L_{3}\sim \mathbf{1}^{j},~~~~E^{c}_{\alpha}\sim \mathbf{1}^{l_{\alpha}}~~~~\text{with}~\alpha=1,2,3$}

The modular invariant superpotential is now given by
\begin{equation}\label{eq:We_5}\mathcal{W}_{E}=\sum_{\alpha=1}^{3}\sum_{\mathbf{r},a,b}g^{E}_{\alpha1\mathbf{r},a}\left[E_{\alpha}^{c}L_{D}Y_{\mathbf{r}a}^{(k_{L_{D}}+k_{E^{c}_{\alpha}})}\right]_{\mathbf{1}}H_{d}+g^{E}_{\alpha2\mathbf{r},b}\left[E_{\alpha}^{c}L_{3}Y_{\mathbf{r}b}^{(k_{L_{3}}+k_{E^{c}_{\alpha}})}\right]_{\mathbf{1}}H_{d}\,.
\end{equation}
The corresponding expressions of the elements in the $M_{E}$ can be obtained as
\begin{eqnarray}\nonumber
  (M_{E})_{\alpha\beta}&=&v_{d}\sum_{a,b,c}\Bigg[ g^{E}_{\alpha 1\mathbf{2},a} \delta_{2,i+l_{\alpha}}(-1)^{\beta+1}Y^{(k_{E^{c}_{\alpha}}+k_{L_{D}})}_{\mathbf{2}a,3-\beta} + g^{E}_{\alpha 1\mathbf{2}',b} \delta_{1,(i+l_{\alpha})|3}(-1)^{\beta+1}Y^{(k_{E^{c}_{\alpha}}+k_{L_{D}})}_{\mathbf{2}'b,3-\beta}\\ \label{eq:ME_6_1}
                       &~&\qquad \quad +g^{E}_{\alpha 1\mathbf{2}'',c} \delta_{0,(i+l_{\alpha})|3}(-1)^{\beta+1}Y^{(k_{E^{c}_{\alpha}}+k_{L_{D}})}_{\mathbf{2}''c,3-\beta}\Bigg]\,,\\ \nonumber
  (M_{E})_{\alpha 3}&=&v_{d}\sum_{a,b,c}\Bigg[ g^{E}_{\alpha 2 \mathbf{1},a} \delta_{0,(j+l_{\alpha})|3}Y^{(k_{E^{c}_{\alpha}}+k_{L_{3}})}_{\mathbf{1}a} + g^{E}_{\alpha 2 \mathbf{1}',b} \delta_{2,(j+l_{\alpha})|3}Y^{(k_{E^{c}_{\alpha}}+k_{L_{3}})}_{\mathbf{1}'b}\\ \label{eq:ME_6_2}
  &~&\qquad \quad +g^{E}_{\alpha 2 \mathbf{1}'',c} \delta_{1,(j+l_{\alpha})|3}Y^{(k_{E^{c}_{\alpha}}+k_{L_{3}})}_{\mathbf{1}''c}\Bigg]\,,
\end{eqnarray}
where $\alpha=1,2,3$ and $\beta=1,2$. In this case, we can obtain five possible texture-zero patterns in $M_{E}$, namely
\begin{equation}
\mathcal{C}_{1}^{(1)}\,,~\mathcal{C}_{2}^{(2)}\,,~\mathcal{C}_{2}^{(3)}\,,~\mathcal{C}_{3}^{(1)}\,,~\mathcal{C}_{4}^{(1)}\,.
\end{equation}

\item{$L_{\alpha}\sim \mathbf{1}^{j_{\alpha}},~~E^{c}\equiv (e^{c},\mu^{c},\tau^{c})^{T}\sim \mathbf{3}$}

The corresponding charged-lepton mass matrix is the transpose of that in Eq.~\eqref{eq:general-ele-lepton}, and no texture zeros can be achieved.

\item{$L_{\alpha}\sim \mathbf{1}^{j_{\alpha}},~~E_{D}^{c}\equiv (e^{c},\mu^{c})^{T}\sim \mathbf{2}^{k},~~E^{c}_{3}\equiv \tau^{c}\sim \mathbf{1}^{l}$}

As in the previous case, the resulting $M_E$ can be obtained by transposing the that of Eqs.~\eqref{eq:ME_6_1} and \eqref{eq:ME_6_2} by switching the transformation properties of LH and RH charged leptons. In this case, the allowed texture zeros are
\begin{equation}\mathcal{C}_{1}^{(1)}\,,~\mathcal{C}_{2}^{(2)}\,,~\mathcal{C}_{2}^{(3)}\,,~\mathcal{C}_{3}^{(2)}\,,~\mathcal{C}_{4}^{(1)}\,.
\end{equation}

\item{$L_{\beta}\sim \mathbf{1}^{j_{\beta}},~~E^{c}_{\alpha}\sim \mathbf{1}^{l_{\alpha}}$}

All lepton multiplets are assigned to singlets of $T'$ and the Yukawa superpotential for the charged-lepton masses reads
\begin{equation}\mathcal{W}_{E}=\sum_{\alpha,\beta=1}^{3}\sum_{\mathbf{r},a}g^{E}_{\alpha\beta \mathbf{r},a}\left(E^{c}_{\alpha}L_{\beta}Y_{\mathbf{r}a}^{(k_{L_{\beta}}+k_{E^{c}_{\alpha}})}\right)_{\mathbf{1}}H_{d}\,.
\end{equation}
The elements of $M_E$ are now given by:
\begin{eqnarray} \nonumber
  (M_{E})_{\alpha\beta}&=& v_d\sum_{a,b,c}\Bigg[g^{E}_{\alpha\beta \mathbf{1},a} \delta_{0,(j_{\beta}+l_{\alpha})|3}Y^{(k_{E^{c}_{\alpha}}+k_{L_{\beta}})}_{\mathbf{1}a} + g^{E}_{\alpha\beta \mathbf{1}',b} \delta_{2,(j_{\beta}+l_{\alpha})|3}Y^{(k_{E^{c}_{\alpha}}+k_{L_{\beta}})}_{\mathbf{1}'b}\\
 \label{eq:We-singlets}   &~&+g^{E}_{\alpha\beta \mathbf{1}'',c}\delta_{1,(j_{\beta}+l_{\alpha})|3}Y^{(k_{E^{c}_{\alpha}}+k_{L_{\beta}})}_{\mathbf{1}''c} \Bigg] \,.
\end{eqnarray}
In this case, the representation and modular weight of each lepton field can be adjusted so that several texture-zero patterns in $M_{E}$ can be obtained. Since we have $S=1$, $T=\omega^{a}$ in the singlet representations $\mathbf{1}^{a}$, the flavour symmetry is essentially the $Z_3$ subgroup generated by $T$ rather than $T'$ for this assignment. As a consequence, a larger number of free parameters would have to be introduced, leading to less predictive power. Moreover, the contributions of all modular forms can be absorbed into the coupling constants, as can be seen from Eq.~\eqref{eq:We-singlets}. Thus, the advantage of the modular symmetry is lost in this case. In the present work, we are mainly concerned on achieving texture-zero patterns with small number of free parameters by taking the most of $T'$ modular symmetry. Therefore, from now on we will not consider the case in which all lepton fields are in singlets of $T'$.
\end{itemize}

\begin{table}[ht!]
\centering
\resizebox{0.87\textwidth}{!}{
\begin{tabular}{|c|c|c|}
\hline  \hline
 $(\psi^{c},\psi)$  & Expressions of Sub-matrix &  Weight and representation assignments \\ \hline
\multirow{24}{*}{$(\mathbf{2}^{i},\mathbf{3})$} & \multirow{2}{*}{$\begin{pmatrix}0~ & ~0 ~& ~0 \\ 0~ & ~0 ~& ~0 \\ \end{pmatrix} $} & $\textcircled{1}~k_{\psi^c}+k_{\psi}\leq 0\,,$ \\
& &$\textcircled{2}~k_{\psi^c}+k_{\psi}=2,4,6,\dots$ \\  \cline{2-3}
& $\begin{pmatrix}0 & -\sqrt{2}g_{1}Y_{\mathbf{2},1}^{(1)} & g_{1}Y_{\mathbf{2},2}^{(1)} \\ \sqrt{2}g_{1}Y_{\mathbf{2},2}^{(1)} & 0 & g_{1}Y_{\mathbf{2},1}^{(1)} \\ \end{pmatrix} $ & $k_{\psi^c}+k_{\psi}= 1,~i=0$ \\   \cline{2-3}
 & $\begin{pmatrix}-\sqrt{2}g_{1}Y_{\mathbf{2},1}^{(1)} &  g_{1}Y_{\mathbf{2},2}^{(1)} & 0  \\ 0 & g_{1}Y_{\mathbf{2},1}^{(1)} & \sqrt{2}g_{1}Y_{\mathbf{2},2}^{(1)} \\ \end{pmatrix} $ & $k_{\psi^c}+k_{\psi}= 1,~i=1$ \\   \cline{2-3}
 & $\begin{pmatrix}g_{1}Y_{\mathbf{2},2}^{(1)} & 0 & -\sqrt{2}g_{1}Y_{\mathbf{2},1}^{(1)} \\ g_{1}Y_{\mathbf{2},1}^{(1)} & \sqrt{2}g_{1}Y_{\mathbf{2},2}^{(1)} & 0\\ \end{pmatrix} $ & $k_{\psi^c}+k_{\psi}= 1,~i=2$ \\   \cline{2-3}
 & $\begin{pmatrix}g_{1}Y_{\mathbf{2}'',2}^{(3)} & -\sqrt{2}g_{2}Y_{\mathbf{2},1}^{(3)} & -\sqrt{2}g_{1}Y_{\mathbf{2}'',1}^{(3)}+g_{2}Y_{\mathbf{2},2}^{(3)} \\ g_{1}Y_{\mathbf{2}'',1}^{(3)}+\sqrt{2}g_{2}Y_{\mathbf{2},2}^{(3)}& \sqrt{2}g_{1}Y_{\mathbf{2}'',2}^{(3)} & g_{2}Y_{\mathbf{2},1}^{(3)} \\ \end{pmatrix} $ & $k_{\psi^c}+k_{\psi}= 3,~i=0$ \\   \cline{2-3}
 & $\begin{pmatrix}-\sqrt{2}g_{2}Y_{\mathbf{2},1}^{(3)} & g_{2}Y_{\mathbf{2},2}^{(3)}-\sqrt{2}g_{1}Y_{\mathbf{2}'',1}^{(3)} & g_{1}Y_{\mathbf{2}'',2}^{(3)} \\ \sqrt{2}g_{1}Y_{\mathbf{2}'',2}^{(3)} & g_{2}Y_{\mathbf{2},1}^{(3)} & \sqrt{2}g_{2}Y_{\mathbf{2},2}^{(3)}+g_{1}Y_{\mathbf{2}'',1}^{(3)} \\ \end{pmatrix} $ & $k_{\psi^c}+k_{\psi}= 3,~i=1$ \\   \cline{2-3}
 & $\begin{pmatrix}g_{2}Y_{\mathbf{2},2}^{(3)}-\sqrt{2}g_{1}Y_{\mathbf{2}'',1}^{(3)} & g_{1}Y_{\mathbf{2}'',2}^{(3)} & -\sqrt{2}g_{2}Y_{\mathbf{2},1}^{(3)} \\ g_{2}Y_{\mathbf{2},1}^{(3)} & \sqrt{2}g_{2}Y_{\mathbf{2},2}^{(3)}+g_{1}Y_{\mathbf{2}'',1}^{(3)} & \sqrt{2}g_{1}Y_{\mathbf{2}'',2}^{(3)} \\ \end{pmatrix} $ & $k_{\psi^c}+k_{\psi}= 3,~i=2$ \\   \cline{2-3}
 & $\begin{pmatrix}g_{1}Y_{\mathbf{2}'',2}^{(5)}-\sqrt{2}g_{2}Y_{\mathbf{2}',1}^{(5)} & g_{2}Y_{\mathbf{2}',2}^{(5)}-\sqrt{2}g_{3}Y_{\mathbf{2},1}^{(5)} & -\sqrt{2}g_{1}Y_{\mathbf{2}'',1}^{(5)}+g_{3}Y_{\mathbf{2},2}^{(5)} \\ g_{1}Y_{\mathbf{2}'',1}^{(5)}+\sqrt{2}g_{3}Y_{\mathbf{2},2}^{(5)} & \sqrt{2}g_{1}Y_{\mathbf{2}'',2}^{(5)}+g_{2}Y_{\mathbf{2}',1}^{(5)} & \sqrt{2}g_{2}Y_{\mathbf{2}',2}^{(5)}+g_{3}Y_{\mathbf{2},1}^{(5)} \\ \end{pmatrix} $ & $k_{\psi^c}+k_{\psi}= 5,~i=0$ \\   \cline{2-3}
 & $\begin{pmatrix}g_{2}Y_{\mathbf{2}',2}^{(5)}-\sqrt{2}g_{3}Y_{\mathbf{2},1}^{(5)} & g_{3}Y_{\mathbf{2},2}^{(5)}-\sqrt{2}g_{1}Y_{\mathbf{2}'',1}^{(5)} & -\sqrt{2}g_{2}Y_{\mathbf{2}',1}^{(5)}+g_{1}Y_{\mathbf{2}'',2}^{(5)} \\ g_{2}Y_{\mathbf{2}',1}^{(5)}+\sqrt{2}g_{1}Y_{\mathbf{2}'',2}^{(5)} & \sqrt{2}g_{2}Y_{\mathbf{2}',2}^{(5)}+g_{3}Y_{\mathbf{2},1}^{(5)} & \sqrt{2}g_{3}Y_{\mathbf{2},2}^{(5)}+g_{1}Y_{\mathbf{2}'',1}^{(5)} \\ \end{pmatrix} $ & $k_{\psi^c}+k_{\psi}= 5,~i=1$ \\   \cline{2-3}
 & $\begin{pmatrix}g_{3}Y_{\mathbf{2},2}^{(5)}-\sqrt{2}g_{1}Y_{\mathbf{2}'',1}^{(5)} & g_{1}Y_{\mathbf{2}'',2}^{(5)}-\sqrt{2}g_{2}Y_{\mathbf{2}',1}^{(5)} & -\sqrt{2}g_{3}Y_{\mathbf{2},1}^{(5)}+g_{2}Y_{\mathbf{2}',2}^{(5)} \\ g_{3}Y_{\mathbf{2},1}^{(5)}+\sqrt{2}g_{2}Y_{\mathbf{2}',2}^{(5)} & \sqrt{2}g_{3}Y_{\mathbf{2},2}^{(5)}+g_{1}Y_{\mathbf{2}'',1}^{(5)} & \sqrt{2}g_{1}Y_{\mathbf{2}'',2}^{(5)}+g_{2}Y_{\mathbf{2}',1}^{(5)} \\ \end{pmatrix} $ & $k_{\psi^c}+k_{\psi}= 5,~i=2$ \\   \hline \hline
 \multirow{26}{*}{$(\mathbf{2}^{i},\mathbf{2}^{k})$} ~&~  \multirow{2}{*}{$\begin{pmatrix} 0 ~&~ 0 \\ 0 ~&~ 0 \\ \end{pmatrix}$}  &  $\textcircled{1}~k_{\psi^{c}}+k_{\psi}<0\,,$ \\
 & & $\textcircled{2}~k_{\psi^{c}}+k_{\psi}=1,3,5,\ldots$ \\ \cline{2-3}

& $  \begin{pmatrix}  0~ &~ g_{1}  \\ -g_{1} ~&~ 0 \\ \end{pmatrix}   $ & $k_{\psi^{c}}+k_{\psi}=0,~k+i=2\,(\text{mod}~3)$\\ \cline{2-3}

& $  \begin{pmatrix}  -g_{1} \sqrt{2}Y^{(2)}_{\mathbf{3},2} & g_{1} Y^{(2)}_{\mathbf{3},3} \\ g_{1} Y^{(2)}_{\mathbf{3},3} & g_{1} \sqrt{2}Y^{(2)}_{\mathbf{3},1} \\ \end{pmatrix}   $ & $k_{\psi^{c}}+k_{\psi}=2,~k+i=0\,(\text{mod}~3)$\\ \cline{2-3}

& $  \begin{pmatrix}  -g_{1} \sqrt{2}Y^{(4)}_{\mathbf{3}, 2} & g_{1} Y^{(4)}_{\mathbf{3}, 3} \\ g_{1} Y^{(4)}_{\mathbf{3}, 3} & g_{1} \sqrt{2}Y^{(4)}_{\mathbf{3}, 1} \\ \end{pmatrix}   $ & $k_{\psi^{c}}+k_{\psi}=4,~k+i=0\,(\text{mod}~3)$\\\cline{2-3}

& $  \begin{pmatrix}  -\sqrt{2}(g_{1} Y^{(6)}_{\mathbf{3}I,2} + g_{2} Y^{(6)}_{\mathbf{3}II,2}) & g_{1} Y^{(6)}_{\mathbf{3}I,3} + g_{2} Y^{(6)}_{\mathbf{3}II,3} \\ g_{1} Y^{(6)}_{\mathbf{3}I,3} + g_{2} Y^{(6)}_{\mathbf{3}II,3} & \sqrt{2}(g_{1} Y^{(6)}_{\mathbf{3}I,1} + g_{2} Y^{(6)}_{\mathbf{3}II,1}) \\ \end{pmatrix}   $ & $k_{\psi^{c}}+k_{\psi}=6,~k+i=0\,(\text{mod}~3)$\\\cline{2-3}

& $  \begin{pmatrix}  -g_{1} \sqrt{2}Y^{(2)}_{\mathbf{3},1} & g_{1} Y^{(2)}_{\mathbf{3},2} \\ g_{1} Y^{(2)}_{\mathbf{3},2} & g_{1} \sqrt{2}Y^{(2)}_{\mathbf{3},3} \\ \end{pmatrix}  $ & $k_{\psi^{c}}+k_{\psi}=2,~k+i=1\,(\text{mod}~3)$\\\cline{2-3}

& $  \begin{pmatrix}  -g_{1} \sqrt{2}Y^{(4)}_{\mathbf{3}, 1} & g_{2}Y^{(4)}_{\mathbf{1}'}+g_{1} Y^{(4)}_{\mathbf{3}, 2} \\ -g_{2}Y^{(4)}_{\mathbf{1}'}+g_{1} Y^{(4)}_{\mathbf{3}, 2} & g_{1} \sqrt{2}Y^{(4)}_{\mathbf{3}, 3} \\ \end{pmatrix}  $ & $k_{\psi^{c}}+k_{\psi}=4,~k+i=1\,(\text{mod}~3)$\\\cline{2-3}

& $  \begin{pmatrix}  -\sqrt{2}(g_{1} Y^{(6)}_{\mathbf{3}I,1}+g_{2} Y^{(6)}_{\mathbf{3}II,1}) & (g_{1} Y^{(6)}_{\mathbf{3}I,2} + g_{2} Y^{(6)}_{\mathbf{3}II,2}) \\ (g_{1} Y^{(6)}_{\mathbf{3}I,2} + g_{2} Y^{(6)}_{\mathbf{3}II,2}) & \sqrt{2}(g_{1} Y^{(6)}_{\mathbf{3}I,3} + g_{2} Y^{(6)}_{\mathbf{3}II,3})\\ \end{pmatrix}  $ & $k_{\psi^{c}}+k_{\psi}=6,~k+i=1\,(\text{mod}~3)$\\\cline{2-3}

& $  \begin{pmatrix}  -g_{1} \sqrt{2}Y^{(2)}_{\mathbf{3},3} & g_{1} Y^{(2)}_{\mathbf{3},1} \\ g_{1} Y^{(2)}_{\mathbf{3},1} & g_{1} \sqrt{2}Y^{(2)}_{\mathbf{3},2} \\ \end{pmatrix}  $ & $k_{\psi^{c}}+k_{\psi}=2,~k+i=2\,(\text{mod}~3)$\\\cline{2-3}

& $  \begin{pmatrix}  -g_{1} \sqrt{2}Y^{(4)}_{\mathbf{3}, 3} & g_{2}Y^{(4)}_{\mathbf{1}}+g_{1} Y^{(4)}_{\mathbf{3}, 1} \\ -g_{2}Y^{(4)}_{\mathbf{1}}+g_{1} Y^{(4)}_{\mathbf{3}, 1} & g_{1} \sqrt{2}Y^{(4)}_{\mathbf{3}, 2} \\ \end{pmatrix}  $ & $k_{\psi^{c}}+k_{\psi}=4,~k+i=2\,(\text{mod}~3)$\\\cline{2-3}

 & $  \begin{pmatrix}  -\sqrt{2}(g_{1} Y^{(6)}_{\mathbf{3}I,3} + g_{2} Y^{(6)}_{\mathbf{3}II,3}) & (g_{1} Y^{(6)}_{\mathbf{3}I,1}+g_{2} Y^{(6)}_{\mathbf{3}II,1}) + g_{3}Y^{(6)}_{\mathbf{1}} \\ (g_{1} Y^{(6)}_{\mathbf{3}I,1}+g_{2} Y^{(6)}_{\mathbf{3}II,1}) - g_{3}Y^{(6)}_{\mathbf{1}} & \sqrt{2}(g_{1} Y^{(6)}_{\mathbf{3}I,2} + g_{2} Y^{(6)}_{\mathbf{3}II,2}) \\ \end{pmatrix}  $ & $k_{\psi^{c}}+k_{\psi}=6,~k+i=2\,(\text{mod}~3)$\\
\hline \hline
\multirow{12}{*}{$(\mathbf{2}^{i},\mathbf{1}^{j})$} & \multirow{4}{*}{$(0,0)^{T}$}  &  $ \textcircled{1}~k_{\psi^{c}}+k_{\psi}<0,$  \\
 & &  $\textcircled{2} ~k_{\psi^{c}}+k_{\psi}=0,2,4,6,\ldots,$ \\
& & $\textcircled{3} ~k_{\psi^{c}}+k_{\psi}=1,~i+j=0,1\,(\text{mod}~3)\,,$  \\
& & $\textcircled{4} ~k_{\psi^{c}}+k_{\psi}=3,~i+j=1\,(\text{mod}~3)\,.$  \\ \cline{2-3}
 & $(g Y^{(3)}_{\mathbf{2}'',2}, - g Y^{(3)}_{\mathbf{2}'',1})^{T}$ & $k_{\psi^{c}}+k_{\psi}=3,~~i+j=0\,(\text{mod}~3)$  \\ \cline{2-3}
 & $(g Y^{(5)}_{\mathbf{2}'',2}, - g Y^{(5)}_{\mathbf{2}'',1})^{T}$ & $k_{\psi^{c}}+k_{\psi}=5,~~i+j=0\,(\text{mod}~3)$  \\ \cline{2-3}
 & $(g Y^{(5)}_{\mathbf{2}',2}, - g Y^{(5)}_{\mathbf{2}',1})^{T}$ & $k_{\psi^{c}}+k_{\psi}=5,~~i+j=1\,(\text{mod}~3)$ \\ \cline{2-3}
 & $(g Y^{(1)}_{\mathbf{2},2}, - g Y^{(1)}_{\mathbf{2},1})^{T}$ & $k_{\psi^{c}}+k_{\psi}=1,~~i+j=2\,(\text{mod}~3)$  \\ \cline{2-3}
 & $(g Y^{(3)}_{\mathbf{2},2}, - g Y^{(3)}_{\mathbf{2},1})^{T}$ & $k_{\psi^{c}}+k_{\psi}=3,~~i+j=2\,(\text{mod}~3)$  \\ \cline{2-3}
 & $(g Y^{(5)}_{\mathbf{2}, 2}, - g Y^{(5)}_{\mathbf{2}, 1})^{T}$ & $k_{\psi^{c}}+k_{\psi}=5,~~i+j=2\,(\text{mod}~3)$ \\
\hline \hline
\end{tabular}}
\caption{\label{Tab:Submatrix}Possible submatrix forms for the charged-lepton mass matrix up to weight-6 modular forms, for the case where the RH charged lepton field $\psi^c$ transform as $\mathbf{2}^i$ and the LH lepton doublet $\psi$ as $\mathbf{3}$, $\mathbf{2}^{k}$ or $\mathbf{1}^{j}$ under $T'$ modular symmetry. Their modular weights are denoted by $k_{\psi^{c}}$ and $k_{\psi}$, respectively. The parameters $g_i$ are coupling constants. If the representation assignments of $\psi$ and $\psi^c$ are exchanged, the corresponding submatrix is the transpose of the original one. }
\end{table}


\section{\label{sec:neutrino_sector}Texture zeros in the neutrino mass matrix}

To carry out a full analysis of lepton models with texture-zero patterns realised by modular $T'$ flavour symmetry, we now turn our attention to the neutrino sector. Given that the particle nature of neutrinos is still unclear, we will consider both scenarios in which neutrinos are Dirac or Majorana particles. In the latter case, we explore two possibilities depending on wether neutrino masses are generated via: $i)$ an effective dimension-five Weinberg operator (without specifying the underlying full theory) or $ii)$ a minimal type-I seesaw model.

\subsection{\label{subsec:Dirac-neutrinos}Dirac neutrinos}

If neutrinos are Dirac particles, three RH (singlet) neutrino fields $N^c$ are necessary, leading to neutrino mass terms similar to the charged-lepton ones. As in Section~\ref{sec:charged_lepton_sector}, we can obtain the possible texture-zero patterns for the neutrino mass matrix by considering all different representation assignment of the lepton fields $L$ and $N^{c}$. Given that neutrino oscillation data requires that at least two neutrinos are massive, the rank of the neutrino mass matrix can be either three or two, in contrast with rank three of the charged-lepton mass matrix. Hence, the set of texture-zero patterns for the Dirac neutrino mass matrix $M_D$ is larger than for $M_E$, as can be seen in Table~\ref{tab:classes_Dirac_neutrino_mass_matrix}. Besides those textures given in Table~\ref{tab:classes_charged_lepton_mass_matrix} for $M_E$, we find the following additional ones which correspond to $M_D$ of rank two:
\begin{equation}\mathcal{D}_{3}^{(4)}\,,\mathcal{D}_{3}^{(5)}\,,\mathcal{D}_{4}^{(4)}\,,\mathcal{D}_{4}^{(5)}\,,\mathcal{D}_{4}^{(6)}\,,\mathcal{D}_{5}^{(1)}\,,\mathcal{D}_{5}^{(2)}\,,\mathcal{D}_{5}^{(3)}\,,\mathcal{D}_{5}^{(4)}\,,\mathcal{D}_{5}^{(5)}\,,\mathcal{D}_{5}^{(6)}\,,\mathcal{D}_{6}^{(1)}\,,\mathcal{D}_{6}^{(2)}\,,\mathcal{D}_{7}^{(1)}\,.
\end{equation}
Similarly to the charged-lepton sector, both $\mathcal{D}_{6}^{(3)}$ and $\mathcal{D}_{7}^{(1)}$ lead to degeneracy between the two nonvanishing neutrino masses and, consequently, will not be considered from now on.

At this point, it is worth stressing that Majorana mass terms for the RH neutrinos can be forbidden by properly choosing the modular weights of the $N^c$ fields, so that the Diracness of light neutrinos stems automatically from modular symmetry. For instance, let us consider a case with the LH leptons and the RH neutrinos transforming as
\begin{equation}
  L=(L_{1},L_{2},L_{3})^{T}\sim \mathbf{3}\,,\quad N^{c}_{D}\equiv (N_{1}^{c},N_{2}^{c})^{T}\sim \mathbf{2}^{k}\,,\quad N^{c}_{3}\sim \mathbf{1}^{l}\,.
\end{equation}
Then, the superpotential for Dirac neutrino masses can be written as 
\begin{equation}
\mathcal{W}_{\nu}^{D}=\sum_{\mathbf{r},a,b}g^{\nu}_{\mathbf{r},a}\left[N_{D}^{c}LY^{(k_{1})}_{\mathbf{r}_{a}}\right]_{\mathbf{1}}H_{u}+g^{\nu}_{\mathbf{r},b}N_{3}^{c}LY^{(k_{2})}_{\mathbf{3}b}H_{u}\,.
\end{equation}
Modular invariance requires the weights of the involved modular forms to satisfy
\begin{equation}
k_{1}=k_{L}+k_{N_{D}^{c}}\,,\qquad k_{2}=k_{L}+k_{N_{3}^{c}}\,,
\end{equation}
which, for any given values of $k_{1}$ and $k_{2}$, implies
\begin{equation}
k_{L}=k_{1}-k_{N_{D}^{c}}\,,\qquad k_{N^{c}_{3}}=k_{N_{D}^{c}}+ k_{2}-k_{1}\,.
\end{equation}
Notice that the value of $k_{N_{D}^{c}}$ is free and, thus, we can always choose a negative value for $k_{N_{D}^{c}}$, so that  $k_{N_{3}^{c}}$ is negative as well. If modular weights of all RH neutrino fields are negative, modular-invariant Majorana mass terms of the type
\begin{equation}
\mathcal{W}_{N}^{M}\sim N_{i}^{c}N_{j}^{c}Y^{(k_{N_{i}^{c}}+k_{N_{j}^{c}})}\,,~~~i,j=D, 3\,,
\end{equation}
cannot be written since there are no modular forms of negative modular weights at level $N$. The above arguments hold for any other representation assignment of the lepton fields. In short, the modular symmetry can enforce light neutrinos to be Dirac particles if the modular weights of the RH neutrinos are properly (not uniquely) assigned.

\begin{table}[t!]
 \centering
 \begin{tabular}{|l|} \hline\hline
$ \mathcal{D}_{1}^{(1)}:\left( \begin{array}{ccc} 0 & \times & \times \\
                 \times & \times & \times \\
                 \times & \times & \times \end{array} \right)$\\  \hline
$\mathcal{D}_{2}^{(1)}:\left( \begin{array}{ccc} 0 & \times & \times \\
                 \times & 0 & \times \\
                 \times & \times & \times \end{array} \right)\,,\quad
\mathcal{D}_{2}^{(2)}:\left( \begin{array}{ccc} \times & \times & \times \\
                 \times & \times & \times \\
                 0 & 0 & \times \end{array} \right)\,,\quad
\mathcal{D}_{2}^{(3)}:\left( \begin{array}{ccc} \times & \times & 0 \\
                 \times & \times & 0 \\
                 \times & \times & \times \end{array} \right)$\\ \hline
$\mathcal{D}_{3}^{(1)}: \left( \begin{array}{ccc} \times & \times & 0 \\
                 \times & \times & \times \\
                 0 & 0 & \times \end{array} \right)\,,\quad
\mathcal{D}_{3}^{(2)}:\left( \begin{array}{ccc} \times & 0 & \times \\
                 \times & 0 & \times \\
                 0 & \times & \times \end{array} \right)\,,\quad
\mathcal{D}_{3}^{(3)}:\left( \begin{array}{ccc} 0 & \times & \times \\
                 \times & 0 & \times \\
                 \times & \times & 0 \end{array} \right)\,, \quad
\mathcal{D}_{3}^{(4)}:\left( \begin{array}{ccc} \times & \times & \times \\
                 \times & \times & \times \\
                 0 & 0 & 0 \end{array} \right)$\\
$\mathcal{D}_{3}^{(5)}:\left( \begin{array}{ccc} \times & \times & 0 \\
                 \times & \times & 0 \\
                 \times & \times & 0 \end{array} \right)$ \\ \hline
$\mathcal{D}_{4}^{(1)}:\left( \begin{array}{ccc} \times & \times & 0 \\
                 \times & \times & 0 \\
                 0 & 0 & \times \end{array} \right)\,,\quad
\mathcal{D}_{4}^{(2)}:\left( \begin{array}{ccc} 0 & \times & 0 \\
                 \times & 0 & 0 \\
                 \times & \times & \times \end{array} \right)\,,\quad
\mathcal{D}_{4}^{(3)}:\left( \begin{array}{ccc} 0 & \times & \times \\
                 \times & 0 & \times \\
                 0 & 0 & \times \end{array} \right)\,,\quad
\mathcal{D}_{4}^{(4)}:\left( \begin{array}{ccc} 0 & 0 & \times \\
                 0 & 0 & \times \\
                 \times & \times & \times \end{array} \right)$ \\
$\mathcal{D}_{4}^{(5)}:\left( \begin{array}{ccc} \times & \times & 0 \\
                 \times & \times & \times \\
                 0 & 0 & 0 \end{array} \right)\,,\quad
\mathcal{D}_{4}^{(6)}:\left( \begin{array}{ccc} \times & \times & 0 \\
                 \times & \times & 0 \\
                 0 & \times & 0 \end{array} \right)$ \\ \hline
$\mathcal{D}_{5}^{(1)}:\left( \begin{array}{ccc} \times & \times & 0 \\
                 \times & \times & 0 \\
                 0 & 0 & 0 \end{array} \right)\,,\quad
\mathcal{D}_{5}^{(2)}:\left( \begin{array}{ccc} 0 & 0 & \times \\
                 0 & 0 & \times \\
                 \times & \times & 0 \end{array} \right)\,,\quad
\mathcal{D}_{5}^{(3)}:\left( \begin{array}{ccc} 0 & \times & \times \\
                 \times & 0 & \times \\
                 0 & 0 & 0 \end{array} \right)\,,\quad
\mathcal{D}_{5}^{(4)}:\left( \begin{array}{ccc} 0 & \times & 0 \\
                 \times & 0 & 0 \\
                 \times & \times & 0 \end{array} \right)$\\
$\mathcal{D}_{5}^{(5)}:\left( \begin{array}{ccc} 0 & 0 & 0 \\
                 0 & 0 & \times \\
                 \times & \times & \times \end{array} \right)\,,\quad
\mathcal{D}_{5}^{(6)}:\left( \begin{array}{ccc} 0 & 0 & \times \\
                 0 & 0 & \times \\
                 0 & \times & \times \end{array} \right)$ \\ \hline
$\mathcal{D}_{6}^{(1)}:\left( \begin{array}{ccc} 0 & 0 & 0 \\
                 0 & 0 & \times \\
                 \times & \times & 0 \end{array} \right)\,,\quad
\mathcal{D}_{6}^{(2)}:\left( \begin{array}{ccc} 0 & 0 & \times \\
                 0 & 0 & \times \\
                 0 & \times & 0 \end{array} \right)\,,\quad
\mathcal{D}_{6}^{(3)}:\left( \begin{array}{ccc} 0 & \times & 0 \\
                 \times & 0 & 0 \\
                 0 & 0 & \times \end{array} \right)$\\ \hline
$\mathcal{D}_{7}^{(1)}:\left( \begin{array}{ccc} 0 & \times & 0 \\
                 \times & 0 & 0 \\
                 0 & 0 & 0 \end{array} \right)$\\
   \hline \hline
\end{tabular}
\caption{\label{tab:classes_Dirac_neutrino_mass_matrix}Texture-zero patterns for the Dirac neutrino mass matrix $M_D$ which can be realised from $T'$ modular symmetry, up to row and column permutations.}
\end{table}

\subsection{Majorana neutrinos}
\label{sec:Majorana_neutrinos}

For Majorana neutrinos, we consider two distinct mass generation mechanisms: the effective Weinberg operator and the type-I seesaw mechanism. In the same token of the discussions for charged leptons and Dirac neutrinos, in Table~\ref{tab:classes_Majorana_neutrino_mass_matrix} we present all possible texture-zero for the effective Majorana neutrino mass matrix $M_{\nu}$ which can be obtained from $T'$ modular symmetry. Note that $\mathcal{W}_{4}^{(1)}$, $\mathcal{W}_{4}^{(4)}$ and $\mathcal{W}_{5}^{(1)}$ lead to two degenerate neutrino masses, which is excluded by neutrino oscillation data. In the following, we first discuss the textures originated from the effective Weinberg operator, and then turn to the type-I seesaw case.

\begin{table}[t!]
 \centering
 \begin{tabular}{|l|} \hline\hline
   $\mathcal{W}_{1}^{(1)}:\left( \begin{array}{ccc} \times & \times & \times \\
                 \times & \times & \times \\
                 \times & \times & 0 \end{array} \right)\,,\quad
\mathcal{W}_{1}^{(2)}:\left( \begin{array}{ccc} \times & 0 & \times \\
                 0 & \times & \times \\
                               \times & \times & \times \end{array} \right)$ \\ \hline
   $\mathcal{W}_{2}^{(1)}:\left( \begin{array}{ccc} 0 & \times & \times \\
                 \times & 0 & \times \\
                 \times & \times & \times \end{array} \right)\,,\quad
\mathcal{W}_{2}^{(2)}:\left( \begin{array}{ccc} \times & \times & 0 \\
                 \times & \times & 0 \\
                 0 & 0 & \times \end{array} \right)$ \\ \hline
   $\mathcal{W}_{3}^{(1)}:\left( \begin{array}{ccc} 0 & 0 & \times \\
                 0 & 0 & \times \\
                 \times & \times & \times \end{array} \right)\,,\quad
\mathcal{W}_{3}^{(2)}:\left( \begin{array}{ccc} \times & \times & 0 \\
                 \times & \times & 0 \\
                 0 & 0 & 0 \end{array} \right)$ \\ \hline
   $\mathcal{W}_{4}^{(1)}:\left( \begin{array}{ccc} 0 & 0 & \times \\
                 0 & 0 & \times \\
                 \times & \times & 0 \end{array} \right)\,,\quad
\mathcal{W}_{4}^{(2)}:\left( \begin{array}{ccc} 0 & \times & 0 \\
                 \times & \times & 0 \\
                 0 & 0 & 0 \end{array} \right)\,,\quad \mathcal{W}_{4}^{(3)}:\left( \begin{array}{ccc} \times & 0 & 0 \\
                 0 & \times & 0 \\
                 0 & 0 & 0 \end{array} \right)\,,\quad \mathcal{W}_{4}^{(4)}:\left( \begin{array}{ccc} \times & 0 & 0 \\
                 0 & 0 & \times \\
                 0 & \times & 0 \end{array} \right)$ \\ \hline
   $\mathcal{W}_{5}^{(1)}:\left( \begin{array}{ccc} 0 & \times & 0 \\
                 \times & 0 & 0 \\
                 0 & 0 & 0 \end{array} \right)$ \\ \hline\hline
\end{tabular}
\caption{\label{tab:classes_Majorana_neutrino_mass_matrix}Texture-zero patterns for the Majorana neutrino mass matrix $M_\nu$ which can be realised by $T'$ modular symmetry, up to row and column permutations. Notice that only $\mathcal{W}_{1}^{(1)}$\,, $\mathcal{W}_{2}^{(2)}$\,, $\mathcal{W}_{3}^{(1)}$\,, $\mathcal{W}_{3}^{(2)}$\,, $\mathcal{W}_{4}^{(1)}$ and $\mathcal{W}_{4}^{(4)}$ can be obtained if neutrino masses are described by the Weinberg operator. On the other hand, all the above textures except $\mathcal{W}_{4}^{(4)}$ can be achieved if neutrino masses are generated via the minimal type I seesaw mechanism. }
\end{table}

\subsubsection{Neutrino masses via Weinberg operator}

The most general modular-invariant Weinberg operator of neutrino masses can be written as
\begin{equation}
  \mathcal{W}_{\nu}\sim \frac{1}{\Lambda}L_{i}L_{j}Y^{(k_{L_{i}}+k_{L_{j}})}H_{u}H_{u}\,,
\end{equation}
where $\Lambda$ denotes the new physics scale where lepton number is violated by two units, and the Higgs field $H_{u}$ is assumed to be invariant singlet of $T'$ with a vanishing modular weight. In this case, the neutrino mass matrix $M_{\nu}$ only depends on the $T'$ representation assignments and modular weights of the LH lepton fields. In the following, we will discuss all possible choices according to Eq.~\eqref{eq:assign_L} and infer about the properties of the resulting $M_{\nu}$ in each case.

\begin{itemize}
\item{$L\equiv (L_{1},L_{2},L_{3})^{T} \sim \mathbf{3}$}

In the case that the three LH lepton fields are assigned to a triplet of $T'$, the effective  dimension-five terms in the superpotential can be written as
\begin{eqnarray}\label{eq:Wnu_general}
  \mathcal{W}_{\nu}=\sum_{\mathbf{r},a}\frac{g_{\mathbf{r},a}^{\nu}}{\Lambda}\big[ L LY_{\mathbf{r}a}^{(2k_{L})}\big]_{\mathbf{1}}H_u H_u\,,
\end{eqnarray}
where $k_{L}$ is the modular weight of $L$. The explicit form of the $M_{\nu}$ elements is 
\begin{eqnarray}\nonumber
\label{eq:general-ele-neutrino-Wein}  (M_{\nu})_{\alpha\beta}&=&\frac{v_{u}^{2}}{\Lambda}\sum_{a,b,c,d} \Big[ g^{\nu}_{\mathbf{3},a}(3\delta_{\alpha\beta}-1)Y^{(2k_{L})}_{\mathbf{3}a,3-(\alpha+\beta)|3}+g^{\nu}_{\mathbf{1},b}\delta_{2,(\alpha+\beta)|3}Y^{(2k_{L})}_{\mathbf{1}b}\\
&~& \qquad \qquad~~ +g^{\nu}_{\mathbf{1}',c}\delta_{1,(\alpha+\beta)|3}Y^{(2k_{L})}_{\mathbf{1'}c}+g^{\nu}_{\mathbf{1}'',d}\delta_{0,(\alpha+\beta)|3}Y^{(2k_{L})}_{\mathbf{1''}d}\Big]\,,
\end{eqnarray}
where $\alpha,\beta=1,2,3$. 
The above expression reveals that $M_\nu$ has no zero elements if $k_{L}>0$. In the case of $k_{L}=0$, only the pattern $\mathcal{W}_{4}^{(4)}$ of Table~\ref{tab:classes_Majorana_neutrino_mass_matrix} can be realised, but it leads to two degenerate masses.

\item{$L_{D}\equiv (L_{1},L_{2})^{T}\sim \mathbf{2}^{i},~~L_{3}\sim \mathbf{1}^{j}$}

If the three lepton doublets transform as the direct sum of doublet and singlet of $T'$, the most general form of $\mathcal{W}_{\nu}$ is
\begin{eqnarray}\nonumber
\mathcal{W}_{\nu}&=&\frac{1}{\Lambda}\sum_{a,b,c}  \Bigg\{g^{\nu}_{1\mathbf{r},a}\left[L_DL_DY_{\mathbf{r}a}^{(2k_{L_{D}})}\right]_{\mathbf{1}}+g^{\nu}_{2\mathbf{r},b} \left[L_DL_3Y_{\mathbf{r}b}^{(k_{L_{D}}+k_{L_{3}})}\right]_{\mathbf{1}}\\
  &~&\qquad\qquad+g^{\nu}_{3\mathbf{r},c} \left[L_3L_3Y^{(2k_{L_{3}})}_{\mathbf{r}c}\right]_{\mathbf{1}}\Bigg\}H_{u}H_{u}\,.
\end{eqnarray}
from which we can read out the expressions for the $M_\nu$ matrix elements
\begin{eqnarray}
 \hskip-0.2in  (M_{\nu})_{\alpha\beta}&=&\sum_{a}\frac{v_{u}^{2}}{\Lambda}\Bigg[ g^{\nu}_{1\mathbf{3},a} (\sqrt{2})^{\delta_{\alpha\beta}}(-1)^{\delta_{2,\alpha+\beta}}Y^{(2k_{L_D})}_{\mathbf{3}a,3-(2i-\alpha-\beta)|3} \Bigg]\,,\\
\nonumber  (M_{\nu})_{\alpha 3}&=&\sum_{a,b,c}\frac{v_{u}^{2}}{\Lambda}\Bigg[ g^{E}_{2\mathbf{2},a} \delta_{2,(i+j)|3}(-1)^{\alpha+1}Y^{(k_{L_{D}}+k_{L_{3}})}_{\mathbf{2}a,3-\alpha}+g^{E}_{2\mathbf{2}',b}
\delta_{1,(i+j)|3}(-1)^{\alpha+1}Y^{(k_{L_{D}}+k_{L_{3}})}_{\mathbf{2}'b,3-\alpha}\\
&~&\qquad \qquad +g^{E}_{2\mathbf{2}'',c} \delta_{0,(i+j)|3}(-1)^{\alpha+1}Y^{(k_{L_{D}}+k_{L_{3}})}_{\mathbf{2}''c,3-\alpha}\Bigg]\,,\\
(M_{\nu})_{3 3}&=&\sum_{a,b,c}\frac{v_{u}^{2}}{\Lambda}\Bigg[ g^{E}_{3\mathbf{1},a} \delta_{0,(2j)|3}Y^{(2k_{L_{3}})}_{\mathbf{1}a} + g^{E}_{3\mathbf{1}',} \delta_{2,(2j)|3}Y^{(2k_{L_{3}})}_{\mathbf{1}'b}+g^{E}_{3\mathbf{1}'',c} \delta_{1,(2j)|3}Y^{(2k_{L_{3}})}_{\mathbf{1}''c}\Bigg]\,,
\end{eqnarray}
with $\alpha,\beta=1,2$. Going through all possible values of $i$, $j$ and of the modular weights $k_{L_{D}}$ and $k_{L_{3}}$, we find that the following five texture-zero patterns can be obtained,
\begin{equation}\mathcal{W}_{1}^{(1)}\,,\mathcal{W}_{2}^{(2)}\,,\mathcal{W}_{3}^{(1)}\,,\mathcal{W}_{3}^{(2)}\,,\mathcal{W}_{4}^{(1)}\,,
\end{equation}
as shown in Table~\ref{tab:classes_Majorana_neutrino_mass_matrix}.

\item{$L_{\alpha}\sim \mathbf{1}^{j_{\alpha}}$}

If all LH leptons transform as one-dimensional representations of $T'$, we have
\begin{equation}
\mathcal{W}_{\nu}=\sum_{\alpha,\beta=1}^{3}\sum_{a}\frac{g^{\nu}_{\alpha\beta \mathbf{r},a}}{\Lambda}\left[L_{\alpha}L_{\beta}Y_{\mathbf{r}a}^{(k_{L_{\alpha}}+k_{L_{\beta}})}\right]_{\mathbf{1}}H_{u}H_{u}\,,
\end{equation}
and the general expression for the neutrino mass matrix is
\begin{eqnarray} \nonumber
  (M_{\nu})_{\alpha\beta}&=& \frac{(1+\delta_{\alpha\beta})v_{u}^{2}}{2\Lambda}  \sum_{a,b,c}\Bigg[ g^{\nu}_{\alpha\beta\mathbf{1},a}\delta_{0,(j_{\alpha}+j_{\beta})|3}Y^{(k_{L_{\alpha}}+k_{L_{\beta}})}_{\mathbf{1}a} + g^{\nu}_{\alpha\beta\mathbf{1}',b}\delta_{2,(j_{\alpha}+j_{\beta})|3}Y^{(k_{L_{\alpha}}+k_{L_{\beta}})}_{\mathbf{1}'b}\\
  &~&\qquad \qquad \qquad+g^{\nu}_{\alpha\beta\mathbf{1}'',c}\delta_{1,(j_{\alpha}+j_{\beta})|3}Y^{(k_{L_{\alpha}}+k_{L_{\beta}})}_{\mathbf{1}''c} \Bigg]\,.
\end{eqnarray}
Analogously to the charged-lepton sector (see the discussion at the end of Section~\ref{sec:charged_lepton_sector}), we shall not consider this case here since it is less constrained by modular symmetry and, in general, more free parameters would be required. Nevertheless, some texture zeros can be realised~\cite{Zhang:2019ngf}.

\end{itemize}

\subsubsection{\label{sec:SS_neutrino_mass}Neutrino masses via type-I seesaw mechanism}

If neutrino masses are generated through the type-I seesaw mechanism, at least two RH neutrino fields are required to accommodate present neutrino oscillation data, namely, three nonzero lepton mixing angles and two mass-squared differences. In this section, we will consider the minimal seesaw model with two RH neutrinos~\cite{King:1999mb,Frampton:2002qc} for which the Dirac (RH Majorana) neutrino mass matrix $M_{D}$ ($M_{N}$) is a $2\times 3$ ($2\times 2$ symmetric) matrix. The RH neutrinos can transform as either doublet or singlet of $T'$, i.e.,
\begin{equation}\label{eq:RH_neutrinos}
  N^{c}\equiv (N_{1}^{c},N_{2}^{c})^{T} \sim \mathbf{2}^{k}\,,\quad \text{or}
  \quad N^{c}_{\alpha}\sim \mathbf{1}^{l_{\alpha}}\,,
\end{equation}
with $\alpha=1, 2$.

\begin{itemize}

\item{$N^{c}\sim\mathbf{2}^{k}$}

In this case, the mass term of the RH neutrinos can be written as
\begin{equation}\label{eq:Wnu_M_1}
\mathcal{W}_\nu^{N} = \sum_{\mathbf{r},a} g^{N}_{\mathbf{r},a} \Lambda\left[ N^c N^{c} Y_{\mathbf{r}a}^{(2k_{N^{c}})} \right]_{\mathbf{1}}\,,
\end{equation}
where $ g^{N}_{\mathbf{r},a}$ are coupling constants and $k_{N^{c}}$ is the modular weight of $N^{c}$.
From Table~\ref{Tab:Submatrix}, we can read out the matrix element of $M_{N}$:
\begin{equation}
  (M_{N})_{\alpha\beta}=\sum_{a} g^{N}_{\mathbf{3},a} (\sqrt{2})^{\delta_{\alpha\beta}}(-1)^{\delta_{2,\alpha+\beta}}Y^{(2k_{N^c})}_{\mathbf{3}a,3-(2k-\alpha-\beta)|3}\,,
\end{equation}
where $\alpha,\beta=1,2$.

\item{$N^{c}_{\alpha}\sim \mathbf{1}^{l_{\alpha}}$}

The RH neutrino fields transform as  $T'$ singlets, and their mass terms read
\begin{equation}
  \mathcal{W}_{\nu}^{N}=\sum _{\alpha,\beta=1}^{2}\sum_{\mathbf{r},a} g^{N}_{\alpha\beta \mathbf{r},a}\Lambda\left[  N^{c}_{\alpha}N^{c}_{\beta}Y_{\mathbf{r}a}^{(k_{N_{\alpha}^{c}}+k_{N_{\beta}^{c}})}\right]_{\mathbf{1}}\,.
\end{equation}
The explicit form of the elements of $M_{N}$ is
\begin{eqnarray}\nonumber
  (M_{N})_{\alpha\beta}&=&\frac{(1+\delta_{\alpha\beta})\Lambda}{2} \sum_{a,b,c}\Bigg[  g^{M}_{\alpha\beta\mathbf{1},a}\delta_{0,(l_{\alpha}+l_{\beta})|3}Y^{(k_{N_{\alpha}^{c}}+k_{N_{\beta}^{c}})}_{\mathbf{1}a} + g^{M}_{\alpha\beta\mathbf{1}',b}\delta_{2,(l_{\alpha}+l_{\beta})|3}Y^{(k_{N_{\alpha}^{c}}+k_{N_{\beta}^{c}})}_{\mathbf{1}'b}\\
  &~&\qquad \qquad \qquad+g^{M}_{\alpha\beta\mathbf{1}'',c}\delta_{1,(l_{\alpha}+l_{\beta})|3}Y^{(k_{N_{\alpha}^{c}}+k_{N_{\beta}^{c}})}_{\mathbf{1}''c} \Bigg]\,.
  \label{eq:MNSStext}
\end{eqnarray}
\end{itemize}
Considering that the RH neutrinos can not be massless, we find that $M_N$ can be of one of the following types,
\begin{eqnarray} \nonumber
  &&\mathfrak{N}^{(1)}_{0}:\left( \begin{array}{cc} \times & \times \\ \times & \times \end{array} \right)\,,\\ \nonumber
  &&\mathfrak{N}^{(1)}_{1}: \left( \begin{array}{cc} 0 & \times  \\ \times & \times  \\ \end{array} \right),\quad
\mathfrak{N}^{(2)}_{1}: \left( \begin{array}{cc} \times & \times  \\ \times & 0  \\ \end{array} \right),\quad \mathfrak{N}^{(3)}_{1}:\left( \begin{array}{cc} \times & 0  \\ 0 & \times  \\ \end{array} \right),\\  \label{eq:TZ_MN}
&& \mathfrak{N}^{(1)}_{2}:\left( \begin{array}{cc} 0 & \times  \\ \times & 0  \\ \end{array} \right),
\label{eq:MDSStext}
\end{eqnarray}
up to row and column permutations. All the above textures can be achieved with the singlet assignments $N^{c}_{\alpha}\sim \mathbf{1}^{l_{\alpha}}$, while the doublet assignment $N^{c}\sim\mathbf{2}^{k}$ can only give rise to $\mathfrak{N}^{(1)}_{0}$. Notice that we have included the pattern $\mathfrak{N}^{(1)}_{0}$ which does not exhibit any zero elements. This is due to the fact that we focus on the texture zero of the effective neutrino mass matrix $M_{\nu}$ which is given by the seesaw formula $M_{\nu}=-M^{T}_{D}M_{N}^{-1}M_{D}$. Thus, even if there is no zero elements in $M_{N}$, $M_{\nu}$ can still have texture zeros, as long as the Dirac neutrino mass matrix $M_{D}$ takes suitable form.

The Dirac neutrino mass term arises from the Yukawa couplings of LH leptons and RH neutrinos. For some representation and weight assignments of these fields, texture zeros in the Dirac neutrino mass matrix $M_{D}$ can be obtained. With two RH neutrinos, we find that $M_{D}$ can take the following six zero patterns,
\begin{eqnarray}\nonumber
&&  \mathfrak{D}^{(1)}_{1}:\left( \begin{array}{ccc} \times & \times & \times \\
                               \times & \times & 0 \end{array} \right)\,,\\ \nonumber
&&  \mathfrak{D}^{(1)}_{2}:\left( \begin{array}{ccc} 0 & \times & \times \\
                               \times & 0 & \times \end{array} \right)\,,\quad
 \mathfrak{D}^{(2)}_{2}:\left( \begin{array}{ccc} \times & \times & \times \\
                               0 & 0 & \times \end{array} \right)\,,\quad
 \mathfrak{D}^{(3)}_{2}:\left( \begin{array}{ccc} \times & \times & 0 \\
                               \times & \times & 0 \end{array} \right)\,,\\ \nonumber
&&\mathfrak{D}^{(1)}_{3}:\left( \begin{array}{ccc} \times & \times & 0 \\
                               0 & 0 & \times \end{array} \right)\,,\\ \label{eq:TZ_MD}
&&\mathfrak{D}^{(1)}_{4}:\left( \begin{array}{ccc} 0 & \times & 0 \\
                               \times & 0 & 0 \end{array} \right)\,,
\end{eqnarray}
up to row and column permutations. We have omitted the cases in which one row or two columns of $M_D$ are vanishing, since they would lead to two massless neutrinos which are not compatible with experimental data. If all elements of $M_D$ are non-vanishing, $M_{\nu}$ will not have texture zeros too and, consequently, we do not consider this case in the following. We analyse the $M_D$ patterns stemming from all possible lepton field representation assignments.
\begin{itemize}

\item{$L\equiv (L_{1},L_{2},L_{3})^{T}\sim \mathbf{3},~~N^{c}\sim \mathbf{2}^{k}$}

If the $L$ and $N^c$ fields transform as the triplet and doublet of $T'$, respectively,  the superpotential for the Dirac neutrino Yukawa couplings is
\begin{equation}
\mathcal{W}_\nu^{D} = \sum_{\mathbf{r},a} g^{D}_{\mathbf{r},a} \left[ N^c L Y_{\mathbf{r}a}^{(k_{L}+k_{N^{c}})} \right]_{\mathbf{1}}H_u\,,
\label{eq:Wnu_D_1}
\end{equation}
where $g^{D}_{\mathbf{r},a}$ are coupling constants. From Table~\ref{Tab:Submatrix}, we can read out the elements of $M_D$ from those given for $M_E$ in Eq.~\eqref{eq:exp_U} by performing the replacements $g^{E}\rightarrow g^{D}$, $k_{E_{D}^{c}}\rightarrow k_{N^{c}}$ and $v_d\rightarrow v_u$. We find that only texture $\mathfrak{D}_{2}^{(1)}$ can be achieved when $k_{L}+k_{N^{c}}=1$.

\item{$L_{D}\equiv (L_{1},L_{2})^{T}\sim \mathbf{2}^{i},~~L_{3} \sim \mathbf{1}^{j},~~N^{c}\sim \mathbf{2}^{k}$}

In this case, 
\begin{equation}\mathcal{W}_{\nu}^{D}=\sum_{\mathbf{r},a,b}g_{1\mathbf{r},a}^{D}\left[N^{c}L_{D}Y^{(k_{L_{D}}+k_{N^{c}})}_{\mathbf{r}a}\right]_{\mathbf{1}}H_{u}+g_{2\mathbf{r},b}^{D}\left[N^{c}L_{3}Y^{(k_{L_{3}}+k_{N^{c}})}_{\mathbf{r}b}\right]_{\mathbf{1}}H_{u}\,. \label{eq:Wnu_D_2}
  \end{equation}
Once more, using the results for $M_E$ given in Eq.~\eqref{eq:W_E_5}, we can obtain the $M_{D}$ elements from Eqs.~\eqref{eq:exp_S} and \eqref{eq:exp_C} with the replacement $g^{E}\rightarrow g^{D}$, $k_ {E_{D}^{c}}\rightarrow k_{N^{c}}$ and $v_{d}\rightarrow v_{u}$. Doing so, $M_{D}$ can take the following three types of texture zeros,
\begin{equation}
  \mathfrak{D}_{2}^{(1)}\,,~~\mathfrak{D}_{2}^{(3)}\,,~~\mathfrak{D}_{4}^{(1)}\,.
\end{equation}

\item{$L_{\beta}\sim \mathbf{1}^{j_{\beta}},\quad  N^{c}\sim \mathbf{2}^{k}$}

In case all the LH leptons transform as singlets of $T'$, we have
\begin{equation}
\label{eq:Wnu_D_3}
\mathcal{W}_{\nu}^{D}=\sum_{\beta=1}^{3}\sum_{\mathbf{r},a}g^{D}_{\beta\mathbf{r},a}\left[N^{c}L_{\beta}Y_{\mathbf{r}_{a}}^{(k_{L_{\beta}}+k_{N^{c}})}\right]_{\mathbf{1}}H_{u}\,,
\end{equation}
and the elements of $M_{D}$ read
\begin{eqnarray}\nonumber
(M_{D})_{\alpha\beta}&=&v_{u}\sum_{a,b,c}\Bigg[ g^{D}_{\mathbf{2},a} \delta_{2,(k+j)|3}(-1)^{\alpha+1}Y^{(k_{L_{\beta}}+k_{N^{c}})}_{\mathbf{2}a,3-\alpha} + g^{D}_{\mathbf{2'},b} \delta_{1,(k+j)|3}(-1)^{\alpha+1}Y^{(k_{L_{\beta}}+k_{N^{c}})}_{\mathbf{2}'b,3-\alpha}\\
&~&\qquad \quad +g^{D}_{\mathbf{2}'',c} \delta_{0,(k+j)|3}(-1)^{\alpha+1}Y^{(k_{L_{\beta}}+k_{N^{c}})}_{\mathbf{2}''c,3-\alpha}\Bigg]\;,\; \alpha=1,2\;,\beta=1,2,3\,.
\end{eqnarray}
We find that there is only one allowed texture-zero pattern for $M_{D}$, which is $\mathfrak{D}_{2}^{(3)}$ (up to row and column permutations).

\item{$L\equiv (L_{1},L_{2},L_{3})^{T}\sim \mathbf{3},~~N^{c}_{\alpha}\sim \mathbf{1}^{l_{\alpha}}$}

With the LH doublet (RH neutrino) fields transforming as triplet (singlets) of $T'$:
\begin{equation}
\label{eq:Wnu_D_4}\mathcal{W}_{\nu}^{D}=\sum_{\alpha=1}^{2}\sum_{\mathbf{r},a}g^{D}_{\alpha\mathbf{r},a}\left[N_{\alpha}^{c}LY_{\mathbf{3}a}^{(k_{L}+k_{N^{c}_{\alpha}})}\right]_{\mathbf{1}}H_{u}\,,
\end{equation}
and the general form of $M_D$ can be extracted from Eq.~\eqref{eq:general-ele-lepton} with the replacements $g^{E}\rightarrow g^{D}$, $k_{E_{\alpha}^{c}}\rightarrow k_{N_{\alpha}^c}$ and $v_d\rightarrow v_u$.
One row of $M_D$ may vanish in this case, and at least two light neutrino would be massless. 

\item{$L_{D}\equiv (L_{1},L_{2})^{T}\sim \mathbf{2}^{i},~~ L_{3} \sim \mathbf{1}^{j},~~N_{\alpha}^{c}\sim \mathbf{1}^{l_{\alpha}}$}

Assigning two lepton doublets to a doublet of $T'$ and the remaining fields to singlets, leads to the superpotential
\begin{equation}
\label{eq:Wnu_D_5}
\mathcal{W}_{\nu}^{D}=\sum_{\alpha=1}^{2}\sum_{\mathbf{r},a,b}g^{D}_{\alpha1\mathbf{r},a}\left[N_{\alpha}^{c}L_{D}Y_{\mathbf{r}a}^{(k_{L_{D}}+k_{N^{c}_{\alpha}})}\right]_{\mathbf{1}}H_{u}+g^{D}_{\alpha2\mathbf{r},b}\left[N_{\alpha}^{c}
L_{3}Y_{\mathbf{r}b}^{(k_{L_{3}}+k_{N^{c}_{\alpha}})}\right]_{\mathbf{1}}H_{u}\,,
\end{equation}
which is analogous to $\mathcal{W}_{E}$ given in Eq.~\eqref{eq:We_5}.  The corresponding expressions of $M_{D}$ can be obtained from $M_{E}$ in Eqs.~\eqref{eq:ME_6_1} and \eqref{eq:ME_6_2} by performing the same replacements as in the previous case. Consequently, the texture-zero patterns
\begin{equation}
\mathfrak{D}_{1}^{(1)}\,,~~\mathfrak{D}_{2}^{(2)}\,,~~\mathfrak{D}_{2}^{(3)}\,,~~\mathfrak{D}_{3}^{(1)}\,,
\end{equation}
can be reproduced.

\item{$L_{\beta}\sim \mathbf{1}^{j_{\beta}},~~N^{c}_{\alpha}\sim \mathbf{1}^{l_{\alpha}}$}

If all lepton multiplets transform as one-dimensional representations of $T'$, then
\begin{equation}
\mathcal{W}_{\nu}^{D}=\sum_{\alpha=1}^{2}\sum_{\beta=1}^{3}\sum_{\mathbf{r},a}g^{D}_{\alpha\beta\mathbf{r},a}\left[N^{c}_{\alpha}L_{\beta}Y_{\mathbf{r}a}^{(k_{L_{\beta}}+k_{N^{c}_{\alpha}} )}\right]_{\mathbf{1}}H_{u}\,.
\end{equation}
The corresponding $M_{D}$ can be obtained from Eq.~\eqref{eq:We-singlets}. As explained at the end of Section~\ref{sec:charged_lepton_sector}, we will not discuss this representation assignments. 
\end{itemize}

So far, we have discussed separately the texture-zero patterns of the Dirac neutrino mass matrix $M_D$ and the RH Majorana neutrino mass matrix $M_N$ for the case of two-RH neutrino fields. The possible $M_{N}$ and $M_{D}$ textures were summarised in \eqref{eq:TZ_MN} and \eqref{eq:TZ_MD}, respectively. The effective light neutrino mass matrix $M_{\nu}$ is given by  the well-known seesaw formula $M_{\nu}=-M^{T}_{D}M_{N}^{-1}M_{D}$ and, combining all possible $\mathfrak{D}$ and $\mathfrak{N}$ structures for $M_D$ and $M_N$, ten possible texture-zero patterns arise for $M_{\nu}$. These are summarised in Table~\ref{Table:seesaw_Mnu}, from which one sees that some $M_{\nu}$ textures can be realised from different $(M_{D},M_{N})$ pairs. For the sake of completeness, we will consider all possible $(M_{D},M_{N})$ combinations in the following. \\

\begin{table}[t!]
 \centering
 \begin{tabular}{|c|c|c|} \hline\hline
  $M_{D}$ & $M_{N}$ & $M_{\nu}$\\
   \hline
 $\mathfrak{D}_{1}^{(1)}$, $\mathfrak{D}_{2}^{(1)}$, $\mathfrak{D}_{3}^{(1)}$ & $\mathfrak{N}_{1}^{(1)}$ &   \multirow{3}{*}{$\mathcal{W}_{1}^{(1)}$}\\ \cline{1-2}
 $\mathfrak{D}_{1}^{(1)}$ & $\mathfrak{N}_{2}^{(1)}$ & \\ \cline{1-2}
 $\mathfrak{D}_{2}^{(1)}$ & $\mathfrak{N}_{1}^{(2)}$ & \\ \hline
 $\mathfrak{D}_{2}^{(1)}$ & $\mathfrak{N}_{1}^{(3)}$ &  $\mathcal{W}_{1}^{(2)}$\\ \hline
$\mathfrak{D}_{2}^{(1)}$ & $\mathfrak{N}_{2}^{(1)}$ &  $\mathcal{W}_{2}^{(1)}$ \\ \hline
$\mathfrak{D}_{3}^{(1)}$ & $\mathfrak{N}_{1}^{(3)}$ & $\mathcal{W}_{2}^{(2)}$ \\ \hline
 $\mathfrak{D}_{2}^{(2)}$ & $\mathfrak{N}_{1}^{(2)}$, $\mathfrak{N}_{2}^{(1)}$ & \multirow{2}{*}{$\mathcal{W}_{3}^{(1)}$}\\ \cline{1-2}
$\mathfrak{D}_{3}^{(1)}$ & $\mathfrak{N}_{1}^{(2)}$ & \\ \hline
 $\mathfrak{D}_{2}^{(3)}$ & $\mathfrak{N}_{0}^{(1)}$, $\mathfrak{N}_{1}^{(1)}$, $\mathfrak{N}_{1}^{(2)}$, $\mathfrak{N}_{1}^{(3)}$, $\mathfrak{N}_{2}^{(1)}$ &    \multirow{2}{*}{$\mathcal{W}_{3}^{(2)}$} \\ \cline{1-2}
   $\mathfrak{D}_{4}^{(1)}$ & $\mathfrak{N}_{0}^{(1)}$ &\\ \hline
   $\mathfrak{D}_{3}^{(1)}$ & $\mathfrak{N}_{2}^{(1)}$ & $\mathcal{W}_{4}^{(1)}$ \\ \hline
   $\mathfrak{D}_{4}^{(1)}$ & $\mathfrak{N}_{1}^{(1)}$, $\mathfrak{N}_{1}^{(2)}$ &  $\mathcal{W}_{4}^{(2)}$ \\ \hline
   $\mathfrak{D}_{4}^{(1)}$ & $\mathfrak{N}_{1}^{(3)}$ &  $\mathcal{W}_{4}^{(3)}$ \\ \hline
   $\mathfrak{D}_{4}^{(1)}$ & $\mathfrak{N}_{2}^{(1)}$ & $\mathcal{W}_{5}^{(1)}$ \\
   \hline \hline
 \end{tabular}
 \caption{\label{Table:seesaw_Mnu}Textures for the effective neutrino mass matrix $M_{\nu}$ in the type-I seesaw mechanism with two RH neutrino fields. $\mathfrak{D}_{1}^{(1)}-\mathfrak{D}_{4}^{(1)}$ are the texture-zero patterns for the Dirac neutrino mass matrix $M_{D}$ and $\mathfrak{N}_{0}^{(1)}-\mathfrak{N}_{2}^{(1)}$ are those for the RH Majorana neutrino mass matrix $M_{N}$ -- see \eqref{eq:TZ_MN} and \eqref{eq:TZ_MD}. The explicit forms of the resulting $M_\nu$ textures $\mathcal{W}_{1}^{(1)}-\mathcal{W}_{5}^{(1)}$ are given in Table~\ref{tab:classes_Majorana_neutrino_mass_matrix}.}
\end{table}

Before proceeding to the phenomenological analysis of lepton models based on the texture-zero patterns discussed in the previous sections, it is worth summarising our findings up to this point. As shown in Sections~\ref{sec:charged_lepton_sector} and \ref{sec:neutrino_sector}, texture zeros of fermion mass matrices can be naturally implemented in the context of modular flavour symmetries, being the nonvanishing elements of those matrices correlated as shown, for instance, in Table~\ref{Tab:Submatrix}. Thus, the resulting lepton models are expected to be much more predictive than, e.g., models based in Abelian symmetries. In Tables~\ref{tab:classes_charged_lepton_mass_matrix}, \ref{tab:classes_Dirac_neutrino_mass_matrix} and \ref{tab:classes_Majorana_neutrino_mass_matrix}, we have listed all possible texture-zero patterns for the charged-lepton, Dirac neutrino and Majorana neutrino mass matrices, respectively, which are obtainable from the $T'$ modular symmetry. Notice that one could get alternative texture-zero patterns from finite modular groups other than $T'$. It is remarkable that some texture zeros of lepton mass matrices listed in Ref.~\cite{Ludl:2014axa} can be reproduced from the $T'$ modular symmetry, as shown in Table~\ref{tab:correspondence_classes_charged_lepton_mass_matrix}. However, some others cannot be realised in the framework of $T'$ modular group, as it is the case of texture ``$5_{1}^{(l)}$" of Ref.~\cite{Ludl:2014axa}. Notice that one texture of ours may correspond to several ones of~\cite{Ludl:2014axa} since our lepton mass matrices are defined up to row and column permutation due to the freedom of representation assignment of matter fields. Moreover, the patterns with one or two zero elements in $M_{E}$ and one zero element in $M_{D}$ were not studied in~\cite{Ludl:2014axa}. Thus, $\mathcal{C}_{1}^{(1)}$, $\mathcal{C}_{2}^{(1)}$, $\mathcal{C}_{2}^{(2)}$, $\mathcal{C}_{2}^{(3)}$ and $\mathcal{D}_{1}^{(1)}$ discussed in this work have no correspondence in~\cite{Ludl:2014axa}.

\begin{table}[t!]
 \centering
 \resizebox{1.0\textwidth}{!}{
 \begin{tabular}{|rc|rc|} \hline\hline
$\mathcal{C}_{3}^{(1)}:$&$ ~3_{1}^{(l)}$&$\mathcal{D}_{4}^{(6)}:$&$ ~4_{6}^{(\nu_{D})},~4_{10}^{(\nu_{D})},~4_{14}^{(\nu_{D})}~4_{20}^{(\nu_{D})},~4_{25}^{(\nu_{D})},~4_{27}^{(\nu_{D})}$\\
$\mathcal{C}_{3}^{(2)}:$&$ ~3_{2}^{(l)}$&$\mathcal{D}_{5}^{(1)}:$&$ ~5_{3}^{(\nu_{D})},~5_{7}^{(\nu_{D})},~5_{9}^{(\nu_{D})}$\\
$\mathcal{C}_{3}^{(3)}:$&$ ~3_{3}^{(l)}$&$\mathcal{D}_{5}^{(2)}:$&$ ~5_{12}^{(\nu_{D})},~5_{20}^{(\nu_{D})},~5_{25}^{(\nu_{D})}$ \\
 $\mathcal{C}_{4}^{(1)}:$&$ ~4_{4}^{(l)}$&$\mathcal{D}_{5}^{(3)}:$&$ ~5_{4}^{(\nu_{D})},~5_{5}^{(\nu_{D})},~5_{8}^{(\nu_{D})}$\\
$\mathcal{C}_{4}^{(2)}:$&$ ~4_{1}^{(l)}$&$\mathcal{D}_{5}^{(4)}:$&$ ~5_{13}^{(\nu_{D})},~5_{17}^{(\nu_{D})},~5_{24}^{(\nu_{D})}$\\
$\mathcal{C}_{4}^{(3)}:$&$ ~4_{2}^{(l)}$&$\mathcal{D}_{5}^{(5)}:$&$ ~5_{1}^{(\nu_{D})},~5_{2}^{(\nu_{D})},~5_{6}^{(\nu_{D})}$\\
$\mathcal{C}_{6}^{(1)}:$&$ ~6_{1}^{(l)}$&$\mathcal{D}_{5}^{(6)}:$&$ ~5_{10}^{(\nu_{D})},~5_{11}^{(\nu_{D})},~5_{19}^{(\nu_{D})}~5_{21}^{(\nu_{D})},~5_{26}^{(\nu_{D})},~5_{27}^{(\nu_{D})}$ \\
   \cline{1-2}
$\mathcal{D}_{2}^{(1)}:$&$ ~2_{2}^{(\nu_{D})}$&$\mathcal{D}_{6}^{(1)}:$&$ ~6_{3}^{(\nu_{D})},~6_{5}^{(\nu_{D})},~6_{7}^{(\nu_{D})}$\\
$\mathcal{D}_{2}^{(2)}:$&$ ~2_{1}^{(\nu_{D})}$&$\mathcal{D}_{6}^{(2)}:$&$ ~6_{10}^{(\nu_{D})},~6_{11}^{(\nu_{D})},~6_{12}^{(\nu_{D})},6_{14}^{(\nu_{D})},~6_{15}^{(\nu_{D})},~6_{16}^{(\nu_{D})}$\\
$\mathcal{D}_{2}^{(3)}:$&$ ~2_{3}^{(\nu_{D})}$&$\mathcal{D}_{6}^{(3)}:$&$ ~6_{13}^{(\nu_{D})}$ \\
$\mathcal{D}_{3}^{(1)}:$&$ ~3_{3}^{(\nu_{D})},~3_{5}^{(\nu_{D})},~3_{10}^{(\nu_{D})}$&$\mathcal{D}_{7}^{(1)}:$&$ ~7_{1}^{(\nu_{D})},~7_{2}^{(\nu_{D})},~7_{3}^{(\nu_{D})}$\\ \cline{3-4}
$\mathcal{D}_{3}^{(2)}:$&$ ~3_{8}^{(\nu_{D})},~3_{9}^{(\nu_{D})},~3_{11}^{(\nu_{D})},~3_{13}^{(\nu_{D})},~3_{16}^{(\nu_{D})},~3_{17}^{(\nu_{D})}$&$\mathcal{W}_{1}^{(1)}:$&$ ~1_{1}^{(\nu_{L})},~1_{2}^{(\nu_{L})},~1_{3}^{(\nu_{L})}$\\
$\mathcal{D}_{3}^{(3)}:$&$ ~3_{12}^{(\nu_{D})}$&$\mathcal{W}_{1}^{(2)}:$&$ ~1_{4}^{(\nu_{L})},~1_{5}^{(\nu_{L})},~1_{6}^{(\nu_{L})}$ \\
$\mathcal{D}_{3}^{(4)}:$&$ ~3_{1}^{(\nu_{D})}$&$\mathcal{W}_{2}^{(1)}:$&$ ~2_{1}^{(\nu_{L})},~2_{2}^{(\nu_{L})},~2_{3}^{(\nu_{L})}$\\
$\mathcal{D}_{3}^{(5)}:$&$ ~3_{7}^{(\nu_{D})},~3_{15}^{(\nu_{D})},~3_{18}^{(\nu_{D})}$&$\mathcal{W}_{2}^{(2)}:$&$ ~2_{13}^{(\nu_{L})},~2_{14}^{(\nu_{L})},~2_{15}^{(\nu_{L})}$\\
$\mathcal{D}_{4}^{(1)}:$&$ ~4_{12}^{(\nu_{D})},~4_{18}^{(\nu_{D})},~4_{21}^{(\nu_{D})}$&$\mathcal{W}_{3}^{(1)}:$&$ ~3_{2}^{(\nu_{L})},~3_{5}^{(\nu_{L})},~3_{10}^{(\nu_{L})}$\\
$\mathcal{D}_{4}^{(2)}:$&$ ~4_{5}^{(\nu_{D})},~4_{9}^{(\nu_{D})},~4_{17}^{(\nu_{D})}$&$\mathcal{W}_{3}^{(2)}:$&$ ~3_{11}^{(\nu_{L})},~3_{16}^{(\nu_{L})},~3_{19}^{(\nu_{L})}$\\
$\mathcal{D}_{4}^{(3)}:$&$ ~4_{7}^{(\nu_{D})},~4_{16}^{(\nu_{D})},~4_{26}^{(\nu_{D})}$&$\mathcal{W}_{4}^{(2)}:$&$ ~4_{1}^{(\nu_{L})},~4_{2}^{(\nu_{L})},~4_{3}^{(\nu_{L})},~4_{4}^{(\nu_{L})},~4_{5}^{(\nu_{L})},~4_{6}^{(\nu_{L})}$\\
$\mathcal{D}_{4}^{(4)}:$&$ ~4_{4}^{(\nu_{D})},~4_{13}^{(\nu_{D})},~4_{24}^{(\nu_{D})}$&$\mathcal{W}_{4}^{(3)}:$&$ ~4_{7}^{(\nu_{L})},~4_{8}^{(\nu_{L})},~4_{9}^{(\nu_{L})}$\\
$\mathcal{D}_{4}^{(5)}:$&$ ~4_{1}^{(\nu_{D})},~4_{2}^{(\nu_{D})},~4_{3}^{(\nu_{D})}$& &\\
   \hline\hline
 \end{tabular}}
\caption{\label{tab:correspondence_classes_charged_lepton_mass_matrix}
Correspondence between our texture-zero patterns $\mathcal{C}$, $\mathcal{D}$ and $\mathcal{W}$ and those of Ref.~\cite{Ludl:2014axa}. Since the assignment of the three generations of lepton fields under $T'$ modular symmetry can be freely exchanged, the lepton mass matrices are determined up to independent row and column permutations. Consequently, one texture zero of ours could correspond to several textures of Ref.~\cite{Ludl:2014axa}.  }
\end{table}

\clearpage

\section{Testing the full set of TZLMs}
\label{sec:lepton_models}

After having analysed separately the realisation of texture zeros in the charged-lepton and neutrino mass matrices with $T'$ modular forms up to weight 6,  we now combine them with the purpose of studying their phenomenology. In practice, we will test all possible texture-zero pairs $(M_{E},M_{\nu})$, where $M_{\nu}$ can be either Dirac or Majorana neutrino mass matrix, taking into account the results of the previous sections. Since the LH charged-lepton and neutrino fields belong to $SU(2)_{L}$ doublets, they should transform in the same way under modular symmetry, i.e., they should share the same modular weight and representation assignments. Each of these pairs correspond to a different {\em texture-zero lepton model} or TZLM. In this paper we are interested in those TZLMs which are, in some sense, predictive. The set of observables against which the test will be performed contains essentially nine experimentally-measured lepton-mass and mixing parameters, namely
\begin{equation}
  m_{e},~m_{\mu},~m_{\tau},~\Delta m_{21}^{2},~\Delta m_{31}^{2},~\theta_{12},~\theta_{13},~\theta_{23},~\delta_{CP}\,,
  \label{eq:observables}
\end{equation}
where $m_{e}$, $m_{\mu}$, $m_{\tau}$ are three charged-lepton masses, $\Delta m_{21}^{2}$, $\Delta m_{31}^{2}$ are two neutrino mass-squared differences, $\theta_{12}$, $\theta_{13}$, $\theta_{23}$ are three lepton mixing angles and $\delta_{CP}$ is the Dirac CP-violation phase. Note that the present statistical significance of the $\delta_{CP}$ measurement is rather weak, and the preferred value of $\delta_{CP}$ from global data analyses should be taken with a grain of salt~\cite{Esteban:2020cvm}. 

Given the above set of observables, we shall focus on TZLMs with nine or less real free input parameters. Notice that, in general, all coupling constants appearing in previous sections can be complex. Still, some of them can be made real by rephasing fields. Thus, for a given TZLM, the number of input parameters is counted after removing all unphysical complex phases. Interestingly, one can further constrain the number of inputs, and thus increase the TZLM predictive power, by imposing the generalized CP symmetry (gCP). In the context of modular symmetries, it has been found that the gCP acts on the complex modulus as
$\tau\xrightarrow[]{\mathcal{CP}} -\tau^{*}$, up to modular transformations~\cite{Acharya:1995ag,Dent:2001cc,Giedt:2002ns,Baur:2019kwi,Novichkov:2019sqv}.
In the basis where both modular generators $S$ and $T$ are represented by symmetric matrices in all irreducible representations, gCP reduces to the canonical CP transformation~\cite{Novichkov:2019sqv}. Hence, all coupling constants are forced to be real if the CG coefficients are real in the symmetric basis. As shown in Appendix~\ref{app:Tp_group}, we are indeed working in the symmetric basis of the $T'$ group with real CG coefficients. Consequently, imposing gCP in our TZLMs amounts to considering all coupling real and, as a result, the number of free parameters is reduced. Given this interesting possibility, we will consider both cases with and without gCP. In all cases we require the total number of real input parameters to be less than or equal to nine. For the three different neutrino-mass generation mechanisms considered in this work we have:\\

{\bf Dirac neutrinos:} From Tables~\ref{tab:classes_charged_lepton_mass_matrix} and \ref{tab:classes_Dirac_neutrino_mass_matrix}, we see the $T'$ modular symmetry can realise 11 (25) distinct texture-zero patterns for $M_{E}$ ($M_{D}$). As pointed out in Sections~\ref{sec:charged_lepton_sector} and \ref{sec:neutrino_sector}, texture $\mathcal{C}_{6}^{(1)}$ for $M_{E}$ and $\mathcal{D}_{6}^{(3)}$ and $\mathcal{D}_{7}^{(1)}$ for $M_{D}$, lead to degenerate mass eigenvalues, which is not compatible with experimental data. Thus, they will be neglected hereafter. It is straightforward to conclude that there are $10\times 23=230$ pairs $(M_{E},M_{D})$. Considering that the LH neutrinos and charged leptons should transform in the same way under modular symmetry, and sticking to our classification of a predictive TZLM, we find that 136 out of those 230 classes of $(M_{E},M_{D})$ can be realised from $T'$ modular symmetry. If gCP is imposed, the number increases to 174. The allowed patterns of $(M_{E},M_{D})$ in $T'$ modular symmetry are summarised in Tables~\ref{tab:Dirac_pair_summary_NO} and \ref{tab:Dirac_pair_summary_IO} (more details on the contents of these tables will be given in the next section).\\

{\bf Majorana neutrinos:} For Majorana neutrinos, we have obtained 11 $M_\nu$ texture-zero patterns, as summarised in Table~\ref{tab:classes_Majorana_neutrino_mass_matrix}. Textures $\mathcal{W}_{4}^{(1)}$, $\mathcal{W}_{4}^{(4)}$ and $\mathcal{W}_{5}^{(1)}$ predict two degenerate neutrino masses, thus they will be excluded in the following analysis. If light neutrino masses are described by the Weinberg operator, combining the possible constructions in the neutrino and charged-lepton sectors, we find that 29 pairs $(M_{E},M_{\nu})$ can be achieved with $T'$ modular symmetry without gCP, and two additional ones are allowed if gCP symmetry is imposed. These findings are summarized in Table~\ref{tab:Majorana_WO_pair_summary_NO_IO}. For neutrino masses generated via the minimal type-I seesaw mechanism (see Section~\ref{sec:SS_neutrino_mass}), the possible $M_{D}$ and $M_{N}$ textures, as well as the resulting patterns for $M_{\nu}$, are presented in Table~\ref{Table:seesaw_Mnu}. Notice that some $M_{\nu}$ given by seesaw formula can be realised from more than one $(M_{D},M_{N})$. We find 35 (36) possible pairs if gCP is not (is) included in the $T'$ modular models, which are listed in Table~\ref{tab:Majorana_SS_pair_summary_NO_IO}.

\subsection{Numerical analysis and TZLM predictions}
\label{sec:numerical_analysis}

In order to quantitatively estimate how well the different TZLMs describe the data, we perform a $\chi^2$ analysis to find out the best fit values of the input parameters for each model, as well as the predictions for lepton mass and mixing parameters. As common practice, we consider both normal ordering (NO) and inverted ordering (IO) for the neutrino mass spectrum, depending on whether $m_1 < m_2 < m_3$ or $m_3 < m_1 < m_2$, respectively. The $\chi^{2}$ function is defined as usual:
\begin{equation}
\chi^2=\sum^{n}_{i=1}\left(\frac{P_i(x_1, x_2,\ldots, x_m)-\mu_i}{\sigma_i}\right)^2\,,
\end{equation}
where $P_i$ are the predictions of a given TZLM for the physical observables $\theta_{12}$, $\theta_{13}$, $\theta_{23}$, $\delta_{CP}$, $m_{e}/m_{\mu}$, $m_{\mu}/m_{\tau}$ and $\Delta m_{21}^{2}/\Delta m_{31}^{2}$. These are nontrivial functions of the free input parameters in each TZLM. $\mu_i$ and $\sigma_i$ denote the best-fit values and standard deviations of the corresponding quantities extracted from global analysis of the data -- see Table~\ref{tab:experimental_data}.

We adopt the standard parametrization of the lepton mixing matrix~\cite{ParticleDataGroup:2022pth},
\begin{equation}\label{eq:PMNS}
U=\left(\begin{array}{ccc}
c_{12}c_{13}  &   s_{12}c_{13}   &   s_{13}e^{-i\delta_{CP}}  \\
-s_{12}c_{23}-c_{12}s_{13}s_{23}e^{i\delta_{CP}}   &  c_{12}c_{23}-s_{12}s_{13}s_{23}e^{i\delta_{CP}}  &  c_{13}s_{23}  \\
s_{12}s_{23}-c_{12}s_{13}c_{23}e^{i\delta_{CP}}   & -c_{12}s_{23}-s_{12}s_{13}c_{23}e^{i\delta_{CP}}  &  c_{13}c_{23}
\end{array}\right)\text{diag}(1,e^{i\frac{\alpha_{21}}{2}},e^{i\frac{\alpha_{31}}{2}})\,,
\end{equation}
where $c_{ij}\equiv \cos\theta_{ij}$, $s_{ij}\equiv \sin\theta_{ij}$, $\delta_{CP}$ is the Dirac CP violation phase, and $\alpha_{21,31}$, are Majorana CP phases. If the lightest neutrino is massless, there is a single Majorana phase $\phi$ and the diagonal phase matrix in the above equation can be replaced by $\text{diag}(1,e^{i\phi/2},1)$. Information on the Majorana phases could be potentially extracted from  neutrinoless double beta decay ($0\nu\beta\beta$) experiments. If only light neutrino masses provide contributions to  $0\nu\beta\beta$, the decay amplitude depends on the effective Majorana neutrino mass $m_{\beta\beta}$,
\begin{equation}
  m_{\beta\beta}=|m_{1}\cos^{2}\theta_{12}\cos^{2}\theta_{13}+m_{2}\sin^{2}\theta_{12}\cos^{2}\theta_{13}e^{i\alpha_{21}}+m_{3}\sin^{2}\theta_{13}e^{i(\alpha_{31}-2\delta_{CP})}|\,,
\end{equation}
which, in case the lightest neutrino is massless, reduces to
\begin{equation}
  m_{\beta\beta}=\left\{ \begin{aligned} &|m_{2}\sin^{2}\theta_{12}\cos^{2}\theta_{13}e^{i\phi}+m_{3}\sin^{2}\theta_{13}e^{-i2\delta_{CP}}|,& m_{1}=0\;({\rm NO})\,,
      \\ &|m_{1}\cos^{2}\theta_{12}\cos^{2}\theta_{13}+m_{2}\sin^{2}\theta_{12}\cos^{2}\theta_{13}e^{i\phi}|\,, & m_{3}=0\;({\rm IO})\,.\end{aligned} \right.
\end{equation}

\begin{table}[t!]
 \centering
   \begin{tabular}{|c|c|c|c|c|} \hline\hline
     \multirow{2}{*}{ Observables} & \multicolumn{2}{c|}{Normal Ordering} & \multicolumn{2}{c|}{Inverted Ordering}  \\
     \cline{2-5}
       & $\text{bfp}\pm 1\sigma$ & $3\sigma$ range & $\text{bfp}\pm 1\sigma$ & $3\sigma$ range  \\
     \hline
     $\sin^{2}\theta_{12}$ & $0.304_{-0.012}^{+0.012}$ & $0.269\rightarrow 0.343$ & $0.304_{-0.012}^{+0.013}$ & $0.269\rightarrow 0.343$ \\
     \hline
     $\sin^{2}\theta_{13}$ & $0.02246_{-0.00062}^{+0.00062}$ & $0.02060\rightarrow 0.02435$ & $0.02241_{-0.00062}^{+0.00074}$ & $0.02055\rightarrow 0.02457$ \\
     \hline
      $\sin^{2}\theta_{23}$ & $0.450_{-0.016}^{+0.019}$ & $0.408\rightarrow 0.603$ & $0.570_{-0.022}^{+0.016}$ & $0.410\rightarrow 0.613$ \\
     \hline
      $\delta_{\text{CP}}/^\circ$ & $230_{-25}^{+36}$ & $144\rightarrow 350$ & $278_{-30}^{+22}$ & $194\rightarrow 345$ \\
     \hline
     $\dfrac{\Delta m_{21}^{2}}{10^{-5} \mathrm{eV}^{2}}$ & $7.42_{-0.20}^{+0.21}$ & $6.82\rightarrow 8.04$ & $7.42_{-0.20}^{+0.21}$ & $6.82\rightarrow 8.04$ \\
     \hline
     $\dfrac{\Delta m_{3l}^{2}}{10^{-3} \mathrm{eV}^{2}}$ & $2.515_{-0.028}^{+0.028}$ & $2.430\rightarrow 2.593$ & $-2.490_{-0.028}^{+0.026}$ & $2.410\rightarrow 2.574$ \\ \hline
     & \multicolumn{4}{c|}{$\text{bfv}\pm 1\sigma$ }\\ \hline
     $m_{e}/m_{\mu}$ & \multicolumn{4}{c|}{$0.004737\pm 0.000040$} \\ \hline
     $m_{\mu}/m_{\tau}$ & \multicolumn{4}{c|}{$0.05857 \pm 0.00047$} \\
     \hline
     $m_{\tau}/\text{GeV}$ & \multicolumn{4}{c|}{$1.30234$} \\
     \hline
     \hline
 \end{tabular}
\caption{\label{tab:experimental_data} Allowed ranges for the neutrino oscillation parameters obtained from global analysis of the data, and values of the charged-lepton mass ratios. Here, we use the NuFIT v5.1 results with Super-Kamiokanda atmospheric data~\cite{Esteban:2020cvm}. Note that $\Delta m_{3l}^{2}\equiv \Delta m_{31}^{2}>0$ for NO and $\Delta m_{3l}^{2}\equiv \Delta m_{32}^{2}<0$ for IO. The charged-lepton masses are taken from~\cite{Antusch:2013jca} with $\tan\beta=10$ and SUSY-breaking scale $M_{\text{SUSY}}=10~\text{TeV}$.}
\end{table}

In this work we will use the NuFIT v5.1 results~\cite{Esteban:2020cvm} for the allowed ranges of the neutrino oscillation parameters (see also Refs.~\cite{deSalas:2020pgw} and \cite{Capozzi:2021fjo}). The charge-lepton masses will enter in the $\chi^2$ function in the form of their ratios, being the best fit values (bfv) and $1\sigma$ errors taken from Ref.~\cite{Antusch:2013jca}. For $\chi^2$ minimisation and consequent determination of the inputs which best fit the data, we use the CERN package  \texttt{TMinuit}~\cite{minuit}. The input value of the modulus $\tau$ will be a random complex number in the fundamental domain $\mathcal{F}: |\text{Re}\tau|\leq\frac{1}{2}$, $\text{Im}\tau>0$ and $|\tau|\geq1$. The absolute values and phases of all coupling constants are free to vary in the ranges $[0, 10^6]$ and $[0, 2\pi)$, respectively. As a general criterion, we will consider that a given TZLM is compatible with experimental data if the predicted values of the neutrino mass and mixing parameters at the minimum of the $\chi^{2}$ are within the $3\sigma$ ranges given in Table~\ref{tab:experimental_data}. For the charged-lepton masses, the model's best-fit values should not deviate from the experimental central values by more than $3\sigma$. After setting the general grounds of out numerical analysis, we now discuss its results for all allowed TZLMs identified in Section~\ref{sec:lepton_models}.

\subsubsection{Dirac neutrinos}
\label{sec:Drac_neutrinos_fitting}

\begin{table}[t!]
 \centering
 \resizebox{1.0\textwidth}{!}{
\begin{tabular}{|c|c|c|c|c|c|c|c|c|c|c|c|} \hline\hline
\diagbox{$M_{D}$}{$M_{E}$}& $\mathcal{C}_{1}^{(1)}$ & $\mathcal{C}_{2}^{(1)}$ & $\mathcal{C}_{2}^{(2)}$ & $\mathcal{C}_{2}^{(3)}$ & $\mathcal{C}_{3}^{(1)}$ & $\mathcal{C}_{3}^{(2)}$ & $\mathcal{C}_{3}^{(3)}$ & $\mathcal{C}_{4}^{(1)}$ & $\mathcal{C}_{4}^{(2)}$ & $\mathcal{C}_{4}^{(3)}$ \\ \hline$\mathcal{D}_{1}^{(1)}$  & \ding{52} (\ding{52}) &- (\ding{52}) & \ding{56} (\ding{52}) & \ding{52} (\ding{52}) &- (\ding{52}) &- (-) & \ding{56} (\ding{56})  & \ding{52} (\ding{52}) & \ding{56} (\ding{56})  & \ding{52} (\ding{52}) \\ \hline$\mathcal{D}_{2}^{(1)}$  & \ding{56} (\ding{52}) & \ding{56} (\ding{52}) & \ding{56} (\ding{52}) & \ding{52} (\ding{52}) & \ding{52} (\ding{52}) &- (-) & \ding{56} (\ding{56})  & \ding{56} (\ding{56})  & \ding{56} (\ding{52}) & \ding{56} (\ding{56})  \\ \hline$\mathcal{D}_{2}^{(2)}$  & \ding{56} (\ding{52}) &- (\ding{52}) & \ding{52} (\ding{52}) & \ding{56} (\ding{52}) &- (\ding{52}) &- (-) & \ding{56} (\ding{56})  & \ding{56} (\ding{52}) & \ding{56} (\ding{56})  & \ding{56} (\ding{56})  \\ \hline$\mathcal{D}_{2}^{(3)}$  & \ding{56} (\ding{52}) &- (\ding{52}) & \ding{56} (\ding{52}) & \ding{56} (\ding{52}) &- (\ding{52}) &- (-) & \ding{56} (\ding{56})  & \ding{52} (\ding{52}) & \ding{56} (\ding{56})  & \ding{56} (\ding{52}) \\ \hline$\mathcal{D}_{3}^{(1)}$  &- (\ding{52}) &- (\ding{56}) &- (\ding{52}) &- (\ding{52}) &- (-) &- (-) &- (\ding{56}) & \ding{52} (\ding{52}) &- (\ding{56}) &- (\ding{52}) \\ \hline$\mathcal{D}_{3}^{(2)}$  &- (-) &- (-) &- (-) &- (-) &- (-) &- (-) &- (-) &- (-) &- (-) &- (-) \\ \hline$\mathcal{D}_{3}^{(3)}$  & \ding{56} (\ding{56})  &- (\ding{52}) & \ding{56} (\ding{52}) & \ding{56} (\ding{56})  &- (\ding{52}) &- (-) & \ding{56} (\ding{56})  & \ding{56} (\ding{52}) & \ding{56} (\ding{56})  &- (-) \\ \hline$\mathcal{D}_{3}^{(4)}$  & \ding{56} (\ding{52}) & \ding{56} (\ding{52}) & \ding{56} (\ding{52}) & \ding{52} (\ding{52}) & \ding{52} (\ding{52}) &- (-) & \ding{56} (\ding{52}) & \ding{56} (\ding{52}) & \ding{56} (\ding{52}) & \ding{52} (\ding{52}) \\ \hline$\mathcal{D}_{3}^{(5)}$  & \ding{56} (\ding{52}) & \ding{56} (\ding{56})  & \ding{56} (\ding{52}) & \ding{52} (\ding{52}) & \ding{52} (\ding{52}) &- (-) & \ding{56} (\ding{56})  & \ding{56} (\ding{56})  & \ding{56} (\ding{56})  & \ding{56} (\ding{56})  \\ \hline$\mathcal{D}_{4}^{(1)}$  & \ding{52} (\ding{52}) & \ding{56} (\ding{56})  & \ding{56} (\ding{52}) & \ding{52} (\ding{52}) & \ding{52} (\ding{52}) &- (-) & \ding{56} (\ding{52}) & \ding{56} (\ding{56})  & \ding{56} (\ding{56})  & \ding{56} (\ding{56})  \\ \hline$\mathcal{D}_{4}^{(2)}$  & \ding{56} (\ding{56})  &- (\ding{56}) & \ding{56} (\ding{52}) & \ding{56} (\ding{56})  &- (\ding{52}) &- (-) & \ding{56} (\ding{56})  & \ding{56} (\ding{56})  & \ding{56} (\ding{56})  &- (-) \\ \hline$\mathcal{D}_{4}^{(3)}$  & \ding{56} (\ding{52}) &- (\ding{56}) & \ding{56} (\ding{52}) & \ding{56} (\ding{52}) &- (\ding{52}) &- (-) &- (-) & \ding{56} (\ding{56})  &- (-) & \ding{56} (\ding{56})  \\ \hline$\mathcal{D}_{4}^{(4)}$  &- (\ding{52}) &- (\ding{52}) &- (\ding{52}) &- (\ding{52}) &- (\ding{52}) &- (-) &- (\ding{56}) & \ding{56} (\ding{52}) &- (\ding{56}) &- (\ding{56}) \\ \hline$\mathcal{D}_{4}^{(5)}$  & \ding{56} (\ding{52}) &- (\ding{52}) & \ding{56} (\ding{52}) & \ding{56} (\ding{52}) &- (\ding{52}) &- (-) & \ding{56} (\ding{56})  & \ding{52} (\ding{52}) & \ding{56} (\ding{56})  & \ding{56} (\ding{56})  \\ \hline$\mathcal{D}_{4}^{(6)}$  &- (-) &- (-) &- (\ding{52}) &- (-) &- (-) &- (-) &- (\ding{52}) & \ding{56} (\ding{56})  &- (-) &- (-) \\ \hline$\mathcal{D}_{5}^{(1)}$  & \ding{52} (\ding{52}) & \ding{56} (\ding{56})  & \ding{56} (\ding{52}) & \ding{52} (\ding{52}) & \ding{52} (\ding{52}) &- (-) & \ding{56} (\ding{56})  & \ding{56} (\ding{56})  & \ding{56} (\ding{56})  & \ding{56} (\ding{56})  \\ \hline$\mathcal{D}_{5}^{(2)}$  & \ding{56} (\ding{52}) & \ding{56} (\ding{56})  & \ding{56} (\ding{52}) & \ding{56} (\ding{56})  & \ding{56} (\ding{56})  &- (-) & \ding{56} (\ding{52}) & \ding{56} (\ding{56})  & \ding{56} (\ding{56})  & \ding{56} (\ding{56})  \\ \hline$\mathcal{D}_{5}^{(3)}$  & \ding{56} (\ding{52}) & \ding{56} (\ding{56})  & \ding{56} (\ding{52}) & \ding{56} (\ding{56})  & \ding{52} (\ding{52}) &- (-) & \ding{56} (\ding{56})  & \ding{56} (\ding{56})  & \ding{56} (\ding{56})  &- (-) \\ \hline$\mathcal{D}_{5}^{(4)}$  & \ding{56} (\ding{52}) &- (\ding{56}) & \ding{56} (\ding{52}) & \ding{56} (\ding{52}) & \ding{56} (\ding{56})  &- (-) &- (-) & \ding{56} (\ding{56})  &- (-) & \ding{56} (\ding{56})  \\ \hline$\mathcal{D}_{5}^{(5)}$  & \ding{56} (\ding{52}) &- (\ding{52}) & \ding{52} (\ding{52}) & \ding{56} (\ding{52}) &- (\ding{52}) &- (-) & \ding{56} (\ding{56})  & \ding{56} (\ding{56})  & \ding{56} (\ding{56})  & \ding{56} (\ding{56})  \\ \hline$\mathcal{D}_{5}^{(6)}$  &- (-) &- (-) &- (\ding{52}) &- (-) &- (-) &- (-) &- (\ding{52}) & \ding{56} (\ding{56})  &- (-) &- (-) \\ \hline$\mathcal{D}_{6}^{(1)}$  & \ding{56} (\ding{52}) & \ding{56} (\ding{56})  & \ding{56} (\ding{52}) & \ding{56} (\ding{56})  & \ding{56} (\ding{56})  &- (-) & \ding{56} (\ding{56})  & \ding{56} (\ding{56})  & \ding{56} (\ding{56})  & \ding{56} (\ding{56})  \\ \hline$\mathcal{D}_{6}^{(2)}$  &- (\ding{52}) &- (-) & \ding{56} (\ding{56})  &- (\ding{52}) &- (-) &- (-) & \ding{56} (\ding{56})  & \ding{56} (\ding{56})  &- (-) &- (-) \\ \hline \hline \hline \end{tabular}}
\caption{\label{tab:Dirac_pair_summary_NO}
Allowed combinations of $(M_{E},M_{D})$ realised by $T'$ modular symmetry with and without gCP for NO neutrino mass spectrum. For all texture pairs the symbol inside (outside) the parenthesis corresponds to the case when gCP is (is not) imposed. Also, ``-" means that a particular configuration cannot be achieved by modular symmetry, while ``\ding{52}" (``\ding{56}" ) identifies those cases which are realised by the modular symmetry and (but) are (are not) phenomenologically viable.}
\end{table}

\begin{table}[t!]
 \centering
 \resizebox{1.0\textwidth}{!}{
\begin{tabular}{|c|c|c|c|c|c|c|c|c|c|c|c|} \hline\hline
\diagbox{$M_{D}$}{$M_{E}$}& $\mathcal{C}_{1}^{(1)}$ & $\mathcal{C}_{2}^{(1)}$ & $\mathcal{C}_{2}^{(2)}$ & $\mathcal{C}_{2}^{(3)}$ & $\mathcal{C}_{3}^{(1)}$ & $\mathcal{C}_{3}^{(2)}$ & $\mathcal{C}_{3}^{(3)}$ & $\mathcal{C}_{4}^{(1)}$ & $\mathcal{C}_{4}^{(2)}$ & $\mathcal{C}_{4}^{(3)}$ \\ \hline$\mathcal{D}_{1}^{(1)}$  & \ding{52} (\ding{52}) &- (\ding{52}) & \ding{52} (\ding{52}) & \ding{52} (\ding{52}) &- (\ding{52}) &- (-) & \ding{56} (\ding{56})  & \ding{52} (\ding{52}) & \ding{56} (\ding{56})  & \ding{52} (\ding{52}) \\ \hline$\mathcal{D}_{2}^{(1)}$  & \ding{56} (\ding{52}) & \ding{56} (\ding{52}) & \ding{56} (\ding{56})  & \ding{56} (\ding{56})  & \ding{56} (\ding{56})  &- (-) & \ding{56} (\ding{56})  & \ding{56} (\ding{56})  & \ding{56} (\ding{52}) & \ding{56} (\ding{56})  \\ \hline$\mathcal{D}_{2}^{(2)}$  & \ding{56} (\ding{52}) &- (\ding{56}) & \ding{52} (\ding{52}) & \ding{56} (\ding{52}) &- (\ding{52}) &- (-) & \ding{56} (\ding{56})  & \ding{56} (\ding{52}) & \ding{56} (\ding{56})  & \ding{56} (\ding{52}) \\ \hline$\mathcal{D}_{2}^{(3)}$  & \ding{56} (\ding{52}) &- (\ding{52}) & \ding{56} (\ding{52}) & \ding{56} (\ding{52}) &- (\ding{52}) &- (-) & \ding{56} (\ding{56})  & \ding{52} (\ding{52}) & \ding{56} (\ding{56})  & \ding{56} (\ding{52}) \\ \hline$\mathcal{D}_{3}^{(1)}$  &- (\ding{52}) &- (\ding{56}) &- (\ding{52}) &- (\ding{56}) &- (-) &- (-) &- (\ding{56}) & \ding{56} (\ding{56})  &- (\ding{56}) &- (\ding{56}) \\ \hline$\mathcal{D}_{3}^{(2)}$  &- (-) &- (-) &- (-) &- (-) &- (-) &- (-) &- (-) &- (-) &- (-) &- (-) \\ \hline$\mathcal{D}_{3}^{(3)}$  & \ding{56} (\ding{56})  &- (\ding{56}) & \ding{56} (\ding{56})  & \ding{56} (\ding{56})  &- (\ding{56}) &- (-) & \ding{56} (\ding{56})  & \ding{56} (\ding{56})  & \ding{56} (\ding{56})  &- (-) \\ \hline$\mathcal{D}_{3}^{(4)}$  & \ding{52} (\ding{52}) & \ding{56} (\ding{52}) & \ding{52} (\ding{52}) & \ding{52} (\ding{52}) & \ding{52} (\ding{52}) &- (-) & \ding{56} (\ding{52}) & \ding{52} (\ding{52}) & \ding{56} (\ding{56})  & \ding{52} (\ding{52}) \\ \hline$\mathcal{D}_{3}^{(5)}$  & \ding{52} (\ding{52}) & \ding{56} (\ding{56})  & \ding{52} (\ding{52}) & \ding{52} (\ding{52}) & \ding{52} (\ding{52}) &- (-) & \ding{56} (\ding{56})  & \ding{56} (\ding{56})  & \ding{56} (\ding{56})  & \ding{56} (\ding{56})  \\ \hline$\mathcal{D}_{4}^{(1)}$  & \ding{56} (\ding{52}) & \ding{56} (\ding{56})  & \ding{56} (\ding{52}) & \ding{52} (\ding{52}) & \ding{52} (\ding{52}) &- (-) & \ding{56} (\ding{56})  & \ding{56} (\ding{56})  & \ding{56} (\ding{56})  & \ding{56} (\ding{56})  \\ \hline$\mathcal{D}_{4}^{(2)}$  & \ding{56} (\ding{56})  &- (\ding{56}) & \ding{56} (\ding{56})  & \ding{56} (\ding{56})  &- (\ding{56}) &- (-) & \ding{56} (\ding{56})  & \ding{56} (\ding{56})  & \ding{56} (\ding{56})  &- (-) \\ \hline$\mathcal{D}_{4}^{(3)}$  & \ding{56} (\ding{56})  &- (\ding{56}) & \ding{56} (\ding{56})  & \ding{56} (\ding{56})  &- (\ding{56}) &- (-) &- (-) & \ding{56} (\ding{56})  &- (-) & \ding{56} (\ding{56})  \\ \hline$\mathcal{D}_{4}^{(4)}$  &- (\ding{56}) &- (\ding{56}) &- (\ding{56}) &- (\ding{56}) &- (\ding{56}) &- (-) &- (\ding{56}) & \ding{56} (\ding{56})  &- (\ding{56}) &- (\ding{56}) \\ \hline$\mathcal{D}_{4}^{(5)}$  & \ding{52} (\ding{52}) &- (\ding{52}) & \ding{52} (\ding{52}) & \ding{56} (\ding{52}) &- (\ding{52}) &- (-) & \ding{56} (\ding{56})  & \ding{52} (\ding{56}) & \ding{56} (\ding{56})  & \ding{56} (\ding{56})  \\ \hline$\mathcal{D}_{4}^{(6)}$  &- (-) &- (-) &- (\ding{52}) &- (-) &- (-) &- (-) &- (\ding{56}) & \ding{56} (\ding{56})  &- (-) &- (-) \\ \hline$\mathcal{D}_{5}^{(1)}$  & \ding{52} (\ding{52}) & \ding{56} (\ding{56})  & \ding{52} (\ding{52}) & \ding{52} (\ding{52}) & \ding{52} (\ding{52}) &- (-) & \ding{56} (\ding{56})  & \ding{56} (\ding{56})  & \ding{56} (\ding{56})  & \ding{56} (\ding{56})  \\ \hline$\mathcal{D}_{5}^{(2)}$  & \ding{56} (\ding{56})  & \ding{56} (\ding{56})  & \ding{56} (\ding{52}) & \ding{56} (\ding{56})  & \ding{56} (\ding{56})  &- (-) & \ding{56} (\ding{56})  & \ding{56} (\ding{56})  & \ding{56} (\ding{56})  & \ding{56} (\ding{56})  \\ \hline$\mathcal{D}_{5}^{(3)}$  & \ding{56} (\ding{52}) & \ding{56} (\ding{56})  & \ding{56} (\ding{52}) & \ding{56} (\ding{56})  & \ding{56} (\ding{56})  &- (-) & \ding{56} (\ding{56})  & \ding{56} (\ding{56})  & \ding{56} (\ding{56})  &- (-) \\ \hline$\mathcal{D}_{5}^{(4)}$  & \ding{56} (\ding{52}) &- (\ding{56}) & \ding{56} (\ding{52}) & \ding{56} (\ding{56})  & \ding{56} (\ding{56})  &- (-) &- (-) & \ding{56} (\ding{56})  &- (-) & \ding{56} (\ding{56})  \\ \hline$\mathcal{D}_{5}^{(5)}$  & \ding{56} (\ding{56})  &- (\ding{56}) & \ding{52} (\ding{56}) & \ding{56} (\ding{56})  &- (\ding{56}) &- (-) & \ding{56} (\ding{56})  & \ding{56} (\ding{56})  & \ding{56} (\ding{56})  & \ding{56} (\ding{56})  \\ \hline$\mathcal{D}_{5}^{(6)}$  &- (-) &- (-) &- (\ding{52}) &- (-) &- (-) &- (-) &- (\ding{56}) & \ding{56} (\ding{56})  &- (-) &- (-) \\ \hline$\mathcal{D}_{6}^{(1)}$  & \ding{56} (\ding{56})  & \ding{56} (\ding{56})  & \ding{56} (\ding{56})  & \ding{56} (\ding{56})  & \ding{56} (\ding{56})  &- (-) & \ding{56} (\ding{56})  & \ding{56} (\ding{56})  & \ding{56} (\ding{56})  & \ding{56} (\ding{56})  \\ \hline$\mathcal{D}_{6}^{(2)}$  &- (\ding{56}) &- (-) & \ding{56} (\ding{56})  &- (\ding{56}) &- (-) &- (-) & \ding{56} (\ding{56})  & \ding{56} (\ding{56})  &- (-) &- (-) \\ \hline \hline \hline \end{tabular}}
\caption{\label{tab:Dirac_pair_summary_IO}The same as in Table~\ref{tab:Dirac_pair_summary_NO} but for IO neutrino mass spectrum.}
\end{table}

For Dirac neutrinos, the allowed combinations of $(M_{E},M_{D})$ realised by $T'$ modular symmetry with and without gCP are presented in Table~\ref{tab:Dirac_pair_summary_NO} (Table~\ref{tab:Dirac_pair_summary_IO}) for NO (IO). For all texture pairs the symbol inside (outside) the parenthesis corresponds to the case when gCP is (is not) imposed. Also, ``-" means that a particular configuration cannot be achieved by modular symmetry, while ``\ding{52}" (``\ding{56}" ) identifies those cases which are realised by the modular symmetry and (but) are (are not) phenomenologically viable. From Table~\ref{tab:Dirac_pair_summary_NO} we can see that for NO and if gCP is not included, 27 out of 136 allowed texture pairs provide a good fit to the experimental data . In Table~\ref{tab:assign_Dirac_NO} of Appendix~\ref{sec:appendix_C}, we show a representative model for each viable $(M_{E},M_{D})$ pair by listing the representation and modular-weight assignments of lepton fields. The corresponding values of $\sin^2\theta_{ij}$, $m_i$ and $\delta_{CP}$ at the best-fit point are given in Table~\ref{tab:prediction_Dirac_NO}. From Table~\ref{tab:assign_Dirac_NO}, we can see that the minimal models have $8$ free real input parameters. In the IO case, there are 34 viable pairs of $(M_{E},M_{D})$, as shown in Table~\ref{tab:Dirac_pair_summary_IO}, and we provide the field assignments of representative models and the corresponding fitting results of lepton parameters in Tables~\ref{tab:assign_Dirac_IO} and \ref{tab:prediction_Dirac_IO}, respectively. All representative models are found to contain either $8$ or $9$ free real parameters. 

When gCP is imposed, there are allowed $(M_{E},M_{D})$ pairs. By performing the $\chi^{2}$ analysis of the corresponding TZLMs, we obtain 97 (88) phenomenologically-viable ones for NO (IO), as presented in Table~\ref{tab:Dirac_pair_summary_NO} (Table~\ref{tab:Dirac_pair_summary_IO}). The details of the representative models and corresponding predictions for each compatible case are displayed in Tables~\ref{tab:assign_Dirac_NO} and \ref{tab:prediction_Dirac_NO} (Tables~\ref{tab:assign_Dirac_IO} and \ref{tab:prediction_Dirac_IO}) for NO (IO). We find that the number of input parameters for representative models are at least 8, as in the case of no gCP.

At this point, it is worth comparing the parameter counting of TZLMs realised by $T'$ modular symmetry with that of texture-zero scenarios where nonvanishing entries in the mass matrices are not correlated. We have seen that, in the case of Dirac neutrinos, and demanding the number of inputs not to exceed 9, viable TZLMs require either $8$ or $9$ real parameters. We emphasize that this number is much smaller than what is typically found with uncorrelated nonvanishing matrix elements. This is apparent in Tables~\ref{tab:assign_Dirac_NO} and \ref{tab:assign_Dirac_IO} where we compare the number of free parameters ``$\#\text{P}$'' for each TZLM realisable with $T'$ modular symmetry (third column) with the same number for the same matrix textures but with uncorrelated nonzero entries in the mass matrices (``$\#\text{P}_{0}$'' in the second column). For all cases $\#\text{P}\ll\#\text{P}_{0}$, showing that modular symmetry increases drastically the predictive power of TZLMs. As an example, in the case 
$(M_{E},M_{D}) \sim (C_{2}^{(3)},D_{5}^{(1)})$, we find that $\#\text{P}_{0}=20$, while the same pair is realised in a specific $T'$ modular model with only $8$ free parameters -- see line 17 of Table~\ref{tab:assign_Dirac_NO}. As will become clear in the following, this is a general feature which is present also for Majorana neutrinos.

\subsubsection{Majorana neutrinos}

We now repeat the analysis presented in the previous section but for TZLMs with Majorana neutrinos. Both cases of $M_\nu$ generated via a Weinberg operator and via the minimal type-I seesaw mechanism will be considered. Each model is now classified according to the textures of its corresponding $(M_E,M_\nu)$ pair.
\\

{\bf Weinberg operator:} For neutrino masses generated by a Weinberg operator, and without imposing gCP, there are 29 texture pairs $(M_{E},M_{\nu})$ which can be realised with $T'$ modular symmetry, as shown in Table~\ref{tab:Majorana_WO_pair_summary_NO_IO} for both NO and IO neutrino masses. By performing their $\chi^2$ analysis, we find that, out of those 29 models, only 5 are compatible with experimental data for NO neutrino masses spectrum (upper rows of the table). As done for Dirac neutrinos, we present in Table~\ref{tab:assign_Majorana_WO_NO} the representation and weight assignments of a representative model for each phenomenologically viable case, while the best fitting results of physical observables and the model predictions are provided in Table~\ref{tab:prediction_Majorana_WO_NO}. For IO neutrino mass spectrum, we obtain 7 phenomenologically viable $(M_{E},M_{\nu})$ pairs, for which the representative models are listed in Table~\ref{tab:assign_Majorana_WO_IO}, and the numerical results in Table~\ref{tab:prediction_Majorana_WO_IO}. Note that the minimal viable NO (IO) model $\mathcal{C}_{4}^{(1)}-\mathcal{W}_{1}^{(1)}$ ($\mathcal{C}_{2}^{(2)}-\mathcal{W}_{3}^{(1)}$) -- see Table~\ref{tab:assign_Majorana_WO_NO} (Table~\ref{tab:assign_Majorana_WO_IO}) -- is realisable with only 7 (8) input parameters to explain all 9 measured observables. 

In case the TZLMs are constructed imposing gCP, 11 out of the 31 allowed $(M_{E},M_{\nu})$ pairs are consistent with the measured data for NO as shown in Table~\ref{tab:Majorana_WO_pair_summary_NO_IO}. The representative models for these cases are given in Table~\ref{tab:assign_Majorana_WO_NO} and the corresponding fitting results in Table~\ref{tab:prediction_Majorana_WO_NO}. The number of input parameters for the representative models vary from 7 to 9. For IO with gCP, we have 10 $(M_{E},M_{\nu})$ pairs compatible with data (see Tables~\ref{tab:Majorana_WO_pair_summary_NO_IO}, \ref{tab:assign_Majorana_WO_IO} and \ref{tab:prediction_Majorana_WO_IO} for the representative models and fitting results).\\

\begin{table}[t!]
 \centering
 \resizebox{1.0\textwidth}{!}{
\begin{tabular}{|c|c|c|c|c|c|c|c|c|c|c|c|} \hline\hline
\multicolumn{11}{|c|}{NO} \\ \hline
\diagbox{$M_{\nu}$}{$M_{E}$} & $\mathcal{C}_{1}^{(1)}$ & $\mathcal{C}_{2}^{(1)}$ & $\mathcal{C}_{2}^{(2)}$ & $\mathcal{C}_{2}^{(3)}$ & $\mathcal{C}_{3}^{(1)}$ & $\mathcal{C}_{3}^{(2)}$ & $\mathcal{C}_{3}^{(3)}$ & $\mathcal{C}_{4}^{(1)}$ & $\mathcal{C}_{4}^{(2)}$ & $\mathcal{C}_{4}^{(3)}$ \\ \hline$\mathcal{W}_{1}^{(1)}$  & \ding{56} (\ding{52}) & \ding{56} (\ding{56})  & \ding{56} (\ding{52}) & \ding{52} (\ding{52}) & \ding{52} (\ding{52}) & -(-) & -(\ding{56}) & \ding{52} (\ding{52}) & -(\ding{56}) & \ding{56} (\ding{56})  \\ \hline$\mathcal{W}_{2}^{(2)}$  & \ding{56} (\ding{52}) & \ding{56} (\ding{56})  & \ding{56} (\ding{52}) & \ding{52} (\ding{52}) & \ding{52} (\ding{52}) & -(-) & -(-) & \ding{56} (\ding{56})  & -(-) & \ding{56} (\ding{56})  \\ \hline$\mathcal{W}_{3}^{(1)}$  & \ding{56} (\ding{52}) & \ding{56} (\ding{56})  & \ding{56} (\ding{52}) & -(-) & \ding{56} (\ding{56})  & -(-) & \ding{56} (\ding{56})  & -(-) & \ding{56} (\ding{56})  & -(-) \\ \hline$\mathcal{W}_{3}^{(2)}$  & \ding{56} (\ding{56})  & \ding{56} (\ding{56})  & \ding{56} (\ding{56})  & \ding{56} (\ding{56})  & \ding{56} (\ding{56})  & -(-) & \ding{56} (\ding{56})  & \ding{56} (\ding{56})  & \ding{56} (\ding{56})  & \ding{56} (\ding{56})  \\ \hline
\multicolumn{11}{|c|}{IO} \\ \hline
\diagbox{$M_{\nu}$}{$M_{E}$} & $\mathcal{C}_{1}^{(1)}$ & $\mathcal{C}_{2}^{(1)}$ & $\mathcal{C}_{2}^{(2)}$ & $\mathcal{C}_{2}^{(3)}$ & $\mathcal{C}_{3}^{(1)}$ & $\mathcal{C}_{3}^{(2)}$ & $\mathcal{C}_{3}^{(3)}$ & $\mathcal{C}_{4}^{(1)}$ & $\mathcal{C}_{4}^{(2)}$ & $\mathcal{C}_{4}^{(3)}$  \\ \hline$\mathcal{W}_{1}^{(1)}$  & \ding{56} (\ding{52}) & \ding{56} (\ding{56})  & \ding{56} (\ding{52}) & \ding{52} (\ding{52}) & \ding{52} (\ding{52}) &- (-) &- (\ding{56}) & \ding{52} (\ding{52}) &- (\ding{56}) & \ding{56} (\ding{56})  \\ \hline$\mathcal{W}_{2}^{(2)}$  & \ding{56} (\ding{52}) & \ding{56} (\ding{56})  & \ding{56} (\ding{52}) & \ding{52} (\ding{52}) & \ding{52} (\ding{52}) &- (-) &- (-) & \ding{56} (\ding{56})  &- (-) & \ding{56} (\ding{56})  \\ \hline$\mathcal{W}_{3}^{(1)}$  & \ding{56} (\ding{56})  & \ding{56} (\ding{56})  & \ding{52} (\ding{56}) &- (-) & \ding{52} (\ding{56}) &- (-) & \ding{56} (\ding{56})  &- (-) & \ding{56} (\ding{56})  &- (-) \\ \hline$\mathcal{W}_{3}^{(2)}$  & \ding{56} (\ding{52}) & \ding{56} (\ding{56})  & \ding{56} (\ding{56})  & \ding{56} (\ding{56})  & \ding{56} (\ding{56})  &- (-) & \ding{56} (\ding{56})  & \ding{56} (\ding{56})  & \ding{56} (\ding{56})  & \ding{56} (\ding{56})  \\ \hline \hline \end{tabular}}
\caption{\label{tab:Majorana_WO_pair_summary_NO_IO} Same analysis as in Tables~\ref{tab:Dirac_pair_summary_NO} and 
\ref{tab:Dirac_pair_summary_IO} but now for a Majorana neutrino mass matrix $M_\nu$ generated via the Weinberg operator -- see Table~\ref{tab:classes_Majorana_neutrino_mass_matrix} for the general structure of the $\mathcal{W}$ textures. The results for NO (IO) are shown in the upper (lower) part of the table. 
}
\end{table}

{\bf Seesaw mechanism:} When Majorana neutrino masses are generated through the (minimal) type-I seesaw mechanism, we find that 35 $(M_{E},M_{\nu})$ pairs are realised by $T'$ modular symmetry if gCP is not considered -- see Table~\ref{tab:Majorana_SS_pair_summary_NO_IO}. For NO (IO) neutrino mass spectrum, only 6 (11) of these 35 cases are phenomenologically viable. As illustrated in Table~\ref{Table:seesaw_Mnu}, some patterns of $M_{\nu}$ can be realised in more than one way with different $(M_{D},M_{N})$ combinations. For the sake of completeness, we present representative models for distinct $(M_{D},M_{N})$ pairs leading to the same textures of $(M_{E},M_{\nu})$ in Tables~\ref{tab:assign_Majorana_SS_NO} and \ref{tab:assign_Majorana_SS_IO} for NO and IO, respectively. The best-fit values of lepton mass and mixing parameters are instead given in  Tables~\ref{tab:prediction_Majorana_SS_NO} and \ref{tab:prediction_Majorana_SS_IO}. All viable NO models contain $9$ free real parameters, while for IO this number can be $8$ or $9$. With gCP, there are 36 allowed pairs of $(M_{E},M_{\nu})$, being 13 and 15 of them phenomenologically viable for NO and IO, respectively (see Table~\ref{tab:Majorana_SS_pair_summary_NO_IO}). The representative models and fitting results analogous to the previous cases are shown in Tables~\ref{tab:assign_Majorana_SS_NO}-\ref{tab:prediction_Majorana_SS_IO}. Due to the fact that the coupling constants are required to be real by gCP, we can find viable models with $8$ parameters for NO neutrino masses, while this number does not changed for IO when compared with the non-gCP case.\\

We present a grand summary of our results in Table~\ref{tab:chisq_res} where, for the three considered neutrino mass generation mechanisms, we show the number of texture pairs with no more than 9 real input parameters which can be realised via $T'$ modular symmetry (third column) with and without gCP (second column). The number of phenomenologically viable cases (i.e. those which are able to fit the data at the $3\sigma$ level) is shown in the fourth column for both NO and IO. 

\begin{table}[t!]
 \centering
 \resizebox{1.0\textwidth}{!}{
\begin{tabular}{|c|c|c|c|c|c|c|c|c|c|c|c|} \hline\hline
\multicolumn{11}{|c|}{NO} \\ \hline
\diagbox{$M_{\nu}$}{$M_{E}$}  & $\mathcal{C}_{1}^{(1)}$ & $\mathcal{C}_{2}^{(1)}$ & $\mathcal{C}_{2}^{(2)}$ & $\mathcal{C}_{2}^{(3)}$ & $\mathcal{C}_{3}^{(1)}$ & $\mathcal{C}_{3}^{(2)}$ & $\mathcal{C}_{3}^{(3)}$ & $\mathcal{C}_{4}^{(1)}$ & $\mathcal{C}_{4}^{(2)}$ & $\mathcal{C}_{4}^{(3)}$ \\ \hline$\mathcal{W}_{1}^{(1)}$  & \ding{56} (\ding{52}) & \ding{56} (\ding{56})  & \ding{56} (\ding{52}) & \ding{52} (\ding{52}) & \ding{52} (\ding{52}) &- (-) & \ding{56} (\ding{56})  & \ding{56} (\ding{52}) & \ding{56} (\ding{56})  & \ding{56} (\ding{56})  \\ \hline$\mathcal{W}_{1}^{(2)}$  &- (-) &- (-) &- (-) &- (-) &- (-) &- (-) &- (-) &- (-) &- (-) &- (-) \\ \hline$\mathcal{W}_{2}^{(1)}$  &- (-) &- (-) &- (-) &- (-) &- (-) &- (-) &- (-) &- (-) &- (-) &- (-) \\ \hline$\mathcal{W}_{2}^{(2)}$  & \ding{56} (\ding{56})  & \ding{56} (\ding{56})  & \ding{56} (\ding{56})  & \ding{56} (\ding{56})  &- (\ding{56}) &- (-) & \ding{56} (\ding{56})  & \ding{56} (\ding{56})  & \ding{56} (\ding{56})  & \ding{56} (\ding{56})  \\ \hline$\mathcal{W}_{3}^{(1)}$  & \ding{56} (\ding{52}) & \ding{56} (\ding{56})  & \ding{56} (\ding{52}) & \ding{56} (\ding{52}) & \ding{56} (\ding{52}) &- (-) & \ding{56} (\ding{56})  & \ding{56} (\ding{56})  & \ding{56} (\ding{56})  & \ding{56} (\ding{56})  \\ \hline$\mathcal{W}_{3}^{(2)}$  & \ding{52} (\ding{52}) & \ding{56} (\ding{56})  & \ding{52} (\ding{52}) & \ding{52} (\ding{52}) & \ding{52} (\ding{52}) &- (-) & \ding{56} (\ding{56})  & \ding{56} (\ding{56})  & \ding{56} (\ding{56})  & \ding{56} (\ding{56})  \\ \hline$\mathcal{W}_{4}^{(2)}$  &- (-) &- (-) &- (-) &- (-) &- (-) &- (-) &- (-) &- (-) &- (-) &- (-) \\ \hline$\mathcal{W}_{4}^{(3)}$  &- (-) &- (-) &- (-) &- (-) &- (-) &- (-) &- (-) &- (-) &- (-) &- (-) \\ \hline
\multicolumn{11}{|c|}{IO} \\ \hline
\diagbox{$M_{\nu}$}{$M_{E}$}  & $\mathcal{C}_{1}^{(1)}$ & $\mathcal{C}_{2}^{(1)}$ & $\mathcal{C}_{2}^{(2)}$ & $\mathcal{C}_{2}^{(3)}$ & $\mathcal{C}_{3}^{(1)}$ & $\mathcal{C}_{3}^{(2)}$ & $\mathcal{C}_{3}^{(3)}$ & $\mathcal{C}_{4}^{(1)}$ & $\mathcal{C}_{4}^{(2)}$ & $\mathcal{C}_{4}^{(3)}$ \\ \hline$\mathcal{W}_{1}^{(1)}$  & \ding{52} (\ding{52}) & \ding{56} (\ding{56})  & \ding{56} (\ding{52}) & \ding{52} (\ding{52}) & \ding{52} (\ding{52}) &- (-) & \ding{56} (\ding{56})  & \ding{52} (\ding{52}) & \ding{56} (\ding{56})  & \ding{56} (\ding{52}) \\ \hline$\mathcal{W}_{1}^{(2)}$  &- (-) &- (-) &- (-) &- (-) &- (-) &- (-) &- (-) &- (-) &- (-) &- (-) \\ \hline$\mathcal{W}_{2}^{(1)}$  &- (-) &- (-) &- (-) &- (-) &- (-) &- (-) &- (-) &- (-) &- (-) &- (-) \\ \hline$\mathcal{W}_{2}^{(2)}$  & \ding{56} (\ding{56})  & \ding{56} (\ding{56})  & \ding{56} (\ding{56})  & \ding{56} (\ding{56})  &- (\ding{56}) &- (-) & \ding{56} (\ding{56})  & \ding{56} (\ding{56})  & \ding{56} (\ding{56})  & \ding{56} (\ding{56})  \\ \hline$\mathcal{W}_{3}^{(1)}$  & \ding{56} (\ding{52}) & \ding{56} (\ding{56})  & \ding{56} (\ding{52}) & \ding{52} (\ding{52}) & \ding{52} (\ding{52}) &- (-) & \ding{56} (\ding{56})  & \ding{56} (\ding{56})  & \ding{56} (\ding{56})  & \ding{56} (\ding{56})  \\ \hline$\mathcal{W}_{3}^{(2)}$  & \ding{52} (\ding{52}) & \ding{56} (\ding{56})  & \ding{52} (\ding{52}) & \ding{52} (\ding{52}) & \ding{52} (\ding{52}) &- (-) & \ding{56} (\ding{56})  & \ding{56} (\ding{56})  & \ding{56} (\ding{56})  & \ding{56} (\ding{56})  \\ \hline$\mathcal{W}_{4}^{(2)}$  &- (-) &- (-) &- (-) &- (-) &- (-) &- (-) &- (-) &- (-) &- (-) &- (-) \\ \hline$\mathcal{W}_{4}^{(3)}$  &- (-) &- (-) &- (-) &- (-) &- (-) &- (-) &- (-) &- (-) &- (-) &- (-) \\ \hline \hline \end{tabular}}
\caption{\label{tab:Majorana_SS_pair_summary_NO_IO} Same as in Table~\ref{tab:Majorana_WO_pair_summary_NO_IO} for $M_\nu$ generated via (minimal) type-I seesaw mechanism from $M_D$ and $M_N$ textures given in \eqref{eq:MDSStext} and \eqref{eq:MNSStext}, respectively. The $(M_D,M_N) \rightarrow M_\nu$ dictionary is provided in Table~\ref{Table:seesaw_Mnu}.
}
\end{table}

\begin{table}[ht!]
 \centering
   \begin{tabular}{|c|c|c|c|} \hline\hline
     Neutrino nature  &  gCP & Number of textures & Viable \\ \hline
     \multirow{4}{*}{Dirac}  & \multirow{2}{*}{no} & \multirow{2}{*}{136} & 23 (NO) \\ \cline{4-4}
   & & & 27 (IO) \\ \cline{2-4}
                     & \multirow{2}{*}{yes} & \multirow{2}{*}{174} & 97 (NO) \\ \cline{4-4}
                     & & & 57 (IO) \\ \hline
       & \multirow{2}{*}{no} & \multirow{2}{*}{29} & 5 (NO) \\ \cline{4-4}
                    Majorana & & & 7 (IO) \\ \cline{2-4}
                    (Weinberg operator) & \multirow{2}{*}{yes} & \multirow{2}{*}{31} & 11 (NO) \\ \cline{4-4}
                     & & & 10 (IO) \\ \hline
       & \multirow{2}{*}{no} & \multirow{2}{*}{35} & 6 (NO) \\ \cline{4-4}
                   Majorana  & & & 10 (IO) \\ \cline{2-4}
                   (Seesaw mechanism)  & \multirow{2}{*}{yes} & \multirow{2}{*}{36} & 13 (NO) \\ \cline{4-4}
      & & & 14 (IO) \\
     \hline \hline
   \end{tabular}
\caption{\label{tab:chisq_res} Grand summary of the TZLM compatibility analysis. We show the number of texture pairs with no more than 9 real input parameters which can be realised via $T'$ modular symmetry (third column) with and without gCP (second column). The number of phenomenologically viable cases (i.e. those which are able to fit the data at the $3\sigma$ level) is given in the fourth column for both NO and IO.}
\end{table}

\section{Benchmark models}
\label{sec:benchmark_models}

The analysis presented in the previous section provided a complete view of how modular symmetries drastically increase the predictive power of TZLMs. It is obviously impossible to go through all the listed cases in detail and to present a complete graphical treatment of each model predictions. Nevertheless, we believe it is worth providing a small set of benchmark cases where the quality of the results can be appreciated. With this purpose, in the following we select one TZLM realisable with $T'$ modular symmetry for each neutrino mass generation mechanism.

\subsection{Model for Dirac neutrino masses}

For the Dirac neutrino benchmark case, we consider the lepton fields transforming under the $T'$ modular symmetry $T'$ as
  \begin{eqnarray}\nonumber
    &&L_{D}\sim \mathbf{2}\,,~~L_{3}\sim \mathbf{1'}\,,~~e^{c}\sim \mathbf{1}\,,~~\mu^{c}\sim \mathbf{1''}\,,~~\tau^{c}\sim \mathbf{1'}\,,~~N_{D}^{c}\equiv \{N_{1}^{c},N_{2}^{c}\}\sim \mathbf{2}\,,~~N^{c}_{3}\sim \mathbf{1'}\,,
      \end{eqnarray}
      with modular-weight assignments:
       \begin{eqnarray}
    &&k_{L_{D}}=2-x\,,~~k_{L_{3}}=3-x\,,~~k_{e^{c}}=1+x\,,~~k_{\mu^{c}}=1+x\,,~~k_{\tau^{c}}=1+x\,,\\
    &&k_{N_{D}^{c}}=x\,,~~k_{N^{c}_{3}}=1+x\,,
  \end{eqnarray}
  where $x$ is an arbitrary integer. The corresponding modular-invariant superpotentials relevant for charged-lepton and neutrino masses are given by
\begin{eqnarray}\nonumber
  \mathcal{W}_{E}&=&y^{e}_{1}\, e^{c}\left[L_{D}Y^{(3)}_{\mathbf{2''}}\right]_{\mathbf{1}}H_{d} + y^{e}_{2} \,\mu^{c}\left[L_{D}Y^{(3)}_{\mathbf{2}}\right]_{\mathbf{1'}}H_{d} + y^{e}_{3}\, \mu^{c}L_{3}Y^{(4)}_{\mathbf{1}}H_{d} + y^{e}_{4} \,\tau^{c}L_{3}Y^{(4)}_{\mathbf{1'}}H_{d} \,,\\
  \mathcal{W}_{\nu}&=&y^{d}_{1}\left[(L_{D}N^{c})_{\mathbf{3}}Y^{(1)}_{\mathbf{3}}\right]_{\mathbf{1}}H_{u} +y^{d}_{2}L_{3}N_{3}^{c}\,Y^{(4)}_{\mathbf{1'}}H_{u}\,,
\end{eqnarray}
where the $y_k ^{e,d}$ couplings are, in principle, complex. Notice, however, that their phases can be eliminated by rephasing the lepton supermultiplets. As a result, we find that there are 8 real free parameters in this model: 6 real coupling constants ($y_{1-4}^e$ and $y_{1,2}^d$), and the complex modulus $\tau$. The charged-lepton and neutrino mass matrices read
\begin{eqnarray}
  M_{E}=\left(
\begin{array}{ccc}
 y^{e}_{1}Y_{\mathbf{2''},2}^{(3)} & -y^{e}_{1}Y_{\mathbf{2''},1}^{(3)} & 0 \medskip\\
 -y^{e}_{2} Y_{\mathbf{2},2}^{(3)} & y^{e}_{2} Y_{\mathbf{2},1}^{(3)} & y^{e}_{3} Y_{\mathbf{1}}^{(4)} \medskip\\
 0 & 0 & y^{e}_{4} Y_{\mathbf{1'}}^{(4)} \\
\end{array}
\right)v_{d}\,,\quad
M_{D}=\left(
\begin{array}{ccc}
 -y^{d}_{1}Y_{\mathbf{3},2}^{(2)} & \frac{y^{d}_{1}Y_{\mathbf{3},3}^{(2)}}{\sqrt{2}} & 0 \medskip\\
 \frac{y^{d}_{1}Y_{\mathbf{3},3}^{(2)}}{\sqrt{2}} & y^{d}_{1}Y_{\mathbf{3},1}^{(2)} & 0 \medskip\\
 0 & 0 & y^{d}_{2} Y_{\mathbf{1'}}^{(4)} \\
\end{array}
\right)v_{u}\,,
\end{eqnarray}
which correspond to the texture-zero pattern $\mathcal{C}_{3}^{(1)}-\mathcal{D}_{4}^{(1)}$ for $(M_{E},M_{D})$ is  -- see Tables~\ref{tab:classes_charged_lepton_mass_matrix} and ~\ref{tab:classes_Dirac_neutrino_mass_matrix} (this TZLM is also presented in Table~\ref{tab:assign_Dirac_NO}). We perform a global fit to the lepton experimental data and, for normally-ordered neutrino masses, the values of the input parameters at the best-fit point are
\begin{equation}
\hskip-0.2in \begin{gathered}
  \langle\tau\rangle=-0.30546 + 1.05008i\,,\quad y^{e}_{2}/y^{e}_{1}= 10.6173 \,,\quad y^{e}_{3}/y^{e}_{1}=21.5185\,,\quad  y^{e}_{4}/y^{e}_{1}=0.011167\,, \\ y^{d}_{2}/y^{d}_{1}=3.02666\,, \quad y^{e}_{1}v_{d} = 1.68246~\text{GeV}\,,\quad y^{d}_{1}v_{u} = 558.654~\text{meV}\,,
\end{gathered}
\end{equation}
to which correspond
\begin{eqnarray}
\nonumber&& \sin^{2}\theta_{12}=0.3043\,,\quad \sin^{2}\theta_{13}=0.02244\,,\quad \sin^{2}\theta_{23}=0.4509\,,\quad \delta_{CP}=208.4^{\circ}\,,\\
\nonumber&& m_e/m_{\mu}=0.004737\,,\quad m_{\mu}/m_{\tau}=0.05857\,,\quad \frac{\Delta m_{21}^{2}}{\Delta m_{31}^{2}} = 0.02956\,,\\
&&  m_1=39.43~\text{meV}\,,\quad m_2=40.36~\text{meV}\,,\quad m_3=63.75~\text{meV}\,,
\end{eqnarray}
with $\chi^{2}_{\rm min}=0.75$. Notice that, in this case, all best-fit values  lie in the $1\sigma$ experimental ranges. Using the numerical package \texttt{MultiNest}~\cite{Feroz:2007kg,Feroz:2008xx}, we scan the parameter space of the model, and require all observables to be in the $3\sigma$ regions allowed by data. We find that the three mixing angles can nearly take any values within their experimental $3\sigma$ regions, while the Dirac CP phase $\delta_{CP}$ is sharply predicted to be in the narrow range $\delta_{CP}\in [201^{\circ},215^{\circ}]$.

For inverted neutrino masses, the best-fit values of the input parameters are
\begin{equation}
\begin{gathered}
 \langle\tau\rangle=0.21972 + 1.08073i\,,\quad y^{e}_{2}/y^{e}_{1}= 9.51186 \,,\quad y^{e}_{3}/y^{e}_{1}=27.9507\,,\quad y^{e}_{4}/y^{e}_{1}=0.018387\,, \\
y^{d}_{2}/y^{d}_{1}=5.23555\,,\quad y^{e}_{1}v_{d} = 1.65508~\text{GeV}\,,\quad y^{d}_{1}v_{u} = 533.206~\text{meV}\,.
\end{gathered}
\end{equation}
and the corresponding values of lepton mass and mixing parameters:
\begin{eqnarray}
\nonumber&& \sin^{2}\theta_{12}=0.3048\,,\quad \sin^{2}\theta_{13}=0.02234\,,\quad \sin^{2}\theta_{23}=0.5727\,,\quad \delta_{CP}=315.0^{\circ}\,,\\
\nonumber&& m_e/m_{\mu}=0.004737\,,\quad m_{\mu}/m_{\tau}=0.05857\,,\quad \frac{\Delta m_{21}^{2}}{\Delta m_{31}^{2}} = 0.02980\,,\\
&&  m_1=60.72~\text{meV}\,,\quad m_2=61.32~\text{meV}\,,\quad m_3=35.65~\text{meV}\,,
\end{eqnarray}
with $\chi^{2}_{\rm min}=2.87$. Similarly to the NO case, the best fit values of all mass and mixing observables are within the $1\sigma$ experimental ranges, while $\delta_{CP}$ is predicted to be in the $2\sigma$ interval. A scanning of the parameter space shows that the allowed range of $\delta_{CP}$ is $[307^{\circ},345^{\circ}]$ and that the $3\sigma$ regions of all remaining parameters can be achieved.

\subsection{Model for Majorana neutrino masses: Weinberg operator }

As a benchmark TZLM for the case of Majorana neutrino masses generated via the Weinberg operator, we take the $T'$ representation and modular-weight assignments:
  \begin{eqnarray}\nonumber
    &&L_{D}\sim \mathbf{2'}\,,~~L_{3}\sim \mathbf{1'}~~e^{c}\sim \mathbf{1'}\,,~~\mu^{c}\sim \mathbf{1'}\,,~~\tau^{c}\sim \mathbf{1''}\,,\\
    &&k_{L_{D}}=1\,,~~k_{L_{3}}=0\,,~~k_{e^{c}}=2\,,~~k_{\mu^{c}}=0\,,~~k_{\tau^{c}}=0\,.
  \end{eqnarray}
for which the superpotentials $\mathcal{W}_{E,\nu}$ are
\begin{eqnarray}\nonumber
  \mathcal{W}_{E}&=&y^{e}_{1} e^{c}\left[L_{D}Y^{(1)}_{\mathbf{2}}\right]_{\mathbf{1''}}H_{d} + y^{e}_{2} \mu^{c}\left[L_{D}Y^{(3)}_{\mathbf{2}}\right]_{\mathbf{1''}}H_{d}  + y^{e}_{3} \tau^{c}L_3H_{d}\,,\\
  \mathcal{W}_{\nu}&=&\frac{y^{\nu}_{1}}{\Lambda} \left[ (L_{D}L_{D})_{\mathbf{3}_{S}}Y^{(2)}_{\mathbf{3}}\right]_{\mathbf{1}}H_{u}H_{u}+\frac{y^{\nu}_{2}}{\Lambda} L_{3}\left[ L_{D}Y^{(1)}_{\mathbf{2}}\right]_{\mathbf{1''}}H_{u}H_{u}\,,
  \label{eq:BMWO}
\end{eqnarray}
where all coupling constant $y^{e}_{1,2,3}$ and $y^{\nu}_{1,2}$ can be made real by using the freedom of lepton field redefinition. In total, there are only $7$ real free parameters in this model including the real and imaginary parts of $\tau$. The charged-lepton and (Majorana) neutrino mass matrices are
\begin{eqnarray}
  M_{E}=\left(
\begin{array}{ccc}
 -y^{e}_{1}Y_{\mathbf{2},2}^{(1)} & y^{e}_{1}Y_{\mathbf{2},1}^{(1)} & 0 \medskip\\
 -y^{e}_{2} Y_{\mathbf{2},2}^{(3)} & y^{e}_{2} Y_{\mathbf{2},1}^{(3)} & 0 \medskip\\
0 & 0 & y^{e}_{3} \\
\end{array}
\right)v_{d}\,,\quad
  M_{\nu}=\frac{v_{u}^{2}}{\Lambda} \left(
\begin{array}{ccc}
 -y^{\nu}_{1}Y_{\mathbf{3},3}^{(2)} & \dfrac{y^{\nu}_{1}Y_{\mathbf{3},1}^{(2)}}{\sqrt{2}} & -y^{\nu}_{2} Y_{\mathbf{2},2}^{(1)} \\
 \dfrac{y^{\nu}_{1}Y_{\mathbf{3},1}^{(2)}}{\sqrt{2}} & y^{\nu}_{1}Y_{\mathbf{3},2}^{(2)} & y^{\nu}_{2} Y_{\mathbf{2},1}^{(1)} \\
 -y^{\nu}_{2} Y_{\mathbf{2},2}^{(1)} & y^{\nu}_{2} Y_{\mathbf{2},1}^{(1)} & 0 \\
\end{array}
\right)\,,
\end{eqnarray}
 which fit in the texture-zero pattern $\mathcal{C}_{4}^{(1)}-\mathcal{W}_{1}^{(1)}$. The best-fit values of the input parameters are, for this case,
\begin{equation}
\begin{gathered}
 \langle\tau\rangle=-0.18355 + 0.98944\,i\,,\quad y^{e}_{2}/y^{e}_{1}=0.026019 \,,\quad y^{e}_{3}/y^{e}_{1}=6.47820\,, \\
y^{\nu}_{2}/y^{\nu}_{1}=0.41163\,,\quad y^{e}_{1}v_{d} = 201.034~\text{MeV}\,,\quad \frac{y^{\nu}v_{u}^{2}}{\Lambda} = 348.515~\text{meV}\,.
\end{gathered}
\end{equation}

\begin{figure}[t!]
\centering
\includegraphics[width=6.5in]{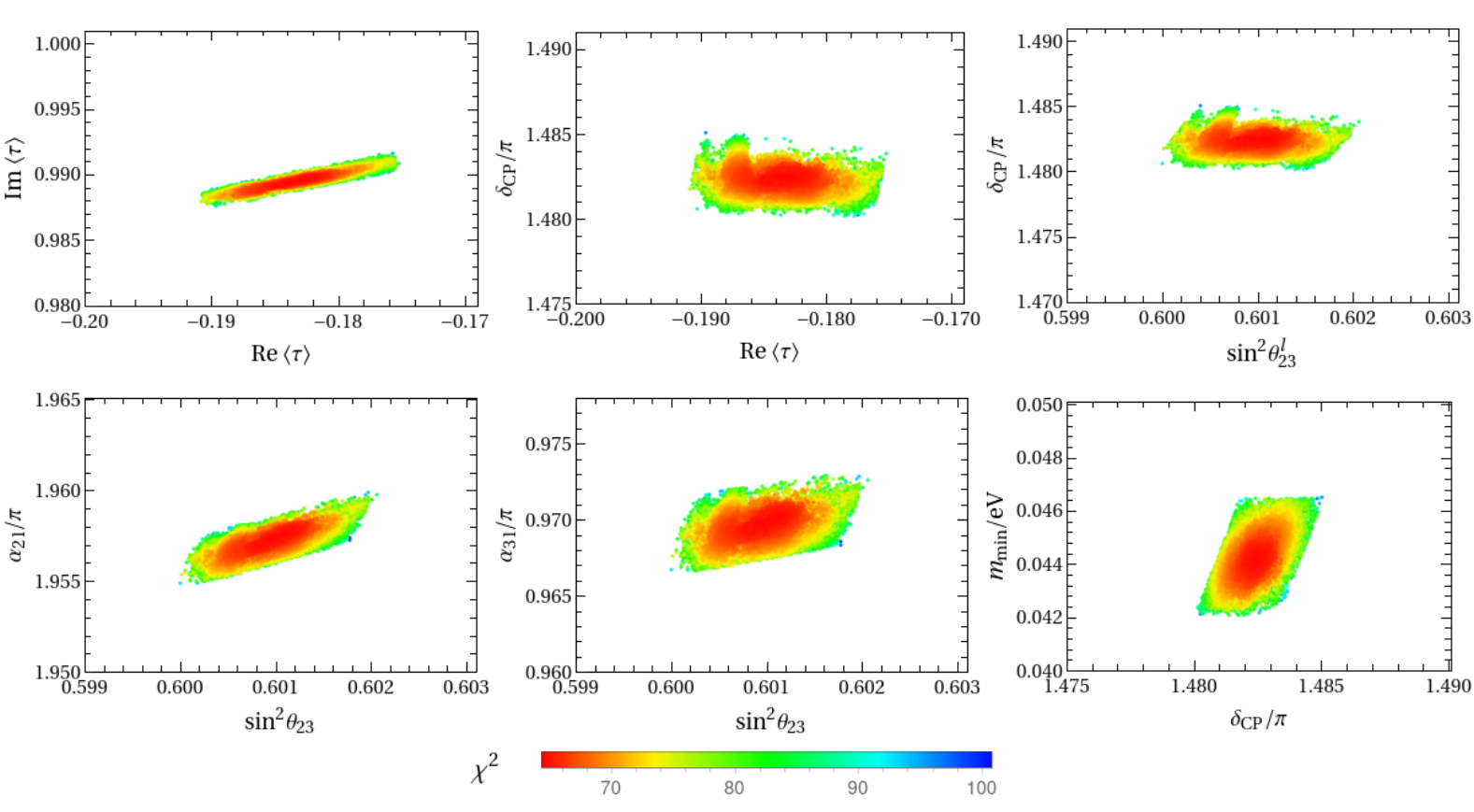}
\caption{Results of the parameter-space scanning for the benchmark model of neutrino masses generated by the Weinberg operator. The $T'$ representation and modular-weight assignments are given in \eqref{eq:BMWO}. We present the scatter plots in distinct bidimensional planes of the neutrino mixing angles, CP phases and the the real/imaginary part of $\langle \tau \rangle$. The colour grading reflects the values of the $\chi^2$ for each point, according to the scale shown at the bottom of the figure.
}
\label{fig:model_WO_NO}
\end{figure}
for which the corresponding values of the lepton observables are
\begin{eqnarray}
\nonumber&& \sin^{2}\theta_{12}=0.3037\,,\quad \sin^{2}\theta_{13}=0.02254\,,\quad \sin^{2}\theta_{23}=0.6010\,,\quad \delta_{CP}=266.8^{\circ}\,,\\
\nonumber&& \alpha_{21}=352.3^{\circ}\,,\quad \alpha_{31}=174.6^{\circ}\,,\quad m_e/m_{\mu}=0.004737\,,\quad m_{\mu}/m_{\tau}=0.05857\,,\quad \frac{\Delta m_{21}^{2}}{\Delta m_{31}^{2}} = 0.02957\,,\\
&&  m_1=44.19~\text{meV}\,,\quad m_2=45.02~\text{meV}\,,\quad m_3=66.80~\text{meV}\,,\quad m_{\beta\beta}=44.86~\text{meV}\,,
\end{eqnarray}
with $\chi^{2}_{\rm min}=64.21$. Almost all observables lie in the $1\sigma$ experimental intervals, except $\sin^2\theta_{23}$ which is close to the $3\sigma$ upper limit. The lightest neutrino mass is predicted to be $44.19~\text{meV}$ and $\sum_{i}m_{i}$ is about $ 156~\text{meV}$. Cosmological data shows that the most stringent bound on the sum of neutrino masses is $\sum_{i}m_{i}<120~\text{meV}$ at $95\%$~C.L. from the Planck Collaboration results~\cite{Planck:2018vyg}. Thus, the present benchmark would be excluded by this data since the prediction $\sum_{i}m_{i}\simeq 156~\text{meV}$ exceeds that bound. However, as is well known, cosmological bounds on $\sum_{i}m_{i}$ significantly depend on the data sets that need to be combined in order to break the degeneracies of the many cosmological parameters~\cite{Planck:2018vyg}. In fact, combining the baryon acoustic oscillation (BAO) data with the cosmic microwave background (CMB) lensing reconstruction power spectrum, one has $\sum_{i}m_{i}<600~\text{meV}$ and, taking this result, the  model would be still viable. 

The latest bound on the effective Majorana neutrino mass $m_{\beta\beta}$ has been set by the KamLAND-Zen experiment, namely $m_{\beta\beta}<36-156$ meV~\cite{KamLAND-Zen:2022tow}, for which the largest uncertainty arises from the computation of nuclear matrix elements. The model prediction $m_{\beta\beta}=44.86~\text{meV}$ respects this bound. With future large-scale $0\nu\beta\beta$-decay experiments aiming at improving the bound on $m_{\beta\beta}$, the present benchmark would be further scrutinised. For instance, the SNO+ Phase II is expected to reach a sensitivity of $19-46$ meV~\cite{SNO:2015wyx}, which is nearly the same as the one foreseen by the LEGEND experiment  (15-50~meV) by operating 1000 kg of detectors for 10 years~\cite{LEGEND:2017cdu}. nEXO, the successor of EXO-200, will be able to probe $m_{\beta\beta}$ down to $5.7-17.7$~meV after 10 years of data taking~\cite{nEXO:2017nam}. 

In Figure~\ref{fig:model_WO_NO}, we show the correlations among the input free parameters, neutrino masses and mixing parameters predicted in this model. We can see that the complex modulus $\tau$ scatter in a very narrow region. The predictions for the atmospheric mixing angle $\theta_{23}$ and the three CP violation phases are very sharp. In particular, the allowed range of $\delta_{CP}$ is very close to $3\pi/2$. In conclusion, this model is very predictive, since it is able to describe the 12 masses and mixing parameters for the NO case  with only 7 input real parameters. The IO neutrino mass spectrum cannot be accommodated in this case.

\subsection{Model for Majorana neutrino masses: seesaw mechanism }

\begin{figure}[t!]
\centering
\includegraphics[width=6.5in]{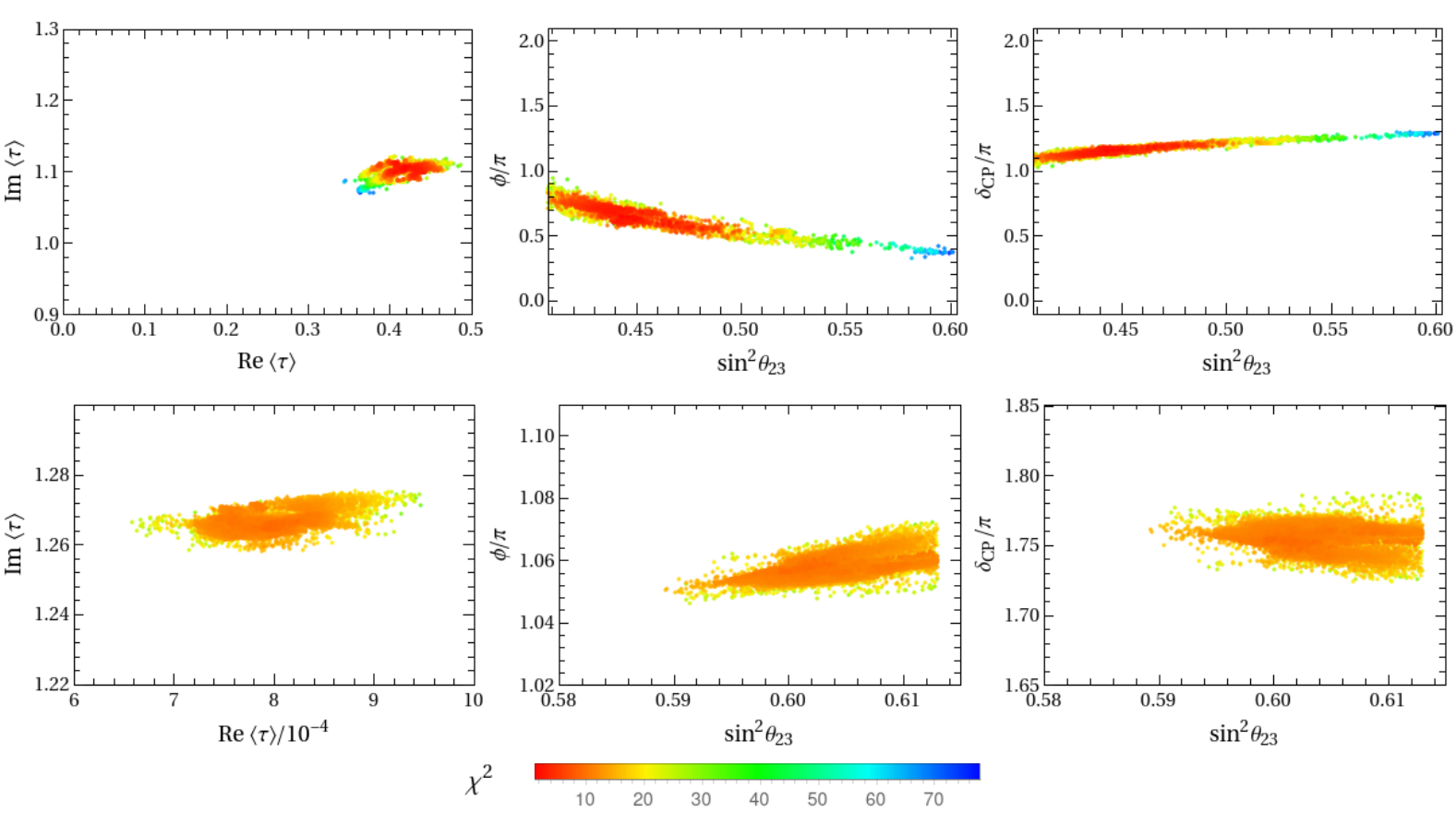}
\caption{Values of the complex modulus $\tau$ compatible with experimental data, and correlations between $\sin^2\theta_{23}$ and the CP-violating phases $\delta_{CP}$ and $\phi$. The upper (lower) panels correspond to the case in which the neutrino mass specturm is NO (IO).
}
\label{fig:model_SS_NO_IO}
\end{figure}

In the last benchmark TZLM  with neutrino masses are generated via the seesaw mechanism with imposed gCP. The transformation properties and modular weights of the lepton fields are,
  \begin{eqnarray}\nonumber
    &&L_{D}\sim \mathbf{2}\,,~~L_{3}\sim \mathbf{1}~~e^{c}\sim \mathbf{1''}\,,~~\mu^{c}\sim \mathbf{1}\,,~~\tau^{c}\sim \mathbf{1}\,,~~N^{c}\equiv \{N_{1}^{c},N_{2}^{c}\}\sim \mathbf{2}\,,\\
    &&k_{L_{D}}=1\,,~~k_{L_{3}}=-2\,,~~k_{e^{c}}=2\,,~~k_{\mu^{c}}=2\,,~~k_{\tau^{c}}=4\,,~~k_{N^{c}}=3\,,
  \end{eqnarray}
from which the modular-invariant superpotentials
\begin{eqnarray}\nonumber
  \mathcal{W}_{E}&=&y^{e}_{1} e^{c}\left[L_{D}Y^{(3)}_{\mathbf{2}}\right
]_{\mathbf{1'}}H_{d} + y^{e}_{2} \mu^{c}\left[L_{D}Y^{(3)}_{\mathbf{2''}}\right]_{\mathbf{1}}H_{d} + y^{e}_{3} \tau^{c}\left[L_{D}Y^{(5)}_{\mathbf{2''}}\right]_{\mathbf{1}}H_{d} + y^{e}_{4} \mu^{c}L_{3}H_{d} \,,\\
  \mathcal{W}_{\nu}&=&y^{D}_{1}\left[(L_{D}N^{c})_{\mathbf{3}}Y^{(4)}_{\mathbf{3}}\right]_{\mathbf{1}}H_{u} + y^{N}_{1}\Lambda \left[ (N^{c}N^{c})_{\mathbf{3}}Y^{(6)}_{\mathbf{3}I}\right]_{\mathbf{1}} + y^{N}_{2}\Lambda \left[ (N^{c}N^{c})_{\mathbf{3}}Y^{(6)}_{\mathbf{3}II}\right]_{\mathbf{1}}\,.
\end{eqnarray}
can be defined, being all coupling constants real due to gCP. The charged-lepton and neutrino mass matrices take the following form,
\begin{eqnarray} \nonumber
  M_{E}&=&\left(
\begin{array}{ccc}
 -y^{e}_{1}Y_{\mathbf{2},2}^{(3)} ~&~ y^{e}_{1}Y_{\mathbf{2},1}^{(3)} ~&~ 0 \medskip \\
 y^{e}_{2} Y_{\mathbf{2''},2}^{(3)} ~&~ -y^{e}_{2} Y_{\mathbf{2''},1}^{(3)} ~&~ y^{e}_{4} \medskip\\
 y^{e}_{3} Y_{\mathbf{2''},2}^{(5)} ~&~ -y^{e}_{3} Y_{\mathbf{2''},1}^{(5)} ~&~ 0 \\
\end{array}
  \right) v_{d}\,,\quad M_{D}=y^{D}_{1}v_{u}\left(
\begin{array}{ccc}
 -Y_{\mathbf{3},2}^{(4)} & \dfrac{Y_{\mathbf{3},3}^{(4)}}{\sqrt{2}} & 0 \medskip\\
 \dfrac{Y_{\mathbf{3},3}^{(4)}}{\sqrt{2}} & Y_{\mathbf{3},1}^{(4)} & 0 \\
\end{array}
\right)\,,\\
  M_{N}&=&\Lambda \left(
\begin{array}{cc}
 -y^{N}_{2} Y_{\mathbf{3}II,2}^{(6)}-y^{N}_{1}Y_{\mathbf{3}I,2}^{(6)} & \dfrac{y^{N}_{2} Y_{\mathbf{3}II,3}^{(6)}+y^{N}_{1}Y_{\mathbf{3}I,3}^{(6)}}{\sqrt{2}} \medskip\\
 \dfrac{y^{N}_{2} Y_{\mathbf{3}II,3}^{(6)}+y^{N}_{1}Y_{\mathbf{3}I,3}^{(6)}}{\sqrt{2}} & y^{N}_{2} Y_{\mathbf{3}II,1}^{(6)}+y^{N}_{1}Y_{\mathbf{3}I,1}^{(6)} \\
\end{array}
\right)\,.
\end{eqnarray}
According to classification given in Eqs.~\eqref{eq:TZ_MN} and \eqref{eq:TZ_MD}, the above matrices correspond to $\mathfrak{D}_{2}^{3}-\mathfrak{N}_{0}^{1}$, while $M_{\nu}$ follows the $\mathcal{W}_{3}^{(2)}$ pattern, as shown in Table~\ref{Table:seesaw_Mnu}. Moreover, the texture of $M_{E}$ is of type $\mathcal{C}_{2}^{(3)}$. There is a total of $8$ effective real parameters in the model. For neutrino masses with NO, the best-fit values of the input parameters are
\begin{equation}
\begin{gathered}
 \langle\tau\rangle=0.41547 + 1.10335\,i\,,\quad y^{e}_{2}/y^{e}_{1}=6.13351\times 10^{2} \,,\quad y^{e}_{3}/y^{e}_{1}=3.39303\times 10^{2}\,,\quad y^{e}_{4}/y^{e}_{1}=16.5963\,, \\
y^{N}_{2}/y^{N}_{1}=-0.29007\,,\quad y^{e}_{1}v_{d} = 35.6572~\text{MeV}\,,\quad \frac{(y^{D}_{1}v_{u})^{2}}{y^{N}_{1}\Lambda} = 303.537~\text{meV}\,,
\end{gathered}
\end{equation}
being the corresponding lepton observables
\begin{eqnarray}
\nonumber&& \sin^{2}\theta_{12}=0.3046\,,\quad \sin^{2}\theta_{13}=0.02242\,,\quad \sin^{2}\theta_{23}=0.4524\,,\quad \delta_{CP}=209.4^{\circ}\,,\\
\nonumber&& \phi=112.3^{\circ}\,,\quad m_e/m_{\mu}=0.004737\,,\quad m_{\mu}/m_{\tau}=0.05857\,,\quad \frac{\Delta m_{21}^{2}}{\Delta m_{31}^{2}} = 0.02957\,,\\
&&  m_1=0~\text{meV}\,,\quad m_2=8.614~\text{meV}\,,\quad m_3=58.68~\text{meV}\,,\quad m_{\beta\beta}=1.466~\text{meV}\,,
\end{eqnarray}
with $\chi^{2}_{\rm min}=0.70$ (remember that for the minimal seesaw considered in this work the lightest neutrino is massless and there is a single Majorana phase $\phi$). The above best-fit values are in very good agreement with the data. Correlations between the values of $\sin^2\theta_{23}$ and of the CP-violation phases $\delta_{CP}$ and $\phi$ are shown in the upper panels of figure~\ref{fig:model_SS_NO_IO} for NO.

If the neutrino mass spectrum is inverted, the best agreement between model predictions and experimental data is achieved with the following values of the input parameters
\begin{equation}
\begin{gathered}
 \langle\tau\rangle=0.000786 + 1.267779\,i\,,\quad y^{e}_{2}/y^{e}_{1}= 9.451055 \,,\quad y^{e}_{3}/y^{e}_{1}=0.0532938\,,\quad y^{e}_{4}/y^{e}_{1}=0.209372\,, \\
y^{N}_{2}/y^{N}_{1}=0.884872\,,\quad y^{e}_{1}v_{d} = 2.333797~\text{GeV}\,,\quad \frac{(y^{D}_{1}v_{u})^{2}}{y^{N}_{1}\Lambda} = 594.172~\text{meV}\,.
\end{gathered}
\end{equation}
At this best-fit point we have:
\begin{eqnarray}
\nonumber&& \sin^{2}\theta_{12}=0.2999\,,\quad \sin^{2}\theta_{13}=0.02239\,,\quad \sin^{2}\theta_{23}=0.6064\,,\quad \delta_{CP}=315.6^{\circ}\,,\\
\nonumber&& \phi=184.2^{\circ}\,,\quad m_e/m_{\mu}=0.004737\,,\quad m_{\mu}/m_{\tau}=0.05857\,,\quad \frac{\Delta m_{21}^{2}}{\Delta m_{31}^{2}} = 0.03068\,,\\
&&  m_1=48.418~\text{meV}\,,\quad m_2=49.179~\text{meV}\,,\quad m_3=0~\text{meV}\,,\quad m_{\beta\beta}=18.791~\text{meV}\,,
\end{eqnarray}
with $\chi^{2}_{\rm min}=9.30$, being all observables in the experimentally allowed $3\sigma$ intervals. In the lower panels of figure~\ref{fig:model_SS_NO_IO}, we show the regions for the real and imaginary parts of $\tau$ and the correlation among $\theta_{23}$ and two CP-violation phases $\delta_{CP}$ and $\phi$.

\section{Conclusions}
\label{sec:conclusion}

The nature of the fermion flavour pattern in the SM is a great puzzle. With the purpose of reducing the number of free parameters in fermion mass matrices, texture-zero patterns have been widely studied in the literature. In the present work, we have performed a systematic analysis of how texture zeros in lepton mass matrices can be realised in the framework of a $T'$ modular symmetry. We show that, by properly assigning representations and modular weights to the matter fields, texture-zero patterns can be naturally reproduced, each assignment leading to a specific TZLM. In particular, we have considered all cases in which lepton fields are assumed to transform as singlets ($\mathbf{1}$, $\mathbf{1}'$, $\mathbf{1}''$), doublets ($\mathbf{2}$, $\mathbf{2}'$, $\mathbf{2}''$) or triplet $\mathbf{3}$ of $T'$. Both cases of Dirac and Majorana neutrino masses (generated by either the Weinberg operator or via the type-I seesaw mechanism) were investigated. The most general form of the lepton mass matrices for all possible representation assignments were found, namely there are 10 (23) [8] texture-zero patterns for the charged-lepton (Dirac neutrino) [effective Majorana neutrino] mass matrix.

Combining the texture-zero patterns for the charged-lepton and neutrino mass matrices, we have obtained 136 allowed $(M_{E},M_{D})$ pairs which can be realised from $T'$ modular symmetry for Dirac neutrinos. In case of Majorana neutrinos, we have found 29 and 35 pairs of texture zeros in $(M_{E},M_{\nu})$ when neutrinos masses are generated by the Weinberg operator and the type-I seesaw mechanism, respectively. If gCP is introduced, more combinations of texture-zero patterns can be achieved, as shown in Table~\ref{tab:chisq_res}. In order to test whether the obtained texture pairs $(M_{E},M_{D})/(M_{E},M_{\nu})$ can accommodate experimental data, we have performed a $\chi^{2}$ analysis for the corresponding TZLMs. It turns out that only part of those models are compatible with the data, as can be seen in Tables~\ref{tab:Dirac_pair_summary_NO}, \ref{tab:Dirac_pair_summary_IO}, \ref{tab:Majorana_WO_pair_summary_NO_IO} and \ref{tab:Majorana_SS_pair_summary_NO_IO}. For each viable texture pair, we have provided representative models, for which the representation and modular-weight assignments are shown in Tables~\ref{tab:assign_Dirac_NO}, \ref{tab:assign_Dirac_IO}, \ref{tab:assign_Majorana_WO_NO}, \ref{tab:assign_Majorana_WO_IO}, \ref{tab:assign_Majorana_SS_NO} and \ref{tab:assign_Majorana_SS_IO}. The corresponding predictions for the lepton observables are summarised in Tables~\ref{tab:prediction_Dirac_NO}, \ref{tab:prediction_Dirac_IO}, \ref{tab:prediction_Majorana_WO_NO}, \ref{tab:prediction_Majorana_WO_IO}, \ref{tab:prediction_Majorana_SS_NO} and \ref{tab:prediction_Majorana_SS_IO}, respectively. We found that the minimal model requires 7 real parameters to explain all 9 measured observables. To illustrate our findings, we studied in more detail three benchmark TZLMs for Dirac and Majorana neutrinos in Section~\ref{sec:benchmark_models}.

In conclusion, we have shown that several texture-zero patterns for lepton mass matrices previously considered in the literature can be realised in the context of modular flavour symmetries. A considerable fraction of those textures are able to accommodate the experimental data, being some of them predictive in the sense that the corresponding TZLMs contain less free parameters than observables. In comparison with typical analyses of texture zeros in the context of Abelian flavour symmetries, the present approach is much more predicitive due to $T'$ modular symmetry which implies correlations among the nonvanishing elements of the mass matrices. Since said correlations depend on the choice of the modular group, it would be surely interesting to explore other possibilities besides $T'$.

\section*{Acknowledgements}

G.J.D. and J.N.L. acknowledge the National Natural Science Foundation of China under Grant Nos. 11975224, 11835013 and the Key Research Program of the Chinese Academy of Sciences under grant NO. XDPB15. J.N.L. is supported by the Grants No.~NSFC-12147110 and the China Post-doctoral Science Foundation under Grant No. 2021M703106. The work of F.R.J. is supported by Fundação para a Ciência e a Tecnologia (FCT, Portugal) through the projects CFTP-FCT Unit UIDB/00777/2020 and UIDP/00777/2020, and through the CERN/FCT project CERN/FIS-PAR/0004/2019, which are partially funded through POCTI (FEDER), COMPETE, QREN and EU.

\section*{Appendices}

\setcounter{equation}{0}
\renewcommand{\theequation}{\thesection.\arabic{equation}}

\begin{appendix}

\section{\label{app:Tp_group} The $T'$ modular group }

The $T'$ group is the double covering of the tetrahedral group $A_4$. All the elements of $T'$ can be generated by three generators $S$, $T$ and $R$ which obey the following relations~\footnote{Alternatively the $T'$ group can be expressed by only $S$ and $T$ obeying $S^4=T^3=(ST)^3=1$, $S^2T=TS^2$.}~\cite{Liu:2019khw}:
\begin{equation}
S^{2}=R,~~ (ST)^{3}=T^{3}=R^{2}=1,~~RT = TR\,.
\end{equation}
The generator $R$ commutes with all elements of the group, and the center of $T^{\prime}$ is the $Z_2$ subgroup generated by $R$. The 24 elements of $T'$ group belong to 7 conjugacy classes:
\begin{eqnarray}
\nonumber 1 C_1:&&\,1\,, \\
\nonumber 1 C_2:&&\,R\,, \\
\nonumber 6 C_4:&&\,S,\,T^{-1}ST,\,TST^{-1},\,SR,\,T^{-1}STR,\,TST^{-1}R \,, \\
\nonumber 4 C_6:&&\,TR,\,TSR,\,STR,\,T^{-1}ST^{-1}R \,, \\
\nonumber 4 C_3:&&\,T^{-1},\,ST^{-1}R,\,T^{-1}SR,\,TSTR \,, \\
\nonumber 4 C'_3:&&\,T,\,TS,\,ST,\,T^{-1}ST^{-1} \,, \\
\label{T'conj_class} 4C'_6:&&\,ST^{-1},\,T^{-1}S,\,TST,\,T^{-1}R \,.
\end{eqnarray}
The $T'$ group has a triplet representation $\mathbf{3}$ and three singlets representations $\mathbf{1}$, $\mathbf{1}'$ and $\mathbf{1}''$ in common with $A_4$. In addition, it has three two-dimensional spinor representations $\mathbf{2}$, $\mathbf{2}'$ and $\mathbf{2}''$. In our working basis, the generators $S$, and $T$ are represented by the following symmetric and unitary matrices:
\begin{eqnarray}
\label{eq:irr-Tp} \begin{array}{cccc}
\mathbf{1:} & S=1, ~&~ T=1\,, \\
\mathbf{1}':& S=1, & T=\omega\,, \\
\mathbf{1}'': & S=1, & T=\omega^{2} \,,\\
\mathbf{2:} ~&~ S=-\frac{i}{\sqrt{3}}
\begin{pmatrix}
1  ~&  \sqrt{2}  \\
\sqrt{2}  ~&  -1
\end{pmatrix}, & T=\left(\begin{array}{cc}
\omega & 0 \\
0 & 1 \\
\end{array}\right) \,,\\
\mathbf{2}': ~&~ S=-\frac{i}{\sqrt{3}}
\begin{pmatrix}
1  ~&  \sqrt{2}  \\
\sqrt{2}  ~&  -1
\end{pmatrix}, & T=\left(\begin{array}{cc}
\omega^{2} & 0 \\
0 & \omega \\
\end{array}\right) \,,\\
\mathbf{2}'': ~&~ S=-\frac{i}{\sqrt{3}}
\begin{pmatrix}
1  ~&  \sqrt{2}  \\
\sqrt{2}  ~&  -1
\end{pmatrix}, & T=\left(\begin{array}{cc}
1 & 0 \\
0 & \omega^{2} \\
\end{array}\right) \,,\\
\mathbf{3:} & S=\frac{1}{3}\left(\begin{array}{ccc}
-1&2&2\\
2&-1&2\\
2&2&-1
\end{array}\right),
 ~&~ T=\left(
\begin{array}{ccc}
1~&0~&0\\
0~&\omega~&0\\
0~&0~&\omega^2
\end{array}\right) \,,
\end{array}
\end{eqnarray}
with $\omega=e^{i2\pi/3}$. Notice that the two-dimensional representation matrices are related to those of Refs.~\cite{Liu:2019khw,Lu:2019vgm} by a similarity transformation, while the remaining ones are the same. The Kronecker products between different irreducible representations of $T'$ are given by
\begin{eqnarray}
\nonumber&&\mathbf{1}^a\otimes \mathbf{r}^b = \mathbf{r}^b \otimes \mathbf{1}^a= \mathbf{r}^{a+b~(\text{mod}~3)},~~~~~{\rm for}~~\mathbf{r} = \mathbf{1}, \mathbf{2}\,,\\
\nonumber&&\mathbf{1}^a \otimes \mathbf{3} = \mathbf{3}\otimes \mathbf{1}^a=\mathbf{3}\,,\\
\nonumber&&\mathbf{2}^a \otimes \mathbf{2}^b=\mathbf{3}\oplus \mathbf{1}^{a+b+1~(\text{mod}~3)}\,, \\
\nonumber&&\mathbf{2}^a \otimes \mathbf{3} =\mathbf{3} \otimes \mathbf{2}^a = \mathbf{2}\oplus \mathbf{2}'\oplus \mathbf{2}''\,, \\
\label{eq:mult}&&\mathbf{3} \otimes \mathbf{3} = \mathbf{3}_S \oplus \mathbf{3}_A \oplus \mathbf{1} \oplus \mathbf{1}' \oplus \mathbf{1}''\,,
\end{eqnarray}
where $a,b=0, 1, 2$ and we have denoted $\mathbf{1}\equiv\mathbf{1}^0$, $\mathbf{1}'\equiv\mathbf{1}^{1}$, $\mathbf{1}''\equiv\mathbf{1}^{2}$ for singlet representations and $\mathbf{2}\equiv\mathbf{2}^0$, $\mathbf{2}'\equiv\mathbf{2}^{1}$, $\mathbf{2}''\equiv\mathbf{2}^{2}$ for the doublet representations. The notations $\mathbf{3}_S$ and $\mathbf{3}_A$ stand for the symmetric and antisymmetric triplet combinations respectively. In the following, we report the Clebsch-Gordon (CG) coefficients of the $T'$ group in the chosen basis. We shall use $\alpha_i$ ($\beta_i $) to denote the elements of the first (second) representation of the product.
\begin{eqnarray}
\mathbf{1}^a\otimes\mathbf{1}^b &=& \mathbf{1}^{a+b~(\text{mod}~3)} \sim \alpha\beta \,, \\
\mathbf{1}^a\otimes\mathbf{2}^b &=& \mathbf{2}^{a+b~(\text{mod}~3)} \sim \left(\begin{array}{c} \alpha\beta_1\\
\alpha\beta_2 \\ \end{array}\right) \,,\\
  \mathbf{1}\otimes\mathbf{3} &=& \mathbf{3} \sim \left(\begin{array}{c} \alpha\beta_1\\
\alpha\beta_2 \\
\alpha\beta_3 \end{array}\right) \,, \\
\mathbf{1}'\otimes\mathbf{3} &=& \mathbf{3} \sim \left(\begin{array}{c} \alpha\beta_3\\
\alpha\beta_1 \\
\alpha\beta_2 \end{array}\right) \,, \\
\mathbf{1}''\otimes\mathbf{3} &=& \mathbf{3} \sim \left(\begin{array}{c} \alpha\beta_2\\
\alpha\beta_3 \\
\alpha\beta_1 \end{array}\right) \,.
\end{eqnarray}

\begin{eqnarray}
  \mathbf{2}\otimes\mathbf{2}=\mathbf{2}'\otimes \mathbf{2}''=\mathbf{1}'\oplus\mathbf{3} ~&\text{with}&~\left\{
\begin{array}{l}
\mathbf{1}'\sim \alpha_1\beta_2-\alpha_2\beta_1 \\ [0.1in]
\mathbf{3}\sim
\left(\begin{array}{c} \alpha_2\beta_2 \\
\frac{1}{\sqrt{2}}(\alpha_1\beta_2+\alpha_2\beta_1)  \\
-\alpha_1\beta_1 \end{array}\right)
\end{array}
\right. \\
\mathbf{2}\otimes\mathbf{2}'=\mathbf{2}''\otimes\mathbf{2}''=\mathbf{1}''\oplus\mathbf{3}~&\text{with}&~\left\{
\begin{array}{l}
\mathbf{1}''\sim \alpha_1\beta_2-\alpha_2\beta_1\\ [0.1in]
\mathbf{3}\sim
 \left(\begin{array}{c} -\alpha_1\beta_1\\
 \alpha_2\beta_2 \\
 \frac{1}{\sqrt{2}}(\alpha_1\beta_2+\alpha_2\beta_1) \end{array}\right)
\end{array}
\right. \\
\mathbf{2}\otimes\mathbf{2}''=\mathbf{2}'\otimes\mathbf{2}'=\mathbf{1}\oplus \mathbf{3} ~&\text{with}&~\left\{
\begin{array}{l}
\mathbf{1} \sim \alpha_1\beta_2-\alpha_2\beta_1 \\ [0.1in]
\mathbf{3}\sim
\left(\begin{array}{c}  \frac{1}{\sqrt{2}}(\alpha_1\beta_2+\alpha_2\beta_1)\\
 -\alpha_1\beta_1\\
 \alpha_2\beta_2  \end{array}\right)
\end{array}
\right. \\
\mathbf{2}\otimes\mathbf{3}=\mathbf{2}\oplus\mathbf{2}'\oplus\mathbf{2}'' ~&\text{with}&~\left\{
\begin{array}{l}
\mathbf{2}\sim
 \left(\begin{array}{c} \alpha_1\beta_1+\sqrt{2}\alpha_2\beta_2 \\
 -\alpha_2\beta_1+\sqrt{2}\alpha_1\beta_3 \end{array}\right)  \\ [0.1in]
\mathbf{2}'\sim
\left(\begin{array}{c} \alpha_1\beta_2+\sqrt{2}\alpha_2\beta_3 \\
-\alpha_2\beta_2+\sqrt{2}\alpha_1\beta_1  \end{array}\right) \\ [0.1in]
\mathbf{2}''\sim
\left(\begin{array}{c} \alpha_1\beta_3+\sqrt{2}\alpha_2\beta_1 \\
-\alpha_2\beta_3+\sqrt{2}\alpha_1\beta_2  \end{array}\right) \\ [0.1in]
\end{array}
\right. \\
\mathbf{2}'\otimes\mathbf{3}=\mathbf{2}\oplus\mathbf{2}'\oplus\mathbf{2}'' ~&\text{with}&~\left\{
\begin{array}{l}
\mathbf{2}\sim
\left(\begin{array}{c} \alpha_1\beta_3+\sqrt{2}\alpha_2\beta_1 \\
-\alpha_2\beta_3+\sqrt{2}\alpha_1\beta_2  \end{array}\right)  \\ [0.1in]
\mathbf{2}'\sim
 \left(\begin{array}{c} \alpha_1\beta_1+\sqrt{2}\alpha_2\beta_2 \\
 -\alpha_2\beta_1+\sqrt{2}\alpha_1\beta_3 \end{array}\right) \\ [0.1in]
\mathbf{2}''\sim
\left(\begin{array}{c} \alpha_1\beta_2+\sqrt{2}\alpha_2\beta_3 \\
-\alpha_2\beta_2+\sqrt{2}\alpha_1\beta_1  \end{array}\right) \\ [0.1in]
\end{array}
\right. \\
\mathbf{2}''\otimes\mathbf{3}=\mathbf{2}\oplus\mathbf{2}'\oplus\mathbf{2}'' ~&\text{with}&~\left\{
\begin{array}{l}
\mathbf{2}\sim
\left(\begin{array}{c} \alpha_1\beta_2+\sqrt{2}\alpha_2\beta_3 \\
-\alpha_2\beta_2+\sqrt{2}\alpha_1\beta_1  \end{array}\right)  \\ [0.1in]
\mathbf{2}'\sim
\left(\begin{array}{c} \alpha_1\beta_3+\sqrt{2}\alpha_2\beta_1 \\
-\alpha_2\beta_3+\sqrt{2}\alpha_1\beta_2  \end{array}\right) \\ [0.1in]
\mathbf{2}''\sim
 \left(\begin{array}{c} \alpha_1\beta_1+\sqrt{2}\alpha_2\beta_2 \\
 -\alpha_2\beta_1+\sqrt{2}\alpha_1\beta_3 \end{array}\right) \\ [0.1in]
\end{array}
\right. \\
\mathbf{3}\otimes\mathbf{3}=\mathbf{3}_S\oplus\mathbf{3}_A\oplus\mathbf{1}\oplus\mathbf{1}'\oplus\mathbf{1}'' ~&\text{with}&~\left\{
\begin{array}{l}
\mathbf{3}_S\sim
 \left(\begin{array}{c} 2\alpha_1\beta_1 - \alpha_2\beta_3 - \alpha_3\beta_2 \\
 2\alpha_3\beta_3 - \alpha_1\beta_2 - \alpha_2\beta_1  \\
 2\alpha_2\beta_2 - \alpha_1\beta_3 - \alpha_3\beta_1 \end{array}\right) \\ [0.1in]
\mathbf{3}_A\sim
 \left(\begin{array}{c} \alpha_2\beta_3 - \alpha_3\beta_2 \\
\alpha_1\beta_2 - \alpha_2\beta_1  \\
 \alpha_3\beta_1 - \alpha_1\beta_3 \end{array}\right) \\ [0.1in]
\mathbf{1} \sim \alpha_1\beta_1 + \alpha_2\beta_3 + \alpha_3\beta_2 \\ [0.1in]
\mathbf{1}' \sim \alpha_3\beta_3 + \alpha_1\beta_2 + \alpha_2\beta_1 \\ [0.1in]
\mathbf{1}'' \sim \alpha_2\beta_2 + \alpha_1\beta_3 + \alpha_3\beta_1 \\ [0.1in]
\end{array}
\right.
\end{eqnarray}
Note that, in our basis, the representation matrices of $S$ and $T$ are unitary and symmetric in all irreducible representations, and all the CG coefficients of the contractions are real.

\section{\label{subsec:modular_forms_higher}Higher-weight modular forms of level $N=3$}

Higher-weight modular forms can be constructed from tensor product of lower-weight ones. In the following, we will use the weight-1 modular forms $Y_{\mathbf{2}}^{(1)}$ given in Eq.~\eqref{eq:Y1-Y2} and the Clebsch-Gordan coefficients of $T'$ presented in Appendix~\ref{app:Tp_group} to construct weight 2, 3, 4, 5 and 6 modular forms of $T'$ modular group.

The weight-2 modular forms can be generated from the tensor products of two $Y^{(1)}_{\mathbf{2}}$,
\begin{equation}
Y^{(2)}_{\mathbf{3}}=\left(Y^{(1)}_{\mathbf{2}}Y^{(1)}_{\mathbf{2}}\right)_{\mathbf{3}}=\left(Y^2_2,\quad \sqrt{2}Y_1Y_2,\quad -Y^2_1 \right)^T\,,
\label{eq:modfvec1}
\end{equation}
 where $Y_{1}$ and $Y_{2}$ are two components of weight 1 modular forms $Y^{(1)}_{\mathbf{2}}=(Y_{1},Y_{2})^{T}$. Then, we can use the weight-1 and weight-2 modular forms to construct the weight-3 modular forms,
\begin{equation}
\begin{aligned}
&Y^{(3)}_{\mathbf{2}}=\left(Y^{(1)}_{\mathbf{2}}Y^{(2)}_{\mathbf{3}}\right)_{\mathbf{2}}= \left(3Y_1Y^2_2 ,\quad -\sqrt{2}Y^3_1-Y^3_2 \right)^T,\\
&Y^{(3)}_{\mathbf{2''}}=\left(Y^{(1)}_{\mathbf{2}}Y^{(2)}_{\mathbf{3}}\right)_{\mathbf{2''}}=\left(-Y^3_1+\sqrt{2}Y^3_2,\quad 3Y_2Y^2_1 \right)^T \,.
\end{aligned}
\end{equation}
At weight $k=4$, we find five independent modular forms which can be arranged into two singlets $\mathbf{1}$ and $\mathbf{1'}$ and a triplet of $T'$,
\begin{equation}
\begin{aligned}
&\hskip-0.12in Y^{(4)}_{\mathbf{3}}=\left(Y^{(1)}_{\mathbf{2}}Y^{(3)}_{\mathbf{2}}\right)_{\mathbf{3}} = \left(-\sqrt{2}Y^3_1Y_2-Y^4_2,~ -Y^4_1+\sqrt{2}Y_1Y^3_2, ~ -3Y^2_1Y^2_2 \right)^T,\\
&\hskip-0.12in Y^{(4)}_{\mathbf{1'}}=\left(Y^{(1)}_{\mathbf{2}}Y^{(3)}_{\mathbf{2}}\right)_{\mathbf{1'}}=-\sqrt{2}Y^4_1-4Y_1Y^3_2 \,, \\
&\hskip-0.12in Y^{(4)}_{\mathbf{1}}=\left(Y^{(1)}_{\mathbf{2}}Y^{(3)}_{\mathbf{2''}}\right)_{\mathbf{1}}=4Y^3_1Y_2-\sqrt{2}Y^4_2 \,.
\end{aligned}
\end{equation}
Similarly, the independent weight-5 modular forms can be constructed from the tensor products of weight-1 and weight-4 modular forms as follows,
\begin{equation}
\begin{aligned}
&\hskip-0.15in Y^{(5)}_{\mathbf{2}}=\left(Y^{(1)}_{\mathbf{2}}Y^{(4)}_{\mathbf{3}}\right)_{\mathbf{2}}= \left[-2\sqrt{2}Y_{1}^{3}Y_{2}+Y_{2}^{4}\right]\left(Y_{1},\quad Y_{2}\right)^T,\\
&\hskip-0.15in Y^{(5)}_{\mathbf{2'}}=\left(Y^{(1)}_{\mathbf{2}}Y^{(4)}_{\mathbf{3}}\right)_{\mathbf{2'}}=\left[-Y_{1}^{4}-2\sqrt{2}Y_{1}Y_{2}^{3}\right]\left(Y_{1},\quad Y_{2}\right)^T,\\
&\hskip-0.15in Y^{(5)}_{\mathbf{2''}}=\left(Y^{(1)}_{\mathbf{2}}Y^{(4)}_{\mathbf{3}}\right)_{\mathbf{2''}}=\left(5Y^3_1Y^2_2-\sqrt{2}Y^5_2,~ -\sqrt{2}Y^5_1+5Y^2_1Y^3_2 \right)^T \,.
\end{aligned}
\end{equation}
Finally, the linearly independent weight-6 modular forms of level 3 can be decomposed into one singlet $\mathbf{1}$ and two triplets $\mathbf{3}$ under $T'$,
\begin{eqnarray}
\nonumber Y^{(6)}_{\mathbf{3},I}&=& \left(Y^{(1)}_{\mathbf{2}}Y^{(5)}_{\mathbf{2}}\right)_{\mathbf{3}} = \left[-2\sqrt{2}Y_{1}^{3}Y_{2}+Y_{2}^{4}\right] \left( Y_{2}^{2},\quad \sqrt{2}Y_{1}Y_{2},\quad -Y_{1}^{2}\right)^{T},\\
\nonumber Y^{(6)}_{\mathbf{3},II}&=&\left(Y^{(1)}_{\mathbf{2}}Y^{(5)}_{\mathbf{2'}}\right)_{\mathbf{3}}=\left[-Y_{1}^{4}-2\sqrt{2}Y_{1}Y_{2}^{3}\right] \left( -Y_{1}^{2},\quad Y_{2}^{2},\quad \sqrt{2}Y_{1}Y_{2}\right)^{T},\\
Y^{(6)}_{\mathbf{1}}&=&\left(Y^{(1)}_{\mathbf{2}}Y^{(5)}_{\mathbf{2''}}\right)_{\mathbf{1}}=\sqrt{2}Y^6_2-\sqrt{2}Y^6_1+10Y^3_1Y^3_2 \,.
\label{eq:modfvec2}
\end{eqnarray}

\begin{table}[t!]
\centering
\begin{tabular}{|c|c|}
\hline  \hline

Modular weight $k$ & Modular form $Y^{(k)}_{\mathbf{r}}$ \\ \hline

$k=1$ & $Y^{(1)}_{\mathbf{2}}$\\  \hline

$k=2$ & $Y^{(2)}_{\mathbf{3}}$\\ \hline

$k=3$ & $Y^{(3)}_{\mathbf{2}}, Y^{(3)}_{\mathbf{2}''}$\\ \hline

$k=4$ & $Y^{(4)}_{\mathbf{1}}, Y^{(4)}_{\mathbf{1}'}, Y^{(4)}_{\mathbf{3}}$\\ \hline

$k=5$ & $Y^{(5)}_{\mathbf{2}}, Y^{(5)}_{\mathbf{2}'}, Y^{(5)}_{\mathbf{2}''}$\\ \hline

$k=6$ & $Y^{(6)}_{\mathbf{1}}, Y^{(6)}_{\mathbf{3}I}, Y^{(6)}_{\mathbf{3}II}$\\ \hline \hline
\end{tabular}
\caption{\label{Tab:Level3_MM}Summary of modular forms of level 3 up to weight 6, the subscript $\mathbf{r}$ denotes the transformation property under $T'$ modular symmetry. Here $Y^{(6)}_{\mathbf{3}I}$ and $Y^{(6)}_{\mathbf{3}II}$ stand for two linearly independent weight-6 modular forms transforming in the representation $\mathbf{3}$ of $T'$}
\end{table}

We summarize the level 3 modular forms up to weight 6 in Table~\ref{Tab:Level3_MM}.

\section{Representative models of lepton texture zeros}
\label{sec:appendix_C}

In this section, we provide representative models for the phenomenologically viable patterns of texture zeros in the charged-lepton and neutrino mass matrices. Each of these models is chosen from a set of viable models which give the same texture of lepton mass matrices, and it contains minimum number of free real input parameters. Moreover, we also present the corresponding predictions for lepton masses and mixing parameters in each case.

\subsection{Dirac neutrinos}

The representative models of viable textures for the case of Dirac neutrinos are presented in Table~\ref{tab:Dirac_pair_summary_NO}, for which the representation and modular-weight assignments can be found in Table~\ref{tab:assign_Dirac_NO} for NO neutrino masses. The corresponding predictions for lepton observables are collected in Table~\ref{tab:prediction_Dirac_NO}. For IO, the viable textures are given in Table~\ref{tab:Dirac_pair_summary_IO} and the representative models and the best fitting results can be found in Table~\ref{tab:assign_Dirac_IO} and Table~\ref{tab:prediction_Dirac_IO} respectively.

\begin{center}
\begin{small}
 \begin{landscape}
  \setlength\LTcapwidth{\textwidth}
\setlength\LTleft{0.6in}            
\setlength\LTright{0pt}           
\begin{longtable}{|c|c|c|c|c|c|c|c|c|}
   \caption{\label{tab:assign_Dirac_NO}Representative Dirac-neutrino models of the viable texture-zero patterns in $(M_{E},M_{D})$ that can accommodate the experimental data at $3\sigma$ level for NO neutrino mass spectrum. Both cases with and without gCP are considered. The number of real free parameters ``$\#P$" involved in each model is listed in the third column. In the second column, we provide the value of ``$\#P_{0}$" which refers to the number of real free parameters of the lepton mass matrices in the general context of texture zero ansatz where the non-vanishing elements of mass matrices are independent from each other.
} \\
\hline
\multicolumn{9}{|c|}{Dirac without gCP (NO)} \\ \hline
Combinations & $\#\text{P}_{0}$ & $\#\text{P}$ & $\rho_{L}$ & $\rho_{E^{c}}$ & $\rho_{N^{c}}$ & $k_{L}$ & $k_{E^{c}}$ & $k_{N^{c}}$ \\ \hline
\endfirsthead

\multicolumn{9}{c}%
{{\bfseries \tablename\ \thetable{} -- continued from previous page}} \\
\hline
Combinations & $\#\text{P}_{0}$ & $\#\text{P}$ & $\rho_{L}$ & $\rho_{E^{c}}$ & $\rho_{N^{c}}$ & $k_{L}$ & $k_{E^{c}}$ & $k_{N^{c}}$ \\ \hline
\endhead

\hline \multicolumn{9}{|r|}{{continues on next page}} \\ \hline
\endfoot

\hline \hline
\endlastfoot

$\mathcal{C}_{1}^{(1)}-\mathcal{D}_{1}^{(1)}$ & $24$ & $9$ & $ \mathbf{2} \oplus \mathbf{1''}   $ & $\mathbf{2''} \oplus \mathbf{1''}   $ & $\mathbf{2} \oplus \mathbf{1''}   $ & $2 - x, 5 - x $ & $x, -1 + x $ & $x, -1 + x $\\ \hline$\mathcal{C}_{1}^{(1)}-\mathcal{D}_{4}^{(1)}$ & $18$ & $9$ & $ \mathbf{2} \oplus \mathbf{1'}   $ & $\mathbf{2'} \oplus \mathbf{1}   $ & $\mathbf{2} \oplus \mathbf{1'}   $ & $2 - x, 3 - x $ & $2 + x, 3 + x $ & $x, 1 + x $\\ \hline$\mathcal{C}_{2}^{(2)}-\mathcal{D}_{2}^{(2)}$ & $20$ & $9$ & $ \mathbf{2} \oplus \mathbf{1'}   $ & $\mathbf{2''} \oplus \mathbf{1'}   $ & $\mathbf{2'} \oplus \mathbf{1'}   $ & $2 - x, 5 - x $ & $x, -1 + x $ & $x, -1 + x $\\ \hline$\mathcal{C}_{2}^{(3)}-\mathcal{D}_{1}^{(1)}$ & $22$ & $9$ & $ \mathbf{2} \oplus \mathbf{1''}   $ & $\mathbf{2''} \oplus \mathbf{1}   $ & $\mathbf{2} \oplus \mathbf{1''}   $ & $2 - x, 1 - x $ & $x, 3 + x $ & $x, 1 + x $\\ \hline$\mathcal{C}_{2}^{(3)}-\mathcal{D}_{2}^{(1)}$ & $20$ & $9$ & $ \mathbf{2} \oplus \mathbf{1''}   $ & $\mathbf{1''} \oplus \mathbf{1''} \oplus \mathbf{1'}   $ & $\mathbf{3}   $ & $1 - x, 2 - x $ & $2 + x, 4 + x, 4 + x $ & $x $\\ \hline$\mathcal{C}_{2}^{(3)}-\mathcal{D}_{4}^{(1)}$ & $16$ & $8$ & $ \mathbf{2} \oplus \mathbf{1}   $ & $\mathbf{1''} \oplus \mathbf{1} \oplus \mathbf{1}   $ & $\mathbf{2} \oplus \mathbf{1}   $ & $2 - x, 1 - x $ & $-1 + x, 1 + x, 3 + x $ & $x, -1 + x $\\ \hline$\mathcal{C}_{3}^{(1)}-\mathcal{D}_{2}^{(1)}$ & $18$ & $9$ & $ \mathbf{2} \oplus \mathbf{1''}   $ & $\mathbf{1'} \oplus \mathbf{1''} \oplus \mathbf{1'}   $ & $\mathbf{3}   $ & $1 - x, 2 - x $ & $2 + x, 2 + x, 4 + x $ & $x $\\ \hline$\mathcal{C}_{3}^{(1)}-\mathcal{D}_{4}^{(1)}$ & $14$ & $8$ & $ \mathbf{2} \oplus \mathbf{1'}   $ & $\mathbf{1} \oplus \mathbf{1''} \oplus \mathbf{1'}   $ & $\mathbf{2} \oplus \mathbf{1'}   $ & $2 - x, 3 - x $ & $1 + x, 1 + x, 1 + x $ & $x, 1 + x $\\
\end{longtable}
\end{landscape}
\end{small}
\end{center}

\clearpage
\newpage

\begin{center}
\begin{small}
\setlength\LTcapwidth{\textwidth}
\setlength\LTleft{-0.15in}            
\setlength\LTright{0pt}           
 \begin{longtable}{|c|c|c|c|c|c|c|c|c|c|}
\caption{\label{tab:prediction_Dirac_NO}Best-fit values of the lepton mixing and neutrino mass parameters for the representative models of Table~\ref{tab:assign_Dirac_NO}. In all representative models, the predictions of the lepton mass ratios are $m_{e}/m_{\mu}=0.00474$, $m_{\mu}/m_{\tau}=0.0586$ and $\Delta m_{21}^{2}/\Delta m_{31}^{2}=0.030$.} \\
\hline
\multicolumn{9}{|c|}{Dirac without gCP (NO)} \\ \hline
\multirow{2}{*}{Combinations}& \multicolumn{7}{c|}{Predictions for mixing parameters and neutrino masses at best fitting point} & \multirow{2}{*}{$\chi^{2}_{\text{min}}$} \\ \cline{2-8}
 & $\sin^2\theta_{12}$ &$\sin^2\theta_{13}$ &$\sin^2\theta_{23}$&$\delta^{l}_ {CP}/\pi$ & $m_ 1$/meV & $m_ 2$/meV & $m_ 3$/meV &  \\ \hline
\endfirsthead

\multicolumn{9}{c}%
{{\bfseries \tablename\ \thetable{} -- continued from previous page}} \\
\hline
\multirow{2}{*}{Combinations}& \multicolumn{7}{c|}{Predictions for mixing parameters and neutrino masses at best fitting point} & \multirow{2}{*}{$\chi^{2}_{\text{min}}$} \\ \cline{2-8}
 & $\sin^2\theta_{12}$ &$\sin^2\theta_{13}$ &$\sin^2\theta_{23}$&$\delta^{l}_ {CP}/\pi$ & $m_ 1$/meV & $m_ 2$/meV & $m_ 3$/meV &  \\ \hline
\endhead

\hline \multicolumn{9}{|r|}{{continues on next page}} \\ \hline
\endfoot

\hline \hline
\endlastfoot

$\mathcal{C}_{1}^{(1)}-\mathcal{D}_{1}^{(1)}$ & $0.304$ & $0.02246$ & $0.452$ & $1.361$ & $44.054$ & $44.888$ & $66.720$ & $0.187$\\ \hline$\mathcal{C}_{1}^{(1)}-\mathcal{D}_{4}^{(1)}$ & $0.326$ & $0.02262$ & $0.440$ & $1.082$ & $7.927$ & $11.706$ & $50.722$ & $5.844$\\ \hline$\mathcal{C}_{2}^{(2)}-\mathcal{D}_{2}^{(2)}$ & $0.286$ & $0.02247$ & $0.442$ & $1.813$ & $2.357\times 10^{-8}$ & $8.614$ & $50.100$ & $9.709$\\ \hline$\mathcal{C}_{2}^{(3)}-\mathcal{D}_{1}^{(1)}$ & $0.304$ & $0.02242$ & $0.455$ & $1.609$ & $29.860$ & $31.078$ & $58.309$ & $2.812$\\ \hline$\mathcal{C}_{2}^{(3)}-\mathcal{D}_{2}^{(1)}$ & $0.303$ & $0.02246$ & $0.473$ & $1.262$ & $34.122$ & $35.192$ & $60.575$ & $1.479$\\ \hline$\mathcal{C}_{2}^{(3)}-\mathcal{D}_{4}^{(1)}$ & $0.304$ & $0.02245$ & $0.451$ & $1.042$ & $35.717$ & $36.741$ & $61.528$ & $2.879$\\ \hline$\mathcal{C}_{3}^{(1)}-\mathcal{D}_{2}^{(1)}$ & $0.303$ & $0.02246$ & $0.473$ & $1.262$ & $34.122$ & $35.192$ & $60.575$ & $1.479$\\ \hline$\mathcal{C}_{3}^{(1)}-\mathcal{D}_{4}^{(1)}$ & $0.304$ & $0.02244$ & $0.451$ & $1.158$ & $39.428$ & $40.358$ & $63.754$ & $0.749$\\
\end{longtable}
\end{small}
\end{center}

\begin{center}
\begin{small}
 \begin{landscape}
  \setlength\LTcapwidth{\textwidth}
\setlength\LTleft{0.45in}            
\setlength\LTright{0pt}           
 \begin{longtable}{|c|c|c|c|c|c|c|c|c|}
   \caption{\label{tab:assign_Dirac_IO}
   The same as in Table~\ref{tab:assign_Dirac_NO} but for IO neutrino mass spectrum.
 } \\
\hline
\multicolumn{9}{|c|}{Dirac without gCP (IO)} \\ \hline
Combinations & $\#\text{P}_{0}$ & $\#\text{P}$ & $\rho_{L}$ & $\rho_{E^{c}}$ & $\rho_{N^{c}}$ & $k_{L}$ & $k_{E^{c}}$ & $k_{N^{c}}$ \\ \hline
\endfirsthead

\multicolumn{9}{c}%
{{\bfseries \tablename\ \thetable{} -- continued from previous page}} \\
\hline
Combinations & $\#\text{P}_{0}$ & $\#\text{P}$ & $\rho_{L}$ & $\rho_{E^{c}}$ & $\rho_{N^{c}}$ & $k_{L}$ & $k_{E^{c}}$ & $k_{N^{c}}$ \\ \hline
\endhead

\hline \multicolumn{9}{|r|}{{continues on next page}} \\ \hline
\endfoot

\hline \hline
\endlastfoot

$\mathcal{C}_{1}^{(1)}-\mathcal{D}_{1}^{(1)}$ & $24$ & $9$ & $ \mathbf{2} \oplus \mathbf{1''}   $ & $\mathbf{2'} \oplus \mathbf{1}   $ & $\mathbf{2''} \oplus \mathbf{1''}   $ & $2 - x, 5 - x $ & $x, 1 + x $ & $x, -1 + x $\\ \hline$\mathcal{C}_{2}^{(2)}-\mathcal{D}_{1}^{(1)}$ & $22$ & $9$ & $ \mathbf{2} \oplus \mathbf{1''}   $ & $\mathbf{2''} \oplus \mathbf{1'}   $ & $\mathbf{2} \oplus \mathbf{1''}   $ & $2 - x, 5 - x $ & $x, 1 + x $ & $x, 1 + x $\\ \hline$\mathcal{C}_{2}^{(2)}-\mathcal{D}_{2}^{(2)}$ & $20$ & $9$ & $ \mathbf{2} \oplus \mathbf{1'}   $ & $\mathbf{2''} \oplus \mathbf{1'}   $ & $\mathbf{2'} \oplus \mathbf{1'}   $ & $2 - x, 5 - x $ & $x, -1 + x $ & $x, -1 + x $\\ \hline$\mathcal{C}_{2}^{(3)}-\mathcal{D}_{1}^{(1)}$ & $22$ & $9$ & $ \mathbf{2} \oplus \mathbf{1}   $ & $\mathbf{2} \oplus \mathbf{1''}   $ & $\mathbf{2''} \oplus \mathbf{1''}   $ & $2 - x, 1 - x $ & $x, 3 + x $ & $x, -1 + x $\\ \hline$\mathcal{C}_{2}^{(3)}-\mathcal{D}_{4}^{(1)}$ & $16$ & $8$ & $ \mathbf{2} \oplus \mathbf{1''}   $ & $\mathbf{1''} \oplus \mathbf{1''} \oplus \mathbf{1}   $ & $\mathbf{2''} \oplus \mathbf{1'}   $ & $2 - x, 3 - x $ & $-1 + x, 1 + x, 1 + x $ & $x, -3 + x $\\ \hline$\mathcal{C}_{3}^{(1)}-\mathcal{D}_{4}^{(1)}$ & $14$ & $8$ & $ \mathbf{2} \oplus \mathbf{1'}   $ & $\mathbf{1} \oplus \mathbf{1''} \oplus \mathbf{1'}   $ & $\mathbf{2} \oplus \mathbf{1'}   $ & $2 - x, 3 - x $ & $1 + x, 1 + x, 1 + x $ & $x, 1 + x $\\ \hline$\mathcal{C}_{4}^{(1)}-\mathcal{D}_{1}^{(1)}$ & $18$ & $8$ & $ \mathbf{2} \oplus \mathbf{1''}   $ & $\mathbf{1''} \oplus \mathbf{1} \oplus \mathbf{1}   $ & $\mathbf{2''} \oplus \mathbf{1''}   $ & $2 - x, 5 - x $ & $-1 + x, -1 + x, 1 + x $ & $x, -1 + x $\\ \hline$\mathcal{C}_{4}^{(1)}-\mathcal{D}_{2}^{(3)}$ & $16$ & $9$ & $ \mathbf{2} \oplus \mathbf{1'}   $ & $\mathbf{1} \oplus \mathbf{1'} \oplus \mathbf{1}   $ & $\mathbf{1''} \oplus \mathbf{1''} \oplus \mathbf{1}   $ & $1 - x, 2 - x $ & $2 + x, 2 + x, 4 + x $ & $x, 4 + x, 4 + x $\\
\end{longtable}
\end{landscape}
\end{small}
\end{center}

\begin{center}
\begin{small}
\setlength\LTcapwidth{\textwidth}
\setlength\LTleft{-0.12in}            
\setlength\LTright{0pt}           
 \begin{longtable}{|c|c|c|c|c|c|c|c|c|c|}
\caption{\label{tab:prediction_Dirac_IO}
   The same as in Table~\ref{tab:prediction_Dirac_NO} but for IO neutrino mass spectrum and for the representative models of Table~~\ref{tab:assign_Dirac_IO}.
} 
\\
\hline
\multicolumn{9}{|c|}{Dirac without gCP (IO)} \\ \hline
\multirow{2}{*}{Combinations}& \multicolumn{7}{c|}{Predictions for mixing parameters and neutrino masses at best fitting point} & \multirow{2}{*}{$\chi^{2}_{\text{min}}$} \\ \cline{2-8}
 & $\sin^2\theta_{12}$ &$\sin^2\theta_{13}$ &$\sin^2\theta_{23}$&$\delta^{l}_ {CP}/\pi$ & $m_ 1$/meV & $m_ 2$/meV & $m_ 3$/meV &  \\ \hline
\endfirsthead

\multicolumn{9}{c}%
{{\bfseries \tablename\ \thetable{} -- continued from previous page}} \\
\hline
\multirow{2}{*}{Combinations}& \multicolumn{7}{c|}{Predictions for mixing parameters and neutrino masses at best fitting point} & \multirow{2}{*}{$\chi^{2}_{\text{min}}$} \\ \cline{2-8}
 & $\sin^2\theta_{12}$ &$\sin^2\theta_{13}$ &$\sin^2\theta_{23}$&$\delta^{l}_ {CP}/\pi$ & $m_ 1$/meV & $m_ 2$/meV & $m_ 3$/meV &  \\ \hline
\endhead

\hline \multicolumn{9}{|r|}{{continues on next page}} \\ \hline
\endfoot

\hline \hline
\endlastfoot

$\mathcal{C}_{1}^{(1)}-\mathcal{D}_{1}^{(1)}$ & $0.306$ & $0.02241$ & $0.569$ & $1.561$ & $52.681$ & $53.381$ & $18.971$ & $0.033$\\ \hline$\mathcal{C}_{2}^{(2)}-\mathcal{D}_{1}^{(1)}$ & $0.298$ & $0.02244$ & $0.560$ & $1.720$ & $57.720$ & $58.359$ & $30.236$ & $2.537$\\ \hline$\mathcal{C}_{2}^{(2)}-\mathcal{D}_{2}^{(2)}$ & $0.287$ & $0.02243$ & $0.577$ & $1.185$ & $49.151$ & $49.900$ & $1.149\times 10^{-8}$ & $6.894$\\ \hline$\mathcal{C}_{2}^{(3)}-\mathcal{D}_{1}^{(1)}$ & $0.306$ & $0.02241$ & $0.570$ & $1.514$ & $52.591$ & $53.292$ & $18.720$ & $0.051$\\ \hline$\mathcal{C}_{2}^{(3)}-\mathcal{D}_{4}^{(1)}$ & $0.305$ & $0.02234$ & $0.573$ & $1.750$ & $60.717$ & $61.325$ & $35.648$ & $2.868$\\ \hline$\mathcal{C}_{3}^{(1)}-\mathcal{D}_{4}^{(1)}$ & $0.305$ & $0.02234$ & $0.573$ & $1.750$ & $60.718$ & $61.326$ & $35.650$ & $2.869$\\ \hline$\mathcal{C}_{4}^{(1)}-\mathcal{D}_{1}^{(1)}$ & $0.307$ & $0.02242$ & $0.569$ & $1.576$ & $52.700$ & $53.399$ & $19.036$ & $0.138$\\ \hline$\mathcal{C}_{4}^{(1)}-\mathcal{D}_{2}^{(3)}$ & $0.304$ & $0.02241$ & $0.570$ & $1.543$ & $72.865$ & $73.372$ & $53.791$ & $0.000$\\
\end{longtable}
\end{small}
\end{center}

\newpage

\subsection{Majorana neutrinos (Weinberg operator)}

In case neutrinos are Majorana particles and their masses are generated via Weinberg operator, the viable pairs of texture zeros of the lepton mass matrices are summarized in Table~\ref{tab:Majorana_WO_pair_summary_NO_IO}. Here, we present the corresponding representative models for NO and IO neutrino masses spectrum in Tables~\ref{tab:assign_Majorana_WO_NO} and \ref{tab:assign_Majorana_WO_IO}, respectively. The numerical results of these representative models are given in Tables~\ref{tab:prediction_Majorana_WO_NO} and ~\ref{tab:prediction_Majorana_WO_IO}.

\begin{table}[ht!]
 \centering
\begin{tabular}{|c|c|c|c|c|c|c|} \hline\hline
\multicolumn{7}{|c|}{Weinberg operator without gCP (NO)} \\ \hline
  Combinations & $\#\text{P}_{0}$ & $\#\text{P}$ & $\rho_{L}$ & $\rho_{E^{c}}$ & $k_{L}$ & $k_{E^{c}}$ \\ \hline$\mathcal{C}_{2}^{(3)}-\mathcal{W}_{1}^{(1)}$ & $18$ & $9$ & $ \mathbf{2'} \oplus \mathbf{1'}   $ & $\mathbf{1'} \oplus \mathbf{1''} \oplus \mathbf{1''}   $ & $1, 0 $ & $0, 2, 4 $\\ \hline$\mathcal{C}_{2}^{(3)}-\mathcal{W}_{2}^{(2)}$ & $16$ & $9$ & $ \mathbf{2} \oplus \mathbf{1'}   $ & $\mathbf{1''} \oplus \mathbf{1} \oplus \mathbf{1''}   $ & $1, 2 $ & $0, 2, 2 $\\ \hline$\mathcal{C}_{3}^{(1)}-\mathcal{W}_{1}^{(1)}$ & $16$ & $9$ & $ \mathbf{2'} \oplus \mathbf{1''}   $ & $\mathbf{1'} \oplus \mathbf{1'} \oplus \mathbf{1}   $ & $1, 2 $ & $-2, 2, 4 $\\ \hline$\mathcal{C}_{3}^{(1)}-\mathcal{W}_{2}^{(2)}$ & $14$ & $9$ & $ \mathbf{2} \oplus \mathbf{1'}   $ & $\mathbf{1} \oplus \mathbf{1''} \oplus \mathbf{1'}   $ & $1, 2 $ & $2, 2, 2 $\\ \hline$\mathcal{C}_{4}^{(1)}-\mathcal{W}_{1}^{(1)}$ & $15$ & $7$ & $ \mathbf{2'} \oplus \mathbf{1'}   $ & $\mathbf{1''} \oplus \mathbf{1'} \oplus \mathbf{1'}   $ & $1, 0 $ & $0, 0, 2 $\\ \hline
\multicolumn{7}{|c|}{Weinberg operator with gCP (NO)} \\ \hline
Combinations & $\#\text{P}_{0}$ & $\#\text{P}$ & $\rho_{L}$ & $\rho_{E^{c}}$ & $k_{L}$ & $k_{E^{c}}$ \\ \hline$\mathcal{C}_{1}^{(1)}-\mathcal{W}_{1}^{(1)}$ & $20$ & $8$ & $ \mathbf{2'} \oplus \mathbf{1''}   $ & $\mathbf{2'} \oplus \mathbf{1''}   $ & $1, 2 $ & $3, 2 $\\ \hline$\mathcal{C}_{1}^{(1)}-\mathcal{W}_{2}^{(2)}$ & $18$ & $8$ & $ \mathbf{2} \oplus \mathbf{1'}   $ & $\mathbf{2'} \oplus \mathbf{1}   $ & $1, 2 $ & $3, 2 $\\ \hline$\mathcal{C}_{1}^{(1)}-\mathcal{W}_{3}^{(1)}$ & $16$ & $9$ & $ \mathbf{2''} \oplus \mathbf{1}   $ & $\mathbf{1} \oplus \mathbf{1} \oplus \mathbf{1'}   $ & $-1, 2 $ & $2, 4, 4 $\\ \hline$\mathcal{C}_{2}^{(2)}-\mathcal{W}_{1}^{(1)}$ & $18$ & $8$ & $ \mathbf{2'} \oplus \mathbf{1''}   $ & $\mathbf{2'} \oplus \mathbf{1}   $ & $1, 2 $ & $3, 2 $\\ \hline$\mathcal{C}_{2}^{(2)}-\mathcal{W}_{2}^{(2)}$ & $16$ & $9$ & $ \mathbf{2'} \oplus \mathbf{1}   $ & $\mathbf{1''} \oplus \mathbf{1} \oplus \mathbf{1}   $ & $1, 2 $ & $2, 2, 4 $\\ \hline$\mathcal{C}_{2}^{(2)}-\mathcal{W}_{3}^{(1)}$ & $14$ & $9$ & $ \mathbf{2'} \oplus \mathbf{1'}   $ & $\mathbf{1''} \oplus \mathbf{1'} \oplus \mathbf{1''}   $ & $-1, 2 $ & $2, 2, 4 $\\ \hline$\mathcal{C}_{2}^{(3)}-\mathcal{W}_{1}^{(1)}$ & $18$ & $8$ & $ \mathbf{2'} \oplus \mathbf{1''}   $ & $\mathbf{1'} \oplus \mathbf{1''} \oplus \mathbf{1}   $ & $1, 2 $ & $2, 2, 4 $\\ \hline$\mathcal{C}_{2}^{(3)}-\mathcal{W}_{2}^{(2)}$ & $16$ & $8$ & $ \mathbf{2} \oplus \mathbf{1'}   $ & $\mathbf{1''} \oplus \mathbf{1} \oplus \mathbf{1''}   $ & $1, 2 $ & $0, 2, 2 $\\ \hline$\mathcal{C}_{3}^{(1)}-\mathcal{W}_{1}^{(1)}$ & $16$ & $8$ & $ \mathbf{2'} \oplus \mathbf{1''}   $ & $\mathbf{1''} \oplus \mathbf{1'} \oplus \mathbf{1}   $ & $1, 2 $ & $2, 2, 2 $\\ \hline$\mathcal{C}_{3}^{(1)}-\mathcal{W}_{2}^{(2)}$ & $14$ & $8$ & $ \mathbf{2} \oplus \mathbf{1'}   $ & $\mathbf{1''} \oplus \mathbf{1} \oplus \mathbf{1'}   $ & $1, 2 $ & $2, 2, 2 $\\ \hline$\mathcal{C}_{4}^{(1)}-\mathcal{W}_{1}^{(1)}$ & $15$ & $7$ & $ \mathbf{2} \oplus \mathbf{1''}   $ & $\mathbf{2''} \oplus \mathbf{1'}   $ & $1, 0 $ & $3, 0 $\\ \hline \hline \end{tabular}
\caption{\label{tab:assign_Majorana_WO_NO}Representative models of the viable patterns of texture zero in $(M_{E},M_{\nu})$ that can accommodate the experimental data at $3\sigma$ level for NO neutrino mass spectrum. Here neutrinos are Majorana particles and neutrino masses are assumed to be described by the Weinberg operator. Models with and without gCP are considered. The same convention as Table~\ref{tab:assign_Dirac_NO} is adopted.}
\end{table}

\begin{table}[ht!]
 \centering \resizebox{1.0\textwidth}{!}{
\begin{tabular}{|c|c|c|c|c|c|c|c|c|c|c|c|c|} \hline
\multicolumn{12}{|c|}{Weinberg operator without gCP} \\ \hline
\multirow{2}{*}{Combinations}& \multicolumn{10}{c|}{Predictions for mixing parameters and neutrino masses at best fitting point} & \multirow{2}{*}{$\chi^{2}_{\text{min}}$} \\ \cline{2-11}
 & $\sin^2\theta_{12}$ &$\sin^2\theta_{13}$ &$\sin^2\theta_{23}$&$\delta_{CP}/\pi$ & $\alpha_{21}/\pi$  &$\alpha_{31}/\pi$ & $m_1$/meV & $m_2$/meV & $m_3$/meV & $m_{\beta\beta}$/meV &  \\ \hline$\mathcal{C}_{2}^{(3)}-\mathcal{W}_{1}^{(1)}$ & $0.306$ & $0.02246$ & $0.428$ & $1.365$ & $0.891$ & $1.936$ & $130.202$ & $130.487$ & $139.517$ & $50.422$ & $2.028$\\
$\mathcal{C}_{2}^{(3)}-\mathcal{W}_{2}^{(2)}$ & $0.304$ & $0.02244$ & $0.451$ & $1.158$ & $0.492$ & $1.683$ & $39.422$ & $40.352$ & $63.750$ & $28.608$ & $0.748$\\
$\mathcal{C}_{3}^{(1)}-\mathcal{W}_{1}^{(1)}$ & $0.305$ & $0.02246$ & $0.452$ & $0.839$ & $1.595$ & $1.628$ & $20.198$ & $21.958$ & $54.062$ & $18.033$ & $10.017$\\
$\mathcal{C}_{3}^{(1)}-\mathcal{W}_{2}^{(2)}$ & $0.304$ & $0.02244$ & $0.451$ & $1.158$ & $0.584$ & $1.775$ & $39.428$ & $40.358$ & $63.754$ & $25.599$ & $0.749$\\
  $\mathcal{C}_{4}^{(1)}-\mathcal{W}_{1}^{(1)}$ & $0.304$ & $0.02254$ & $0.601$ & $1.482$ & $1.957$ & $0.970$ & $44.194$ & $45.025$ & $66.800$ & $44.865$ & $64.205$\\ \hline
  \multicolumn{12}{|c|}{Weinberg operator with gCP} \\ \hline
\multirow{2}{*}{Combinations}& \multicolumn{10}{c|}{Predictions for mixing parameters and neutrino masses at best fitting point} & \multirow{2}{*}{$\chi^{2}_{\text{min}}$} \\ \cline{2-11}
 & $\sin^2\theta_{12}$ &$\sin^2\theta_{13}$ &$\sin^2\theta_{23}$&$\delta_{CP}/\pi$ & $\alpha_{21}/\pi$  &$\alpha_{31}/\pi$ & $m_1$/meV & $m_2$/meV & $m_3$/meV & $m_{\beta\beta}$/meV &  \\ \hline$\mathcal{C}_{1}^{(1)}-\mathcal{W}_{1}^{(1)}$ & $0.304$ & $0.02246$ & $0.450$ & $1.351$ & $0.308$ & $1.199$ & $18.820$ & $20.698$ & $53.510$ & $17.466$ & $0.133$\\
$\mathcal{C}_{1}^{(1)}-\mathcal{W}_{2}^{(2)}$ & $0.304$ & $0.02244$ & $0.451$ & $1.158$ & $1.594$ & $0.785$ & $39.425$ & $40.355$ & $63.752$ & $32.076$ & $0.749$\\
$\mathcal{C}_{1}^{(1)}-\mathcal{W}_{3}^{(1)}$ & $0.304$ & $0.02246$ & $0.450$ & $1.045$ & $0.240$ & $1.330$ & $0$ & $8.614$ & $50.100$ & $1.436$ & $2.807$\\
$\mathcal{C}_{2}^{(2)}-\mathcal{W}_{1}^{(1)}$ & $0.304$ & $0.02246$ & $0.450$ & $1.351$ & $0.308$ & $1.199$ & $18.820$ & $20.698$ & $53.510$ & $17.466$ & $0.133$\\
$\mathcal{C}_{2}^{(2)}-\mathcal{W}_{2}^{(2)}$ & $0.304$ & $0.02245$ & $0.449$ & $1.128$ & $0.308$ & $0.125$ & $3.919$ & $9.464$ & $50.243$ & $5.614$ & $1.171$\\
$\mathcal{C}_{2}^{(2)}-\mathcal{W}_{3}^{(1)}$ & $0.304$ & $0.02246$ & $0.450$ & $1.045$ & $1.665$ & $0.756$ & $0$ & $8.614$ & $50.100$ & $1.436$ & $2.807$\\
$\mathcal{C}_{2}^{(3)}-\mathcal{W}_{1}^{(1)}$ & $0.304$ & $0.02246$ & $0.451$ & $1.412$ & $1.588$ & $0.708$ & $29.686$ & $30.910$ & $58.235$ & $25.667$ & $0.451$\\
$\mathcal{C}_{2}^{(3)}-\mathcal{W}_{2}^{(2)}$ & $0.304$ & $0.02244$ & $0.451$ & $1.158$ & $0.594$ & $1.785$ & $39.422$ & $40.352$ & $63.750$ & $25.259$ & $0.748$\\
$\mathcal{C}_{3}^{(1)}-\mathcal{W}_{1}^{(1)}$ & $0.304$ & $0.02246$ & $0.450$ & $1.374$ & $1.594$ & $0.697$ & $20.178$ & $21.940$ & $54.009$ & $18.013$ & $0.231$\\
$\mathcal{C}_{3}^{(1)}-\mathcal{W}_{2}^{(2)}$ & $0.304$ & $0.02244$ & $0.451$ & $1.158$ & $0.594$ & $1.785$ & $39.428$ & $40.358$ & $63.754$ & $25.259$ & $0.749$\\
$\mathcal{C}_{4}^{(1)}-\mathcal{W}_{1}^{(1)}$ & $0.304$ & $0.02254$ & $0.601$ & $1.482$ & $1.957$ & $0.970$ & $44.214$ & $45.045$ & $66.813$ & $44.883$ & $64.198$\\ \hline \hline \end{tabular}}
\caption{\label{tab:prediction_Majorana_WO_NO}Best-fit values of the lepton mixing and neutrino mass parameters for the representative models presented in Table~\ref{tab:assign_Majorana_WO_NO}. For all representative models, the predictions of the lepton mass ratios are $m_{e}/m_{\mu}=0.00474$, $m_{\mu}/m_{\tau}=0.0586$ and $\Delta m_{21}^{2}/\Delta m_{31}^{2}=0.030$.}
\end{table}


\begin{table}[ht!]
 \centering
\begin{tabular}{|c|c|c|c|c|c|c|} \hline\hline
\multicolumn{7}{|c|}{Weinberg operator without gCP (IO)} \\ \hline
  Combinations & $\#\text{P}_{0}$ & $\#\text{P}$ & $\rho_{L}$ & $\rho_{E^{c}}$ & $k_{L}$ & $k_{E^{c}}$ \\ \hline$\mathcal{C}_{2}^{(3)}-\mathcal{W}_{1}^{(1)}$ & $18$ & $9$ & $ \mathbf{2} \oplus \mathbf{1''}   $ & $\mathbf{1''} \oplus \mathbf{1''} \oplus \mathbf{1}   $ & $2, 1 $ & $1, 3, 3 $\\ \hline$\mathcal{C}_{2}^{(3)}-\mathcal{W}_{2}^{(2)}$ & $16$ & $9$ & $ \mathbf{2} \oplus \mathbf{1'}   $ & $\mathbf{1''} \oplus \mathbf{1''} \oplus \mathbf{1}   $ & $1, 2 $ & $0, 2, 2 $\\ \hline$\mathcal{C}_{3}^{(1)}-\mathcal{W}_{1}^{(1)}$ & $16$ & $9$ & $ \mathbf{2} \oplus \mathbf{1''}   $ & $\mathbf{1'} \oplus \mathbf{1''} \oplus \mathbf{1}   $ & $2, 1 $ & $-1, -1, 3 $\\ \hline$\mathcal{C}_{3}^{(1)}-\mathcal{W}_{2}^{(2)}$ & $14$ & $9$ & $ \mathbf{2} \oplus \mathbf{1'}   $ & $\mathbf{1} \oplus \mathbf{1''} \oplus \mathbf{1'}   $ & $1, 2 $ & $2, 2, 2 $\\ \hline$\mathcal{C}_{4}^{(1)}-\mathcal{W}_{1}^{(1)}$ & $15$ & $9$ & $ \mathbf{2} \oplus \mathbf{1''}   $ & $\mathbf{1'} \oplus \mathbf{1''} \oplus \mathbf{1}   $ & $3, 0 $ & $0, 0, 0 $\\ \hline$\mathcal{C}_{2}^{(2)}-\mathcal{W}_{3}^{(1)}$ & $14$ & $8$ & $ \mathbf{2'} \oplus \mathbf{1'}   $ & $\mathbf{2''} \oplus \mathbf{1''}   $ & $-1, 2 $ & $3, 2 $\\ \hline$\mathcal{C}_{3}^{(1)}-\mathcal{W}_{3}^{(1)}$ & $12$ & $9$ & $ \mathbf{2''} \oplus \mathbf{1}   $ & $\mathbf{1} \oplus \mathbf{1''} \oplus \mathbf{1'}   $ & $-1, 2 $ & $2, 2, 4 $\\ \hline
\multicolumn{7}{|c|}{Weinberg operator with gCP (IO)} \\ \hline
Combinations & $\#\text{P}_{0}$ & $\#\text{P}$ & $\rho_{L}$ & $\rho_{E^{c}}$ & $k_{L}$ & $k_{E^{c}}$ \\ \hline$\mathcal{C}_{1}^{(1)}-\mathcal{W}_{1}^{(1)}$ & $20$ & $8$ & $ \mathbf{2} \oplus \mathbf{1''}   $ & $\mathbf{2''} \oplus \mathbf{1''}   $ & $2, 3 $ & $2, 1 $\\ \hline$\mathcal{C}_{1}^{(1)}-\mathcal{W}_{2}^{(2)}$ & $18$ & $8$ & $ \mathbf{2} \oplus \mathbf{1'}   $ & $\mathbf{2'} \oplus \mathbf{1}   $ & $1, 2 $ & $3, 2 $\\ \hline$\mathcal{C}_{1}^{(1)}-\mathcal{W}_{3}^{(2)}$ & $16$ & $9$ & $ \mathbf{2'} \oplus \mathbf{1}   $ & $\mathbf{1'} \oplus \mathbf{1} \oplus \mathbf{1''}   $ & $2, 1 $ & $3, 3, 3 $\\ \hline$\mathcal{C}_{2}^{(2)}-\mathcal{W}_{1}^{(1)}$ & $18$ & $8$ & $ \mathbf{2} \oplus \mathbf{1}   $ & $\mathbf{2''} \oplus \mathbf{1}   $ & $2, 1 $ & $2, -1 $\\ \hline$\mathcal{C}_{2}^{(2)}-\mathcal{W}_{2}^{(2)}$ & $16$ & $9$ & $ \mathbf{2} \oplus \mathbf{1}   $ & $\mathbf{1} \oplus \mathbf{1} \oplus \mathbf{1''}   $ & $1, 0 $ & $0, 4, 4 $\\ \hline$\mathcal{C}_{2}^{(3)}-\mathcal{W}_{1}^{(1)}$ & $18$ & $8$ & $ \mathbf{2'} \oplus \mathbf{1'}   $ & $\mathbf{2} \oplus \mathbf{1''}   $ & $1, 0 $ & $3, 4 $\\ \hline$\mathcal{C}_{2}^{(3)}-\mathcal{W}_{2}^{(2)}$ & $16$ & $8$ & $ \mathbf{2} \oplus \mathbf{1'}   $ & $\mathbf{1''} \oplus \mathbf{1''} \oplus \mathbf{1}   $ & $1, 2 $ & $0, 2, 2 $\\ \hline$\mathcal{C}_{3}^{(1)}-\mathcal{W}_{1}^{(1)}$ & $16$ & $8$ & $ \mathbf{2'} \oplus \mathbf{1'}   $ & $\mathbf{1'} \oplus \mathbf{1''} \oplus \mathbf{1''}   $ & $1, 0 $ & $0, 0, 4 $\\ \hline$\mathcal{C}_{3}^{(1)}-\mathcal{W}_{2}^{(2)}$ & $14$ & $8$ & $ \mathbf{2'} \oplus \mathbf{1}   $ & $\mathbf{1'} \oplus \mathbf{1} \oplus \mathbf{1''}   $ & $1, 2 $ & $2, 2, 2 $\\ \hline$\mathcal{C}_{4}^{(1)}-\mathcal{W}_{1}^{(1)}$ & $15$ & $8$ & $ \mathbf{2''} \oplus \mathbf{1'}   $ & $\mathbf{2} \oplus \mathbf{1''}   $ & $2, 1 $ & $2, -1 $\\ \hline \hline \end{tabular}
\caption{\label{tab:assign_Majorana_WO_IO} The same as in Table~\ref{tab:assign_Majorana_WO_NO} but for IO neutrino masses.}
\end{table}

\begin{table}[ht!]
 \centering \resizebox{1.0\textwidth}{!}{
\begin{tabular}{|c|c|c|c|c|c|c|c|c|c|c|c|c|} \hline
     \multicolumn{12}{|c|}{Weinberg operator without gCP} \\ \hline
  \multirow{2}{*}{Combinations}& \multicolumn{10}{c|}{Predictions for mixing parameters and neutrino masses at best fitting point} & \multirow{2}{*}{$\chi^{2}_{\text{min}}$} \\ \cline{2-11}
 & $\sin^2\theta_{12}$ &$\sin^2\theta_{13}$ &$\sin^2\theta_{23}$&$\delta_{CP}/\pi$ & $\alpha_{21}/\pi$  &$\alpha_{31}/\pi$ & $m_1$/meV & $m_2$/meV & $m_3$/meV & $m_{\beta\beta}$/meV &  \\ \hline$\mathcal{C}_{2}^{(3)}-\mathcal{W}_{1}^{(1)}$ & $0.304$ & $0.02241$ & $0.570$ & $1.546$ & $0.663$ & $1.585$ & $49.52884$ & $50.27231$ & $6.10583$ & $29.48120$ & $0.00025$\\
$\mathcal{C}_{2}^{(3)}-\mathcal{W}_{2}^{(2)}$ & $0.305$ & $0.02234$ & $0.573$ & $1.750$ & $1.604$ & $0.724$ & $60.71693$ & $61.32492$ & $35.64755$ & $49.81772$ & $2.86785$\\
$\mathcal{C}_{3}^{(1)}-\mathcal{W}_{1}^{(1)}$ & $0.304$ & $0.02241$ & $0.570$ & $1.546$ & $1.695$ & $0.666$ & $49.52893$ & $50.27241$ & $6.10589$ & $44.10146$ & $0.00031$\\
$\mathcal{C}_{3}^{(1)}-\mathcal{W}_{2}^{(2)}$ & $0.305$ & $0.02234$ & $0.573$ & $1.750$ & $0.869$ & $0.724$ & $60.71837$ & $61.32635$ & $35.65006$ & $24.88093$ & $2.86873$\\
$\mathcal{C}_{4}^{(1)}-\mathcal{W}_{1}^{(1)}$ & $0.303$ & $0.02241$ & $0.505$ & $1.573$ & $1.975$ & $0.977$ & $186.45401$ & $186.65288$ & $179.85843$ & $185.75283$ & $8.89698$\\ $\mathcal{C}_{2}^{(2)}-\mathcal{W}_{3}^{(1)}$ & $0.287$ & $0.02245$ & $0.566$ & $1.155$ & $1.887$ & $0.881$ & $49.154$ & $49.903$ & $0$ & $47.639$ & $7.521$\\
$\mathcal{C}_{3}^{(1)}-\mathcal{W}_{3}^{(1)}$ & $0.303$ & $0.02241$ & $0.574$ & $1.573$ & $1.465$ & $0.416$ & $49.152$ & $49.901$ & $0$ & $35.101$ & $0.116$\\ \hline
     \multicolumn{12}{|c|}{Weinberg operator with gCP (IO)} \\ \hline
\multirow{2}{*}{Combinations}& \multicolumn{10}{c|}{Predictions for mixing parameters and neutrino masses at best fitting point} & \multirow{2}{*}{$\chi^{2}_{\text{min}}$} \\ \cline{2-11}
 & $\sin^2\theta_{12}$ &$\sin^2\theta_{13}$ &$\sin^2\theta_{23}$&$\delta_{CP}/\pi$ & $\alpha_{21}/\pi$  &$\alpha_{31}/\pi$ & $m_1$/meV & $m_2$/meV & $m_3$/meV & $m_{\beta\beta}$/meV &  \\ \hline$\mathcal{C}_{1}^{(1)}-\mathcal{W}_{1}^{(1)}$ & $0.304$ & $0.02241$ & $0.568$ & $1.532$ & $0.007$ & $0.957$ & $49.439$ & $50.184$ & $5.327$ & $48.662$ & $0.011$\\
$\mathcal{C}_{1}^{(1)}-\mathcal{W}_{2}^{(2)}$ & $0.305$ & $0.02234$ & $0.573$ & $1.750$ & $1.577$ & $0.724$ & $60.715$ & $61.323$ & $35.644$ & $48.618$ & $2.867$\\
$\mathcal{C}_{1}^{(1)}-\mathcal{W}_{3}^{(2)}$ & $0.304$ & $0.02241$ & $0.572$ & $1.556$ & $0.011$ & $0.958$ & $49.150$ & $49.899$ & $0$ & $48.265$ & $0.020$\\
$\mathcal{C}_{2}^{(2)}-\mathcal{W}_{1}^{(1)}$ & $0.303$ & $0.02242$ & $0.576$ & $1.590$ & $0.023$ & $0.960$ & $49.154$ & $49.903$ & $0.386$ & $48.253$ & $0.308$\\
$\mathcal{C}_{2}^{(2)}-\mathcal{W}_{2}^{(2)}$ & $0.302$ & $0.02241$ & $0.579$ & $1.599$ & $0.014$ & $0.964$ & $56.358$ & $57.013$ & $27.606$ & $55.733$ & $0.515$\\
$\mathcal{C}_{2}^{(3)}-\mathcal{W}_{1}^{(1)}$ & $0.304$ & $0.02241$ & $0.569$ & $1.545$ & $0.016$ & $0.958$ & $57.111$ & $57.757$ & $29.083$ & $56.600$ & $0.001$\\
$\mathcal{C}_{2}^{(3)}-\mathcal{W}_{2}^{(2)}$ & $0.305$ & $0.02234$ & $0.573$ & $1.750$ & $1.578$ & $0.724$ & $60.717$ & $61.325$ & $35.648$ & $48.625$ & $2.868$\\
$\mathcal{C}_{3}^{(1)}-\mathcal{W}_{1}^{(1)}$ & $0.304$ & $0.02241$ & $0.569$ & $1.546$ & $0.016$ & $0.958$ & $56.986$ & $57.633$ & $28.836$ & $56.473$ & $0.003$\\
$\mathcal{C}_{3}^{(1)}-\mathcal{W}_{2}^{(2)}$ & $0.305$ & $0.02234$ & $0.573$ & $1.750$ & $0.578$ & $0.724$ & $60.718$ & $61.326$ & $35.650$ & $40.133$ & $2.869$\\
$\mathcal{C}_{4}^{(1)}-\mathcal{W}_{1}^{(1)}$ & $0.305$ & $0.02241$ & $0.511$ & $1.438$ & $1.991$ & $0.991$ & $105.933$ & $106.283$ & $93.842$ & $105.620$ & $7.721$\\ \hline \hline \end{tabular}}
\caption{\label{tab:prediction_Majorana_WO_IO}The same as in Table~\ref{tab:prediction_Majorana_WO_NO} but for IO neutrino masses.}
\end{table}

\clearpage
\newpage

\subsection{Majorana Neutrinos (Seesaw mechanism)}

If neutrino masses are described by the type-I seesaw mechanism with two right-handed neutrinos, the texture-zero patterns of $(M_{E},M_{\nu})$ that can explain the experimental data on lepton masses and mixing parameters are presented in Table~\ref{tab:Majorana_SS_pair_summary_NO_IO}. Here, we give the examples of lepton models that can realise those texture zeros. For NO neutrino mass spectrum, the representative models and corresponding predictions are provided in Table~\ref{tab:assign_Majorana_SS_NO} and \ref{tab:prediction_Majorana_SS_NO}. The same is presented for IO in Tables~\ref{tab:assign_Majorana_SS_IO} and ~\ref{tab:prediction_Majorana_SS_IO}.

\begin{table}[ht!]
 \centering
 \resizebox{1.0\textwidth}{!}{
\begin{tabular}{|c|c|c|c|c|c|c|c|c|c|} \hline\hline
   \multicolumn{10}{|c|}{Seesaw mechanism without gCP (NO)} \\ \hline
  \multicolumn{2}{|c|}{Combinations} & $\#\text{P}_{0}$ & $\#\text{P}$ & $\rho_{L}$ & $\rho_{E^{c}}$ & $\rho_{N^{c}}$ & $k_{L}$ & $k_{E^{c}}$ & $k_{N^{c}}$ \\ \hline\multirow{2}{*}{$\mathcal{C}_{2}^{(3)}-\mathcal{W}_{3}^{(2)}$} & $\mathfrak{D}_{2}^{(3)}-\mathfrak{N}_{0}^{(1)}$ & \multirow{2}{*}{$14$} & $9$ & $ \mathbf{2'} \oplus \mathbf{1'}   $ & $\mathbf{1''} \oplus \mathbf{1'} \oplus \mathbf{1'}   $ & $\mathbf{2}   $ & $2, -1 $ & $1, 1, 3 $ & $2 $ \\ \cline{2-2}\cline{4-10}$ $ & $\mathfrak{D}_{4}^{(1)}-\mathfrak{N}_{0}^{(1)}$ & $$ & $9$ & $ \mathbf{2''} \oplus \mathbf{1'}   $ & $\mathbf{1} \oplus \mathbf{1'} \oplus \mathbf{1'}   $ & $\mathbf{2}   $ & $-3, -2 $ & $4, 6, 8 $ & $3 $ \\ \hline\multirow{2}{*}{$\mathcal{C}_{3}^{(1)}-\mathcal{W}_{3}^{(2)}$} & $\mathfrak{D}_{2}^{(3)}-\mathfrak{N}_{0}^{(1)}$ & \multirow{2}{*}{$12$} & $9$ & $ \mathbf{2''} \oplus \mathbf{1'}   $ & $\mathbf{1} \oplus \mathbf{1'} \oplus \mathbf{1''}   $ & $\mathbf{2}   $ & $-1, 0 $ & $4, 4, 4 $ & $3 $ \\ \cline{2-2}\cline{4-10}$ $ & $\mathfrak{D}_{4}^{(1)}-\mathfrak{N}_{0}^{(1)}$ & $$ & $9$ & $ \mathbf{2''} \oplus \mathbf{1'}   $ & $\mathbf{1} \oplus \mathbf{1''} \oplus \mathbf{1'}   $ & $\mathbf{2}   $ & $-3, -2 $ & $4, 6, 6 $ & $3 $ \\ \hline $\mathcal{C}_{1}^{(1)}-\mathcal{W}_{3}^{(2)}$ & $\mathfrak{D}_{2}^{(3)}-\mathfrak{N}_{2}^{(1)}$ & $16$ & $9$ & $ \mathbf{2'} \oplus \mathbf{1}   $ & $\mathbf{1'} \oplus \mathbf{1} \oplus \mathbf{1''}   $ & $\mathbf{1''} \oplus \mathbf{1'}   $ & $0, -1 $ & $1, 5, 5 $ & $3, 3 $ \\ \hline$\mathcal{C}_{2}^{(2)}-\mathcal{W}_{3}^{(2)}$ & $\mathfrak{D}_{2}^{(3)}-\mathfrak{N}_{2}^{(1)}$ & $14$ & $9$ & $ \mathbf{2'} \oplus \mathbf{1}   $ & $\mathbf{1} \oplus \mathbf{1} \oplus \mathbf{1''}   $ & $\mathbf{1'} \oplus \mathbf{1''}   $ & $3, 2 $ & $-2, 2, 2 $ & $0, 0 $ \\ \hline$\mathcal{C}_{2}^{(3)}-\mathcal{W}_{1}^{(1)}$ & $\mathfrak{D}_{1}^{(1)}-\mathfrak{N}_{2}^{(1)}$ & $18$ & $9$ & $ \mathbf{2} \oplus \mathbf{1''}   $ & $\mathbf{1} \oplus \mathbf{1''} \oplus \mathbf{1}   $ & $\mathbf{1} \oplus \mathbf{1}   $ & $4, 3 $ & $-1, -1, 1 $ & $-1, 1 $ \\ \hline$\mathcal{C}_{3}^{(1)}-\mathcal{W}_{1}^{(1)}$ & $\mathfrak{D}_{1}^{(1)}-\mathfrak{N}_{2}^{(1)}$ & $16$ & $9$ & $ \mathbf{2} \oplus \mathbf{1''}   $ & $\mathbf{1} \oplus \mathbf{1''} \oplus \mathbf{1'}   $ & $\mathbf{1} \oplus \mathbf{1}   $ & $4, 5 $ & $-1, -1, -1 $ & $-1, 1 $ \\ \hline
   \multicolumn{10}{|c|}{Seesaw mechanism with gCP (NO)} \\ \hline
  \multicolumn{2}{|c|}{Combinations} & $\#\text{P}_{0}$ & $\#\text{P}$ & $\rho_{L}$ & $\rho_{E^{c}}$ & $\rho_{N^{c}}$ & $k_{L}$ & $k_{E^{c}}$ & $k_{N^{c}}$ \\ \hline\multirow{2}{*}{$\mathcal{C}_{1}^{(1)}-\mathcal{W}_{3}^{(2)}$} & $\mathfrak{D}_{2}^{(3)}-\mathfrak{N}_{0}^{(1)}$ & \multirow{2}{*}{$16$} & $8$ & $ \mathbf{2} \oplus \mathbf{1'}   $ & $\mathbf{2''} \oplus \mathbf{1}   $ & $\mathbf{2}   $ & $-1, -2 $ & $5, 6 $ & $3 $ \\ \cline{2-2}\cline{4-10}$ $ & $\mathfrak{D}_{4}^{(1)}-\mathfrak{N}_{0}^{(1)}$ & $$ & $9$ & $ \mathbf{2''} \oplus \mathbf{1'}   $ & $\mathbf{1} \oplus \mathbf{1'} \oplus \mathbf{1''}   $ & $\mathbf{2}   $ & $-3, -4 $ & $6, 8, 8 $ & $3 $ \\ \hline\multirow{2}{*}{$\mathcal{C}_{2}^{(2)}-\mathcal{W}_{3}^{(2)}$} & $\mathfrak{D}_{2}^{(3)}-\mathfrak{N}_{0}^{(1)}$ & \multirow{2}{*}{$14$} & $9$ & $ \mathbf{2'} \oplus \mathbf{1'}   $ & $\mathbf{1''} \oplus \mathbf{1''} \oplus \mathbf{1'}   $ & $\mathbf{2}   $ & $2, 1 $ & $-1, 3, 3 $ & $2 $ \\ \cline{2-2}\cline{4-10}$ $ & $\mathfrak{D}_{4}^{(1)}-\mathfrak{N}_{0}^{(1)}$ & $$ & $9$ & $ \mathbf{2''} \oplus \mathbf{1'}   $ & $\mathbf{1'} \oplus \mathbf{1''} \oplus \mathbf{1''}   $ & $\mathbf{2}   $ & $-3, -2 $ & $6, 6, 8 $ & $3 $ \\ \hline\multirow{2}{*}{$\mathcal{C}_{2}^{(3)}-\mathcal{W}_{3}^{(2)}$} & $\mathfrak{D}_{2}^{(3)}-\mathfrak{N}_{0}^{(1)}$ & \multirow{2}{*}{$14$} & $8$ & $ \mathbf{2} \oplus \mathbf{1}   $ & $\mathbf{1''} \oplus \mathbf{1} \oplus \mathbf{1}   $ & $\mathbf{2}   $ & $1, -2 $ & $2, 2, 4 $ & $3 $ \\ \cline{2-2}\cline{4-10}$ $ & $\mathfrak{D}_{4}^{(1)}-\mathfrak{N}_{0}^{(1)}$ & $$ & $8$ & $ \mathbf{2''} \oplus \mathbf{1'}   $ & $\mathbf{2} \oplus \mathbf{1''}   $ & $\mathbf{2}   $ & $-3, -4 $ & $7, 8 $ & $3 $ \\ \hline\multirow{2}{*}{$\mathcal{C}_{3}^{(1)}-\mathcal{W}_{3}^{(2)}$} & $\mathfrak{D}_{2}^{(3)}-\mathfrak{N}_{0}^{(1)}$ & \multirow{2}{*}{$12$} & $8$ & $ \mathbf{2'} \oplus \mathbf{1'}   $ & $\mathbf{1''} \oplus \mathbf{1'} \oplus \mathbf{1''}   $ & $\mathbf{2}   $ & $2, 1 $ & $-1, -1, 3 $ & $2 $ \\ \cline{2-2}\cline{4-10}$ $ & $\mathfrak{D}_{4}^{(1)}-\mathfrak{N}_{0}^{(1)}$ & $$ & $8$ & $ \mathbf{2''} \oplus \mathbf{1'}   $ & $\mathbf{1''} \oplus \mathbf{1'} \oplus \mathbf{1''}   $ & $\mathbf{2}   $ & $-3, -4 $ & $4, 6, 8 $ & $3 $ \\ \hline\multirow{2}{*}{$\mathcal{C}_{1}^{(1)}-\mathcal{W}_{1}^{(1)}$} & $\mathfrak{D}_{1}^{(1)}-\mathfrak{N}_{1}^{(1)}$ & \multirow{2}{*}{$20$} & $9$ & $ \mathbf{2} \oplus \mathbf{1'}   $ & $\mathbf{2} \oplus \mathbf{1'}   $ & $\mathbf{1''} \oplus \mathbf{1}   $ & $1, 2 $ & $3, 4 $ & $2, 2 $ \\ \cline{2-2}\cline{4-10}$ $ & $\mathfrak{D}_{1}^{(1)}-\mathfrak{N}_{2}^{(1)}$ & $$ & $9$ & $ \mathbf{2'} \oplus \mathbf{1}   $ & $\mathbf{2} \oplus \mathbf{1''}   $ & $\mathbf{1'} \oplus \mathbf{1''}   $ & $4, 3 $ & $2, -1 $ & $-1, 1 $ \\ \hline$\mathcal{C}_{1}^{(1)}-\mathcal{W}_{3}^{(1)}$ & $\mathfrak{D}_{2}^{(2)}-\mathfrak{N}_{2}^{(1)}$ & $16$ & $9$ & $ \mathbf{2''} \oplus \mathbf{1'}   $ & $\mathbf{2''} \oplus \mathbf{1}   $ & $\mathbf{1'} \oplus \mathbf{1''}   $ & $0, 1 $ & $4, 1 $ & $3, 3 $ \\ \hline\multirow{2}{*}{$\mathcal{C}_{1}^{(1)}-\mathcal{W}_{3}^{(2)}$} & $\mathfrak{D}_{2}^{(3)}-\mathfrak{N}_{1}^{(3)}$ & \multirow{2}{*}{$16$} & $9$ & $ \mathbf{2''} \oplus \mathbf{1'}   $ & $\mathbf{2'} \oplus \mathbf{1}   $ & $\mathbf{1} \oplus \mathbf{1}   $ & $1, -4 $ & $5, 2 $ & $0, 2 $ \\ \cline{2-2}\cline{4-10}$ $ & $\mathfrak{D}_{2}^{(3)}-\mathfrak{N}_{2}^{(1)}$ & $$ & $9$ & $ \mathbf{2'} \oplus \mathbf{1}   $ & $\mathbf{1'} \oplus \mathbf{1} \oplus \mathbf{1''}   $ & $\mathbf{1''} \oplus \mathbf{1'}   $ & $0, -1 $ & $1, 5, 5 $ & $3, 3 $ \\ \hline$\mathcal{C}_{1}^{(1)}-\mathcal{W}_{1}^{(1)}$ & $\mathfrak{D}_{3}^{(1)}-\mathfrak{N}_{1}^{(1)}$ & $20$ & $9$ & $ \mathbf{2''} \oplus \mathbf{1}   $ & $\mathbf{2''} \oplus \mathbf{1'}   $ & $\mathbf{1''} \oplus \mathbf{1'}   $ & $1, 2 $ & $3, 4 $ & $2, 2 $ \\ \hline$\mathcal{C}_{1}^{(1)}-\mathcal{W}_{3}^{(1)}$ & $\mathfrak{D}_{3}^{(1)}-\mathfrak{N}_{1}^{(2)}$ & $16$ & $9$ & $ \mathbf{2''} \oplus \mathbf{1}   $ & $\mathbf{2''} \oplus \mathbf{1'}   $ & $\mathbf{1'} \oplus \mathbf{1''}   $ & $1, 2 $ & $3, 4 $ & $2, 2 $ \\ \hline\multirow{2}{*}{$\mathcal{C}_{2}^{(2)}-\mathcal{W}_{1}^{(1)}$} & $\mathfrak{D}_{1}^{(1)}-\mathfrak{N}_{1}^{(1)}$ & \multirow{2}{*}{$18$} & $9$ & $ \mathbf{2'} \oplus \mathbf{1}   $ & $\mathbf{2''} \oplus \mathbf{1}   $ & $\mathbf{1''} \oplus \mathbf{1'}   $ & $1, 2 $ & $3, 2 $ & $2, 2 $ \\ \cline{2-2}\cline{4-10}$ $ & $\mathfrak{D}_{1}^{(1)}-\mathfrak{N}_{2}^{(1)}$ & $$ & $9$ & $ \mathbf{2} \oplus \mathbf{1''}   $ & $\mathbf{2'} \oplus \mathbf{1'}   $ & $\mathbf{1} \oplus \mathbf{1''}   $ & $2, 3 $ & $2, 1 $ & $1, 3 $ \\ \hline\multirow{2}{*}{$\mathcal{C}_{2}^{(2)}-\mathcal{W}_{3}^{(1)}$} & $\mathfrak{D}_{2}^{(2)}-\mathfrak{N}_{1}^{(2)}$ & \multirow{2}{*}{$14$} & $9$ & $ \mathbf{2} \oplus \mathbf{1}   $ & $\mathbf{2} \oplus \mathbf{1}   $ & $\mathbf{1} \oplus \mathbf{1''}   $ & $-1, 2 $ & $3, 2 $ & $2, 2 $ \\ \cline{2-2}\cline{4-10}$ $ & $\mathfrak{D}_{2}^{(2)}-\mathfrak{N}_{2}^{(1)}$ & $$ & $9$ & $ \mathbf{2''} \oplus \mathbf{1'}   $ & $\mathbf{2''} \oplus \mathbf{1''}   $ & $\mathbf{1'} \oplus \mathbf{1''}   $ & $0, 1 $ & $4, 3 $ & $3, 3 $ \\ \hline$\mathcal{C}_{2}^{(2)}-\mathcal{W}_{3}^{(2)}$ & $\mathfrak{D}_{2}^{(3)}-\mathfrak{N}_{2}^{(1)}$ & $14$ & $9$ & $ \mathbf{2'} \oplus \mathbf{1}   $ & $\mathbf{1} \oplus \mathbf{1} \oplus \mathbf{1''}   $ & $\mathbf{1'} \oplus \mathbf{1''}   $ & $3, 2 $ & $-2, 2, 2 $ & $0, 0 $ \\ \hline$\mathcal{C}_{2}^{(3)}-\mathcal{W}_{1}^{(1)}$ & $\mathfrak{D}_{1}^{(1)}-\mathfrak{N}_{2}^{(1)}$ & $18$ & $9$ & $ \mathbf{2} \oplus \mathbf{1}   $ & $\mathbf{1} \oplus \mathbf{1''} \oplus \mathbf{1}   $ & $\mathbf{1} \oplus \mathbf{1}   $ & $4, 3 $ & $-1, -1, 1 $ & $-1, 1 $ \\ \hline\multirow{3}{*}{$\mathcal{C}_{2}^{(3)}-\mathcal{W}_{3}^{(2)}$} & $\mathfrak{D}_{2}^{(3)}-\mathfrak{N}_{1}^{(1)}$ & \multirow{3}{*}{$14$} & $9$ & $ \mathbf{2} \oplus \mathbf{1''}   $ & $\mathbf{1''} \oplus \mathbf{1} \oplus \mathbf{1}   $ & $\mathbf{1''} \oplus \mathbf{1}   $ & $2, 3 $ & $1, 1, 3 $ & $1, 3 $ \\ \cline{2-2}\cline{4-10}$ $ & $\mathfrak{D}_{2}^{(3)}-\mathfrak{N}_{1}^{(2)}$ & $$ & $9$ & $ \mathbf{2'} \oplus \mathbf{1'}   $ & $\mathbf{2} \oplus \mathbf{1''}   $ & $\mathbf{1'} \oplus \mathbf{1''}   $ & $1, -4 $ & $3, 4 $ & $2, 2 $ \\ \cline{2-2}\cline{4-10} $ $ & $\mathfrak{D}_{2}^{(3)}-\mathfrak{N}_{1}^{(3)}$ & $$ & $9$ & $ \mathbf{2''} \oplus \mathbf{1''}   $ & $\mathbf{2''} \oplus \mathbf{1}   $ & $\mathbf{1} \oplus \mathbf{1}   $ & $1, 0 $ & $3, 4 $ & $0, 2 $ \\ \hline$\mathcal{C}_{2}^{(3)}-\mathcal{W}_{1}^{(1)}$ & $\mathfrak{D}_{3}^{(1)}-\mathfrak{N}_{1}^{(1)}$ & $18$ & $9$ & $ \mathbf{2''} \oplus \mathbf{1}   $ & $\mathbf{1} \oplus \mathbf{1} \oplus \mathbf{1'}   $ & $\mathbf{1''} \oplus \mathbf{1'}   $ & $1, 2 $ & $0, 2, 4 $ & $2, 2 $ \\ \hline$\mathcal{C}_{2}^{(3)}-\mathcal{W}_{3}^{(1)}$ & $\mathfrak{D}_{3}^{(1)}-\mathfrak{N}_{1}^{(2)}$ & $14$ & $9$ & $ \mathbf{2''} \oplus \mathbf{1}   $ & $\mathbf{1} \oplus \mathbf{1} \oplus \mathbf{1'}   $ & $\mathbf{1'} \oplus \mathbf{1''}   $ & $1, 2 $ & $0, 2, 4 $ & $2, 2 $ \\ \hline$\mathcal{C}_{3}^{(1)}-\mathcal{W}_{1}^{(1)}$ & $\mathfrak{D}_{1}^{(1)}-\mathfrak{N}_{2}^{(1)}$ & $16$ & $9$ & $ \mathbf{2'} \oplus \mathbf{1''}   $ & $\mathbf{1} \oplus \mathbf{1'} \oplus \mathbf{1''}   $ & $\mathbf{1'} \oplus \mathbf{1''}   $ & $4, 5 $ & $-1, -1, -1 $ & $-1, 1 $ \\ \hline\multirow{2}{*}{$\mathcal{C}_{3}^{(1)}-\mathcal{W}_{3}^{(2)}$} & $\mathfrak{D}_{2}^{(3)}-\mathfrak{N}_{1}^{(1)}$ & \multirow{2}{*}{$12$} & $9$ & $ \mathbf{2} \oplus \mathbf{1'}   $ & $\mathbf{1'} \oplus \mathbf{1''} \oplus \mathbf{1}   $ & $\mathbf{1} \oplus \mathbf{1}   $ & $2, 3 $ & $1, 1, 3 $ & $1, 3 $ \\ \cline{2-2}\cline{4-10}$ $ & $\mathfrak{D}_{2}^{(3)}-\mathfrak{N}_{1}^{(3)}$ & $ $ & $9$ & $ \mathbf{2''} \oplus \mathbf{1'}   $ & $\mathbf{1''} \oplus \mathbf{1'} \oplus \mathbf{1'}   $ & $\mathbf{1} \oplus \mathbf{1'}   $ & $1, 0 $ & $0, 2, 4 $ & $2, 2 $ \\ \hline$\mathcal{C}_{3}^{(1)}-\mathcal{W}_{1}^{(1)}$ & $\mathfrak{D}_{3}^{(1)}-\mathfrak{N}_{1}^{(1)}$ & $16$ & $9$ & $ \mathbf{2''} \oplus \mathbf{1}   $ & $\mathbf{1} \oplus \mathbf{1''} \oplus \mathbf{1'}   $ & $\mathbf{1''} \oplus \mathbf{1'}   $ & $1, 2 $ & $2, 2, 4 $ & $2, 2 $ \\ \hline$\mathcal{C}_{3}^{(1)}-\mathcal{W}_{3}^{(1)}$ & $\mathfrak{D}_{3}^{(1)}-\mathfrak{N}_{1}^{(2)}$ & $12$ & $9$ & $ \mathbf{2''} \oplus \mathbf{1}   $ & $\mathbf{1''} \oplus \mathbf{1} \oplus \mathbf{1'}   $ & $\mathbf{1'} \oplus \mathbf{1''}   $ & $1, 2 $ & $2, 2, 4 $ & $2, 2 $ \\ \hline$\mathcal{C}_{4}^{(1)}-\mathcal{W}_{1}^{(1)}$ & $\mathfrak{D}_{1}^{(1)}-\mathfrak{N}_{1}^{(1)}$ & $15$ & $9$ & $ \mathbf{2'} \oplus \mathbf{1}   $ & $\mathbf{1} \oplus \mathbf{1'} \oplus \mathbf{1''}   $ & $\mathbf{1''} \oplus \mathbf{1'}   $ & $1, 2 $ & $-2, 0, 4 $ & $2, 2 $ \\ \hline \hline \end{tabular}}
\caption{\label{tab:assign_Majorana_SS_NO}Representative models for the viable patterns of texture zero in $(M_{E},M_{\nu})$ that can accommodate the experimental data at $3\sigma$ level for NO neutrino mass spectrum. Here, neutrinos are Majorana particles and their masses are assumed to be generated by the type-I seesaw mechanism. The models are given without and with gCP. The same convention as Table~\ref{tab:assign_Dirac_NO} is adopted.}
\end{table}

\begin{table}[ht!]
 \centering \resizebox{1.0\textwidth}{!}{
\begin{tabular}{|c|c|c|c|c|c|c|c|c|c|c|c|c|} \hline
   \multicolumn{12}{|c|}{Seesaw mechanism without gCP (NO)} \\ \hline
  \multicolumn{2}{|c|}{\multirow{2}{*}{Combinations}}& \multicolumn{9}{c|}{Predictions for mixing parameters and neutrino masses at best fitting point} & \multirow{2}{*}{$\chi^{2}_{\text{min}}$} \\ \cline{3-11}
\multicolumn{2}{|c|}{} & $\sin^2\theta_{12}$ &$\sin^2\theta_{13}$ &$\sin^2\theta_{23}$&$\delta_{CP}/\pi$ & $\phi/\pi$ & $m_1$/meV & $m_2$/meV & $m_3$/meV & $m_{\beta\beta}$/meV &  \\ \hline\multirow{2}{*}{$\mathcal{C}_{2}^{(3)}-\mathcal{W}_{3}^{(2)}$} & $\mathfrak{D}_{2}^{(3)}-\mathfrak{N}_{0}^{(1)}$ & $0.304$ & $0.02245$ & $0.450$ & $1.288$ & $0.125$ & $0$ & $8.614$ & $50.096$ & $2.101$ & $0.004$ \\ \cline{2-12}$ $ & $\mathfrak{D}_{4}^{(1)}-\mathfrak{N}_{0}^{(1)}$ & $0.318$ & $0.02215$ & $0.428$ & $1.136$ & $1.881$ & $0$ & $8.614$ & $49.724$ & $3.692$ & $4.918$ \\ \hline\multirow{2}{*}{$\mathcal{C}_{3}^{(1)}-\mathcal{W}_{3}^{(2)}$} & $\mathfrak{D}_{2}^{(3)}-\mathfrak{N}_{0}^{(1)}$ & $0.304$ & $0.02239$ & $0.453$ & $1.280$ & $0.634$ & $0$ & $8.614$ & $50.078$ & $1.761$ & $0.039$ \\ \cline{2-12}$ $ & $\mathfrak{D}_{4}^{(1)}-\mathfrak{N}_{0}^{(1)}$ & $0.318$ & $0.02215$ & $0.428$ & $1.136$ & $1.881$ & $0$ & $8.614$ & $49.724$ & $3.693$ & $4.934$ \\ \hline $\mathcal{C}_{1}^{(1)}-\mathcal{W}_{3}^{(2)}$ & $\mathfrak{D}_{2}^{(3)}-\mathfrak{N}_{2}^{(1)}$ & $0.304$ & $0.02245$ & $0.449$ & $1.245$ & $1.350$ & $0$ & $8.614$ & $50.114$ & $3.648$ & $0.064$ \\ \hline$\mathcal{C}_{2}^{(2)}-\mathcal{W}_{3}^{(2)}$ & $\mathfrak{D}_{2}^{(3)}-\mathfrak{N}_{2}^{(1)}$ & $0.303$ & $0.02248$ & $0.455$ & $1.465$ & $0.438$ & $0$ & $8.614$ & $50.054$ & $2.332$ & $0.961$ \\ \hline$\mathcal{C}_{2}^{(3)}-\mathcal{W}_{1}^{(1)}$ & $\mathfrak{D}_{1}^{(1)}-\mathfrak{N}_{2}^{(1)}$ & $0.304$ & $0.02246$ & $0.450$ & $1.304$ & $0.134$ & $0$ & $8.614$ & $50.096$ & $1.961$ & $0.018$ \\ \hline$\mathcal{C}_{3}^{(1)}-\mathcal{W}_{1}^{(1)}$ & $\mathfrak{D}_{1}^{(1)}-\mathfrak{N}_{2}^{(1)}$ & $0.304$ & $0.02246$ & $0.453$ & $1.367$ & $0.006$ & $0$ & $8.614$ & $50.081$ & $1.964$ & $0.235$ \\ \hline
   \multicolumn{12}{|c|}{Seesaw mechanism with gCP (NO)} \\ \hline
  \multicolumn{2}{|c|}{\multirow{2}{*}{Combinations}}& \multicolumn{9}{c|}{Predictions for mixing parameters and neutrino masses at best fitting point} & \multirow{2}{*}{$\chi^{2}_{\text{min}}$} \\ \cline{3-11}
\multicolumn{2}{|c|}{} & $\sin^2\theta_{12}$ &$\sin^2\theta_{13}$ &$\sin^2\theta_{23}$&$\delta_{CP}/\pi$ & $\phi/\pi$ & $m_1$/meV & $m_2$/meV & $m_3$/meV & $m_{\beta\beta}$/meV &  \\ \hline\multirow{2}{*}{$\mathcal{C}_{1}^{(1)}-\mathcal{W}_{3}^{(2)}$} & $\mathfrak{D}_{2}^{(3)}-\mathfrak{N}_{0}^{(1)}$ & $0.300$ & $0.02252$ & $0.458$ & $1.003$ & $1.998$ & $0$ & $8.6139$ & $50.169$ & $3.659$ & $4.214$ \\ \cline{2-12}$ $ & $\mathfrak{D}_{4}^{(1)}-\mathfrak{N}_{0}^{(1)}$ & $0.304$ & $0.02246$ & $0.451$ & $1.108$ & $1.493$ & $0$ & $8.6144$ & $50.095$ & $3.425$ & $1.491$ \\ \hline\multirow{2}{*}{$\mathcal{C}_{2}^{(2)}-\mathcal{W}_{3}^{(2)}$} & $\mathfrak{D}_{2}^{(3)}-\mathfrak{N}_{0}^{(1)}$ & $0.304$ & $0.02246$ & $0.450$ & $1.273$ & $1.566$ & $0$ & $8.6139$ & $50.099$ & $3.637$ & $0.001$ \\ \cline{2-12}$ $ & $\mathfrak{D}_{4}^{(1)}-\mathfrak{N}_{0}^{(1)}$ & $0.304$ & $0.02246$ & $0.451$ & $1.070$ & $1.953$ & $0$ & $8.6144$ & $50.093$ & $3.709$ & $2.234$ \\ \hline\multirow{2}{*}{$\mathcal{C}_{2}^{(3)}-\mathcal{W}_{3}^{(2)}$} & $\mathfrak{D}_{2}^{(3)}-\mathfrak{N}_{0}^{(1)}$ & $0.305$ & $0.02242$ & $0.452$ & $1.163$ & $0.624$ & $0$ & $8.6139$ & $50.064$ & $1.466$ & $0.704$ \\ \cline{2-12}$ $ & $\mathfrak{D}_{4}^{(1)}-\mathfrak{N}_{0}^{(1)}$ & $0.313$ & $0.02227$ & $0.436$ & $1.014$ & $1.987$ & $0$ & $8.6139$ & $49.844$ & $3.748$ & $5.181$ \\ \hline\multirow{2}{*}{$\mathcal{C}_{3}^{(1)}-\mathcal{W}_{3}^{(2)}$} & $\mathfrak{D}_{2}^{(3)}-\mathfrak{N}_{0}^{(1)}$ & $0.298$ & $0.02262$ & $0.456$ & $1.097$ & $0.887$ & $0$ & $8.6139$ & $50.157$ & $1.443$ & $2.088$ \\ \cline{2-12}$ $ & $\mathfrak{D}_{4}^{(1)}-\mathfrak{N}_{0}^{(1)}$ & $0.316$ & $0.02222$ & $0.430$ & $1.014$ & $1.988$ & $0$ & $8.6139$ & $49.791$ & $3.766$ & $6.575$ \\ \hline\multirow{2}{*}{$\mathcal{C}_{1}^{(1)}-\mathcal{W}_{1}^{(1)}$} & $\mathfrak{D}_{1}^{(1)}-\mathfrak{N}_{1}^{(1)}$ & $0.303$ & $0.02246$ & $0.452$ & $1.318$ & $1.028$ & $0$ & $8.614$ & $50.109$ & $3.259$ & $0.051$ \\ \cline{2-12}$ $ & $\mathfrak{D}_{1}^{(1)}-\mathfrak{N}_{2}^{(1)}$ & $0.304$ & $0.02246$ & $0.450$ & $1.302$ & $0.131$ & $0$ & $8.614$ & $50.096$ & $1.986$ & $0.015$ \\ \hline$\mathcal{C}_{1}^{(1)}-\mathcal{W}_{3}^{(1)}$ & $\mathfrak{D}_{2}^{(2)}-\mathfrak{N}_{2}^{(1)}$ & $0.335$ & $0.02273$ & $0.438$ & $1.045$ & $1.019$ & $0$ & $8.614$ & $50.345$ & $1.784$ & $10.717$ \\ \hline\multirow{2}{*}{$\mathcal{C}_{1}^{(1)}-\mathcal{W}_{3}^{(2)}$} & $\mathfrak{D}_{2}^{(3)}-\mathfrak{N}_{1}^{(3)}$ & $0.316$ & $0.02220$ & $0.431$ & $1.015$ & $1.987$ & $0$ & $8.614$ & $49.773$ & $3.769$ & $6.495$ \\ \cline{2-12}$ $ & $\mathfrak{D}_{2}^{(3)}-\mathfrak{N}_{2}^{(1)}$ & $0.304$ & $0.02245$ & $0.449$ & $1.245$ & $1.350$ & $0$ & $8.614$ & $50.114$ & $3.648$ & $0.064$ \\ \hline$\mathcal{C}_{1}^{(1)}-\mathcal{W}_{1}^{(1)}$ & $\mathfrak{D}_{3}^{(1)}-\mathfrak{N}_{1}^{(1)}$ & $0.305$ & $0.02246$ & $0.449$ & $1.190$ & $0.621$ & $0$ & $8.614$ & $50.095$ & $1.440$ & $0.408$ \\ \hline$\mathcal{C}_{1}^{(1)}-\mathcal{W}_{3}^{(1)}$ & $\mathfrak{D}_{3}^{(1)}-\mathfrak{N}_{1}^{(2)}$ & $0.305$ & $0.02246$ & $0.449$ & $1.190$ & $0.621$ & $0$ & $8.614$ & $50.095$ & $1.440$ & $0.408$ \\ \hline\multirow{2}{*}{$\mathcal{C}_{2}^{(2)}-\mathcal{W}_{1}^{(1)}$} & $\mathfrak{D}_{1}^{(1)}-\mathfrak{N}_{1}^{(1)}$ & $0.303$ & $0.02246$ & $0.452$ & $1.318$ & $1.028$ & $0$ & $8.614$ & $50.109$ & $3.259$ & $0.051$ \\ \cline{2-12}$ $ & $\mathfrak{D}_{1}^{(1)}-\mathfrak{N}_{2}^{(1)}$ & $0.299$ & $0.02249$ & $0.448$ & $1.796$ & $1.030$ & $0$ & $8.614$ & $50.261$ & $2.341$ & $6.966$ \\ \hline\multirow{2}{*}{$\mathcal{C}_{2}^{(2)}-\mathcal{W}_{3}^{(1)}$} & $\mathfrak{D}_{2}^{(2)}-\mathfrak{N}_{1}^{(2)}$ & $0.273$ & $0.02248$ & $0.464$ & $1.000$ & $1.000$ & $0$ & $8.614$ & $50.100$ & $1.170$ & $11.456$ \\ \cline{2-12}$ $ & $\mathfrak{D}_{2}^{(2)}-\mathfrak{N}_{2}^{(1)}$ & $0.335$ & $0.02273$ & $0.438$ & $1.045$ & $1.019$ & $0$ & $8.614$ & $50.346$ & $1.784$ & $10.723$ \\ \hline$\mathcal{C}_{2}^{(2)}-\mathcal{W}_{3}^{(2)}$ & $\mathfrak{D}_{2}^{(3)}-\mathfrak{N}_{2}^{(1)}$ & $0.303$ & $0.02248$ & $0.455$ & $1.465$ & $0.438$ & $0$ & $8.614$ & $50.054$ & $2.332$ & $0.961$ \\ \hline$\mathcal{C}_{2}^{(3)}-\mathcal{W}_{1}^{(1)}$ & $\mathfrak{D}_{1}^{(1)}-\mathfrak{N}_{2}^{(1)}$ & $0.304$ & $0.02246$ & $0.450$ & $1.304$ & $0.134$ & $0$ & $8.614$ & $50.096$ & $1.961$ & $0.018$ \\ \hline\multirow{3}{*}{$\mathcal{C}_{2}^{(3)}-\mathcal{W}_{3}^{(2)}$} & $\mathfrak{D}_{2}^{(3)}-\mathfrak{N}_{1}^{(1)}$ & $0.300$ & $0.02266$ & $0.463$ & $1.414$ & $0.103$ & $0$ & $8.614$ & $49.972$ & $1.501$ & $1.208$ \\ \cline{2-12}$ $ & $\mathfrak{D}_{2}^{(3)}-\mathfrak{N}_{1}^{(2)}$ & $0.291$ & $0.02275$ & $0.502$ & $1.676$ & $1.680$ & $0$ & $8.614$ & $49.805$ & $1.324$ & $13.186$ \\ \cline{2-12} $ $ & $\mathfrak{D}_{2}^{(3)}-\mathfrak{N}_{1}^{(3)}$ & $0.313$ & $0.02226$ & $0.436$ & $1.015$ & $1.986$ & $0$ & $8.614$ & $49.861$ & $3.746$ & $5.146$ \\ \hline$\mathcal{C}_{2}^{(3)}-\mathcal{W}_{1}^{(1)}$ & $\mathfrak{D}_{3}^{(1)}-\mathfrak{N}_{1}^{(1)}$ & $0.305$ & $0.02246$ & $0.449$ & $1.189$ & $0.621$ & $0$ & $8.614$ & $50.095$ & $1.440$ & $0.411$ \\ \hline$\mathcal{C}_{2}^{(3)}-\mathcal{W}_{3}^{(1)}$ & $\mathfrak{D}_{3}^{(1)}-\mathfrak{N}_{1}^{(2)}$ & $0.305$ & $0.02246$ & $0.449$ & $1.189$ & $0.621$ & $0$ & $8.614$ & $50.095$ & $1.440$ & $0.411$ \\ \hline$\mathcal{C}_{3}^{(1)}-\mathcal{W}_{1}^{(1)}$ & $\mathfrak{D}_{1}^{(1)}-\mathfrak{N}_{2}^{(1)}$ & $0.304$ & $0.02246$ & $0.453$ & $1.367$ & $0.006$ & $0$ & $8.614$ & $50.081$ & $1.964$ & $0.235$ \\ \hline\multirow{2}{*}{$\mathcal{C}_{3}^{(1)}-\mathcal{W}_{3}^{(2)}$} & $\mathfrak{D}_{2}^{(3)}-\mathfrak{N}_{1}^{(1)}$ & $0.302$ & $0.02250$ & $0.458$ & $1.648$ & $1.772$ & $0$ & $8.614$ & $50.076$ & $1.462$ & $3.638$ \\ \cline{2-12}$ $ & $\mathfrak{D}_{2}^{(3)}-\mathfrak{N}_{1}^{(3)}$ & $0.304$ & $0.02243$ & $0.455$ & $1.738$ & $0.716$ & $0$ & $8.614$ & $50.062$ & $3.544$ & $5.390$ \\ \hline$\mathcal{C}_{3}^{(1)}-\mathcal{W}_{1}^{(1)}$ & $\mathfrak{D}_{3}^{(1)}-\mathfrak{N}_{1}^{(1)}$ & $0.305$ & $0.02246$ & $0.449$ & $1.189$ & $0.621$ & $0$ & $8.614$ & $50.095$ & $1.440$ & $0.411$ \\ \hline$\mathcal{C}_{3}^{(1)}-\mathcal{W}_{3}^{(1)}$ & $\mathfrak{D}_{3}^{(1)}-\mathfrak{N}_{1}^{(2)}$ & $0.305$ & $0.02246$ & $0.449$ & $1.189$ & $0.621$ & $0$ & $8.614$ & $50.095$ & $1.440$ & $0.411$ \\ \hline$\mathcal{C}_{4}^{(1)}-\mathcal{W}_{1}^{(1)}$ & $\mathfrak{D}_{1}^{(1)}-\mathfrak{N}_{1}^{(1)}$ & $0.303$ & $0.02245$ & $0.453$ & $1.386$ & $1.043$ & $0$ & $8.614$ & $50.120$ & $3.544$ & $0.336$ \\ \hline \hline \end{tabular}}
\caption{\label{tab:prediction_Majorana_SS_NO}Best-fit values of the lepton mixing and neutrino mass parameters for the representative models presented in Table~\ref{tab:assign_Majorana_SS_NO}. For all cases, the predictions of the lepton mass ratios are $m_{e}/m_{\mu}=0.00474$, $m_{\mu}/m_{\tau}=0.0586$ and $\Delta m_{21}^{2}/\Delta m_{31}^{2}=0.030$.}
\end{table}

\clearpage
\newpage

\begin{center}
\begin{small}
\setlength\LTcapwidth{\textwidth}
\setlength\LTleft{-0.2in}            
\setlength\LTright{0pt}           
 \begin{longtable}{|c|c|c|c|c|c|c|c|c|c|}
\caption{\label{tab:assign_Majorana_SS_IO}
The same as in Table~\ref{tab:assign_Majorana_SS_NO} but for IO neutrino mass spectrum.
} \\
\hline
\hline \multicolumn{10}{|c|}{Seesaw mechanism without gCP (IO)} \\ \hline
  \multicolumn{2}{|c|}{Combinations} & $\#\text{P}_{0}$ & $\#\text{P}$ & $\rho_{L}$ & $\rho_{E^{c}}$ & $\rho_{N^{c}}$ & $k_{L}$ & $k_{E^{c}}$ & $k_{N^{c}}$ \\ \hline
\endfirsthead

\multicolumn{10}{c}%
{{\bfseries \tablename\ \thetable{} -- continued from previous page}} \\
\hline
  \multicolumn{2}{|c|}{Combinations} & $\#\text{P}_{0}$ & $\#\text{P}$ & $\rho_{L}$ & $\rho_{E^{c}}$ & $\rho_{N^{c}}$ & $k_{L}$ & $k_{E^{c}}$ & $k_{N^{c}}$ \\
\endhead

\hline \multicolumn{10}{|r|}{{continues on next page}} \\ \hline
\endfoot

\hline \hline
\endlastfoot

\multirow{2}{*}{$\mathcal{C}_{1}^{(1)}-\mathcal{W}_{3}^{(2)}$} & $\mathfrak{D}_{2}^{(3)}-\mathfrak{N}_{0}^{(1)}$ & \multirow{2}{*}{$16$} & $8$ & $ \mathbf{2} \oplus \mathbf{1'}   $ & $\mathbf{1} \oplus \mathbf{1''} \oplus \mathbf{1''}   $ & $\mathbf{2}   $ & $1, 2 $ & $2, 2, 4 $ & $1 $ \\ \cline{2-2}\cline{4-10}$ $ & $\mathfrak{D}_{4}^{(1)}-\mathfrak{N}_{0}^{(1)}$ & $$ & $8$ & $ \mathbf{2''} \oplus \mathbf{1}   $ & $\mathbf{1'} \oplus \mathbf{1} \oplus \mathbf{1}   $ & $\mathbf{2}   $ & $-1, 0 $ & $4, 4, 6 $ & $1 $ \\ \hline\multirow{2}{*}{$\mathcal{C}_{2}^{(2)}-\mathcal{W}_{3}^{(2)}$} & $\mathfrak{D}_{2}^{(3)}-\mathfrak{N}_{0}^{(1)}$ & \multirow{2}{*}{$14$} & $8$ & $ \mathbf{2} \oplus \mathbf{1'}   $ & $\mathbf{1'} \oplus \mathbf{1''} \oplus \mathbf{1''}   $ & $\mathbf{2}   $ & $1, 2 $ & $2, 2, 4 $ & $1 $ \\ \cline{2-2}\cline{4-10}$ $ & $\mathfrak{D}_{4}^{(1)}-\mathfrak{N}_{0}^{(1)}$ & $$ & $8$ & $ \mathbf{2''} \oplus \mathbf{1}   $ & $\mathbf{1} \oplus \mathbf{1''} \oplus \mathbf{1}   $ & $\mathbf{2}   $ & $-1, 0 $ & $4, 4, 6 $ & $1 $ \\ \hline\multirow{2}{*}{$\mathcal{C}_{2}^{(3)}-\mathcal{W}_{3}^{(2)}$} & $\mathfrak{D}_{2}^{(3)}-\mathfrak{N}_{0}^{(1)}$ & \multirow{2}{*}{$14$} & $9$ & $ \mathbf{2} \oplus \mathbf{1'}   $ & $\mathbf{1''} \oplus \mathbf{1} \oplus \mathbf{1'}   $ & $\mathbf{2}   $ & $1, -2 $ & $2, 2, 4 $ & $3 $ \\ \cline{2-2}\cline{4-10}$ $ & $\mathfrak{D}_{4}^{(1)}-\mathfrak{N}_{0}^{(1)}$ & $$ & $9$ & $ \mathbf{2''} \oplus \mathbf{1}   $ & $\mathbf{1} \oplus \mathbf{1'} \oplus \mathbf{1}   $ & $\mathbf{2}   $ & $-3, -6 $ & $6, 6, 8 $ & $3 $ \\ \hline\multirow{2}{*}{$\mathcal{C}_{3}^{(1)}-\mathcal{W}_{3}^{(2)}$} & $\mathfrak{D}_{2}^{(3)}-\mathfrak{N}_{0}^{(1)}$ & \multirow{2}{*}{$12$} & $9$ & $ \mathbf{2''} \oplus \mathbf{1'}   $ & $\mathbf{1''} \oplus \mathbf{1'} \oplus \mathbf{1'}   $ & $\mathbf{2}   $ & $-1, -2 $ & $2, 4, 6 $ & $3 $ \\ \cline{2-2}\cline{4-10}$ $ & $\mathfrak{D}_{4}^{(1)}-\mathfrak{N}_{0}^{(1)}$ & $$ & $9$ & $ \mathbf{2''} \oplus \mathbf{1}   $ & $\mathbf{1} \oplus \mathbf{1} \oplus \mathbf{1'}   $ & $\mathbf{2}   $ & $-3, -2 $ & $2, 6, 6 $ & $3 $ \\ \hline$\mathcal{C}_{1}^{(1)}-\mathcal{W}_{1}^{(1)}$ & $\mathfrak{D}_{1}^{(1)}-\mathfrak{N}_{2}^{(1)}$ & $20$ & $8$ & $ \mathbf{2''} \oplus \mathbf{1'}   $ & $\mathbf{2'} \oplus \mathbf{1}   $ & $\mathbf{1''} \oplus \mathbf{1'}   $ & $5, 6 $ & $-3, -2 $ & $0, 0 $ \\ \hline$\mathcal{C}_{1}^{(1)}-\mathcal{W}_{3}^{(2)}$ & $\mathfrak{D}_{2}^{(3)}-\mathfrak{N}_{2}^{(1)}$ & $16$ & $9$ & $ \mathbf{2'} \oplus \mathbf{1}   $ & $\mathbf{1''} \oplus \mathbf{1} \oplus \mathbf{1'}   $ & $\mathbf{1'} \oplus \mathbf{1''}   $ & $5, 6 $ & $-2, 0, 0 $ & $0, 0 $ \\ \hline$\mathcal{C}_{2}^{(2)}-\mathcal{W}_{3}^{(2)}$ & $\mathfrak{D}_{2}^{(3)}-\mathfrak{N}_{2}^{(1)}$ & $14$ & $9$ & $ \mathbf{2'} \oplus \mathbf{1}   $ & $\mathbf{1''} \oplus \mathbf{1} \oplus \mathbf{1}   $ & $\mathbf{1'} \oplus \mathbf{1''}   $ & $5, 6 $ & $-2, -2, 0 $ & $0, 0 $ \\ \hline$\mathcal{C}_{2}^{(3)}-\mathcal{W}_{1}^{(1)}$ & $\mathfrak{D}_{1}^{(1)}-\mathfrak{N}_{2}^{(1)}$ & $18$ & $9$ & $ \mathbf{2'} \oplus \mathbf{1''}   $ & $\mathbf{1'} \oplus \mathbf{1''} \oplus \mathbf{1''}   $ & $\mathbf{1''} \oplus \mathbf{1'}   $ & $3, 4 $ & $0, 0, 2 $ & $0, 0 $ \\ \hline$\mathcal{C}_{2}^{(3)}-\mathcal{W}_{3}^{(1)}$ & $\mathfrak{D}_{2}^{(2)}-\mathfrak{N}_{2}^{(1)}$ & $14$ & $9$ & $ \mathbf{2''} \oplus \mathbf{1'}   $ & $\mathbf{1} \oplus \mathbf{1} \oplus \mathbf{1''}   $ & $\mathbf{1''} \oplus \mathbf{1'}   $ & $0, 1 $ & $1, 3, 5 $ & $3, 3 $ \\ \hline$\mathcal{C}_{3}^{(1)}-\mathcal{W}_{1}^{(1)}$ & $\mathfrak{D}_{1}^{(1)}-\mathfrak{N}_{2}^{(1)}$ & $16$ & $9$ & $ \mathbf{2'} \oplus \mathbf{1''}   $ & $\mathbf{1'} \oplus \mathbf{1} \oplus \mathbf{1''}   $ & $\mathbf{1''} \oplus \mathbf{1'}   $ & $0, 1 $ & $3, 3, 5 $ & $3, 3 $ \\ \hline$\mathcal{C}_{3}^{(1)}-\mathcal{W}_{3}^{(1)}$ & $\mathfrak{D}_{2}^{(2)}-\mathfrak{N}_{2}^{(1)}$ & $12$ & $9$ & $ \mathbf{2} \oplus \mathbf{1'}   $ & $\mathbf{1''} \oplus \mathbf{1'} \oplus \mathbf{1}   $ & $\mathbf{1'} \oplus \mathbf{1''}   $ & $4, 5 $ & $-1, -1, -1 $ & $-1, 1 $ \\ \hline$\mathcal{C}_{4}^{(1)}-\mathcal{W}_{1}^{(1)}$ & $\mathfrak{D}_{1}^{(1)}-\mathfrak{N}_{2}^{(1)}$ & $15$ & $8$ & $ \mathbf{2''} \oplus \mathbf{1'}   $ & $\mathbf{1''} \oplus \mathbf{1} \oplus \mathbf{1'}   $ & $\mathbf{1''} \oplus \mathbf{1'}   $ & $5, 6 $ & $-6, -2, 0 $ & $0, 0 $ \\

\hline \multicolumn{10}{|c|}{Seesaw mechanism with gCP (IO)} \\ \hline
\multicolumn{2}{|c|}{Combinations} & $\#\text{P}_{0}$ & $\#\text{P}$ & $\rho_{L}$ & $\rho_{E^{c}}$ & $\rho_{N^{c}}$ & $k_{L}$ & $k_{E^{c}}$ & $k_{N^{c}}$ \\ \hline\multirow{2}{*}{$\mathcal{C}_{1}^{(1)}-\mathcal{W}_{1}^{(1)}$} & $\mathfrak{D}_{1}^{(1)}-\mathfrak{N}_{1}^{(1)}$ & \multirow{2}{*}{$20$} & $9$ & $ \mathbf{2} \oplus \mathbf{1''}   $ & $\mathbf{2} \oplus \mathbf{1''}   $ & $\mathbf{1''} \oplus \mathbf{1}   $ & $1, 2 $ & $1, 2 $ & $2, 2 $ \\ \cline{2-2}\cline{4-10}$ $ & $\mathfrak{D}_{1}^{(1)}-\mathfrak{N}_{2}^{(1)}$ & $$ & $8$ & $ \mathbf{2''} \oplus \mathbf{1'}   $ & $\mathbf{2'} \oplus \mathbf{1}   $ & $\mathbf{1''} \oplus \mathbf{1'}   $ & $5, 6 $ & $-3, -2 $ & $0, 0 $ \\ \hline$\mathcal{C}_{1}^{(1)}-\mathcal{W}_{3}^{(1)}$ & $\mathfrak{D}_{2}^{(2)}-\mathfrak{N}_{2}^{(1)}$ & $16$ & $9$ & $ \mathbf{2} \oplus \mathbf{1'}   $ & $\mathbf{2'} \oplus \mathbf{1}   $ & $\mathbf{1'} \oplus \mathbf{1''}   $ & $4, 5 $ & $0, -1 $ & $-1, 1 $ \\ \hline\multirow{4}{*}{$\mathcal{C}_{1}^{(1)}-\mathcal{W}_{3}^{(2)}$} & $\mathfrak{D}_{2}^{(3)}-\mathfrak{N}_{1}^{(1)}$ & \multirow{4}{*}{$16$} & $9$ & $ \mathbf{2} \oplus \mathbf{1''}   $ & $\mathbf{2'} \oplus \mathbf{1''}   $ & $\mathbf{1''} \oplus \mathbf{1}   $ & $0, -1 $ & $4, 5 $ & $1, 3 $ \\ \cline{2-2}\cline{4-10}$ $ & $\mathfrak{D}_{2}^{(3)}-\mathfrak{N}_{1}^{(2)}$ & $$ & $9$ & $ \mathbf{2} \oplus \mathbf{1'}   $ & $\mathbf{2''} \oplus \mathbf{1''}   $ & $\mathbf{1} \oplus \mathbf{1''}   $ & $1, 0 $ & $3, 2 $ & $2, 2 $ \\ \cline{2-2}\cline{4-10}$ $ & $\mathfrak{D}_{2}^{(3)}-\mathfrak{N}_{1}^{(3)}$ & $$ & $9$ & $ \mathbf{2''} \oplus \mathbf{1''}   $ & $\mathbf{2} \oplus \mathbf{1'}   $ & $\mathbf{1} \oplus \mathbf{1}   $ & $1, -2 $ & $3, 4 $ & $0, 2 $ \\ \cline{2-2}\cline{4-10}$ $ & $\mathfrak{D}_{2}^{(3)}-\mathfrak{N}_{2}^{(1)}$ & $$ & $9$ & $ \mathbf{2''} \oplus \mathbf{1}   $ & $\mathbf{1} \oplus \mathbf{1} \oplus \mathbf{1''}   $ & $\mathbf{1'} \oplus \mathbf{1''}   $ & $5, 6 $ & $-2, 0, 0 $ & $0, 0 $ \\ \hline$\mathcal{C}_{1}^{(1)}-\mathcal{W}_{1}^{(1)}$ & $\mathfrak{D}_{3}^{(1)}-\mathfrak{N}_{1}^{(1)}$ & $20$ & $9$ & $ \mathbf{2''} \oplus \mathbf{1}   $ & $\mathbf{2''} \oplus \mathbf{1'}   $ & $\mathbf{1''} \oplus \mathbf{1'}   $ & $1, 2 $ & $3, 4 $ & $2, 2 $ \\ \hline$\mathcal{C}_{1}^{(1)}-\mathcal{W}_{3}^{(1)}$ & $\mathfrak{D}_{3}^{(1)}-\mathfrak{N}_{1}^{(2)}$ & $16$ & $9$ & $ \mathbf{2'} \oplus \mathbf{1''}   $ & $\mathbf{2} \oplus \mathbf{1''}   $ & $\mathbf{1} \oplus \mathbf{1''}   $ & $1, 2 $ & $3, 4 $ & $2, 2 $ \\ \hline\multirow{2}{*}{$\mathcal{C}_{2}^{(2)}-\mathcal{W}_{1}^{(1)}$} & $\mathfrak{D}_{1}^{(1)}-\mathfrak{N}_{1}^{(1)}$ & \multirow{2}{*}{$18$} & $9$ & $ \mathbf{2'} \oplus \mathbf{1''}   $ & $\mathbf{2'} \oplus \mathbf{1}   $ & $\mathbf{1''} \oplus \mathbf{1'}   $ & $1, 2 $ & $1, 2 $ & $2, 2 $ \\ \cline{2-2}\cline{4-10}$ $ & $\mathfrak{D}_{1}^{(1)}-\mathfrak{N}_{2}^{(1)}$ & $$ & $9$ & $ \mathbf{2'} \oplus \mathbf{1''}   $ & $\mathbf{2'} \oplus \mathbf{1}   $ & $\mathbf{1'} \oplus \mathbf{1'}   $ & $0, 1 $ & $4, 3 $ & $1, 3 $ \\ \hline\multirow{4}{*}{$\mathcal{C}_{2}^{(2)}-\mathcal{W}_{3}^{(2)}$} & $\mathfrak{D}_{2}^{(3)}-\mathfrak{N}_{1}^{(1)}$ & \multirow{4}{*}{$14$} & $9$ & $ \mathbf{2''} \oplus \mathbf{1'}   $ & $\mathbf{2''} \oplus \mathbf{1''}   $ & $\mathbf{1} \oplus \mathbf{1}   $ & $0, 1 $ & $4, 3 $ & $1, 3 $ \\ \cline{2-2}\cline{4-10}$ $ & $\mathfrak{D}_{2}^{(3)}-\mathfrak{N}_{1}^{(2)}$ & $$ & $9$ & $ \mathbf{2} \oplus \mathbf{1''}   $ & $\mathbf{2'} \oplus \mathbf{1'}   $ & $\mathbf{1} \oplus \mathbf{1''}   $ & $1, 0 $ & $3, 0 $ & $2, 2 $ \\ \cline{2-2}\cline{4-10}$ $ & $\mathfrak{D}_{2}^{(3)}-\mathfrak{N}_{1}^{(3)}$ & $$ & $9$ & $ \mathbf{2''} \oplus \mathbf{1'}   $ & $\mathbf{2} \oplus \mathbf{1''}   $ & $\mathbf{1} \oplus \mathbf{1}   $ & $1, 2 $ & $3, 2 $ & $0, 2 $ \\ \cline{2-2}\cline{4-10} $ $ & $\mathfrak{D}_{2}^{(3)}-\mathfrak{N}_{2}^{(1)}$ & $$ & $9$ & $ \mathbf{2''} \oplus \mathbf{1}   $ & $\mathbf{1''} \oplus \mathbf{1} \oplus \mathbf{1}   $ & $\mathbf{1'} \oplus \mathbf{1''}   $ & $5, 6 $ & $-2, -2, 0 $ & $0, 0 $ \\ \hline$\mathcal{C}_{2}^{(2)}-\mathcal{W}_{1}^{(1)}$ & $\mathfrak{D}_{3}^{(1)}-\mathfrak{N}_{1}^{(1)}$ & $18$ & $9$ & $ \mathbf{2'} \oplus \mathbf{1''}   $ & $\mathbf{2} \oplus \mathbf{1}   $ & $\mathbf{1''} \oplus \mathbf{1}   $ & $1, 2 $ & $3, 2 $ & $2, 2 $ \\ \hline$\mathcal{C}_{2}^{(2)}-\mathcal{W}_{3}^{(1)}$ & $\mathfrak{D}_{3}^{(1)}-\mathfrak{N}_{1}^{(2)}$ & $14$ & $9$ & $ \mathbf{2'} \oplus \mathbf{1''}   $ & $\mathbf{2} \oplus \mathbf{1'}   $ & $\mathbf{1} \oplus \mathbf{1''}   $ & $1, 2 $ & $3, -2 $ & $2, 2 $ \\ \hline\multirow{2}{*}{$\mathcal{C}_{2}^{(3)}-\mathcal{W}_{1}^{(1)}$} & $\mathfrak{D}_{1}^{(1)}-\mathfrak{N}_{1}^{(1)}$ & \multirow{2}{*}{$18$} & $9$ & $ \mathbf{2} \oplus \mathbf{1'}   $ & $\mathbf{2} \oplus \mathbf{1''}   $ & $\mathbf{1''} \oplus \mathbf{1}   $ & $3, 4 $ & $-1, 0 $ & $2, 2 $ \\ \cline{2-2}\cline{4-10}$ $ & $\mathfrak{D}_{1}^{(1)}-\mathfrak{N}_{2}^{(1)}$ & $$ & $9$ & $ \mathbf{2} \oplus \mathbf{1}   $ & $\mathbf{2'} \oplus \mathbf{1}   $ & $\mathbf{1} \oplus \mathbf{1''}   $ & $2, 1 $ & $2, 3 $ & $1, 3 $ \\ \hline$\mathcal{C}_{2}^{(3)}-\mathcal{W}_{3}^{(1)}$ & $\mathfrak{D}_{2}^{(2)}-\mathfrak{N}_{2}^{(1)}$ & $14$ & $9$ & $ \mathbf{2} \oplus \mathbf{1'}   $ & $\mathbf{1''} \oplus \mathbf{1} \oplus \mathbf{1''}   $ & $\mathbf{1''} \oplus \mathbf{1'}   $ & $3, 4 $ & $-2, 0, 2 $ & $0, 0 $ \\ \hline\multirow{4}{*}{$\mathcal{C}_{2}^{(3)}-\mathcal{W}_{3}^{(2)}$} & $\mathfrak{D}_{2}^{(3)}-\mathfrak{N}_{1}^{(1)}$ & \multirow{4}{*}{$14$} & $9$ & $ \mathbf{2} \oplus \mathbf{1''}   $ & $\mathbf{1} \oplus \mathbf{1''} \oplus \mathbf{1'}   $ & $\mathbf{1''} \oplus \mathbf{1}   $ & $0, -1 $ & $3, 5, 5 $ & $1, 3 $ \\ \cline{2-2}\cline{4-10}$ $ & $\mathfrak{D}_{2}^{(3)}-\mathfrak{N}_{1}^{(2)}$ & $$ & $9$ & $ \mathbf{2'} \oplus \mathbf{1''}   $ & $\mathbf{1''} \oplus \mathbf{1'} \oplus \mathbf{1''}   $ & $\mathbf{1'} \oplus \mathbf{1''}   $ & $1, 0 $ & $2, 4, 4 $ & $2, 2 $ \\ \cline{2-2}\cline{4-10} $ $ & $\mathfrak{D}_{2}^{(3)}-\mathfrak{N}_{1}^{(3)}$ & $$ & $9$ & $ \mathbf{2''} \oplus \mathbf{1''}   $ & $\mathbf{2''} \oplus \mathbf{1'}   $ & $\mathbf{1} \oplus \mathbf{1'}   $ & $1, 0 $ & $3, 4 $ & $0, 2 $ \\ \cline{2-2}\cline{4-10}$ $ & $\mathfrak{D}_{2}^{(3)}-\mathfrak{N}_{2}^{(1)}$ & $$ & $9$ & $ \mathbf{2''} \oplus \mathbf{1}   $ & $\mathbf{2} \oplus \mathbf{1}   $ & $\mathbf{1'} \oplus \mathbf{1''}   $ & $5, 0 $ & $1, 0 $ & $0, 0 $ \\ \hline$\mathcal{C}_{2}^{(3)}-\mathcal{W}_{1}^{(1)}$ & $\mathfrak{D}_{3}^{(1)}-\mathfrak{N}_{1}^{(1)}$ & $18$ & $9$ & $ \mathbf{2} \oplus \mathbf{1''}   $ & $\mathbf{1} \oplus \mathbf{1''} \oplus \mathbf{1''}   $ & $\mathbf{1''} \oplus \mathbf{1'}   $ & $1, 2 $ & $2, 2, 4 $ & $2, 2 $ \\ \hline$\mathcal{C}_{2}^{(3)}-\mathcal{W}_{3}^{(1)}$ & $\mathfrak{D}_{3}^{(1)}-\mathfrak{N}_{1}^{(2)}$ & $14$ & $9$ & $ \mathbf{2} \oplus \mathbf{1''}   $ & $\mathbf{1} \oplus \mathbf{1''} \oplus \mathbf{1}   $ & $\mathbf{1'} \oplus \mathbf{1''}   $ & $1, 2 $ & $2, 4, 4 $ & $2, 2 $ \\ \hline$\mathcal{C}_{3}^{(1)}-\mathcal{W}_{1}^{(1)}$ & $\mathfrak{D}_{1}^{(1)}-\mathfrak{N}_{2}^{(1)}$ & $16$ & $9$ & $ \mathbf{2'} \oplus \mathbf{1''}   $ & $\mathbf{1} \oplus \mathbf{1'} \oplus \mathbf{1''}   $ & $\mathbf{1''} \oplus \mathbf{1'}   $ & $0, 1 $ & $3, 3, 5 $ & $3, 3 $ \\ \hline$\mathcal{C}_{3}^{(1)}-\mathcal{W}_{3}^{(1)}$ & $\mathfrak{D}_{2}^{(2)}-\mathfrak{N}_{2}^{(1)}$ & $12$ & $9$ & $ \mathbf{2} \oplus \mathbf{1'}   $ & $\mathbf{1''} \oplus \mathbf{1'} \oplus \mathbf{1}   $ & $\mathbf{1'} \oplus \mathbf{1''}   $ & $4, 5 $ & $-1, -1, -1 $ & $-1, 1 $ \\ \hline\multirow{2}{*}{$\mathcal{C}_{3}^{(1)}-\mathcal{W}_{3}^{(2)}$} & $\mathfrak{D}_{2}^{(3)}-\mathfrak{N}_{1}^{(1)}$ & \multirow{2}{*}{$12$} & $9$ & $ \mathbf{2} \oplus \mathbf{1}   $ & $\mathbf{1} \oplus \mathbf{1''} \oplus \mathbf{1''}   $ & $\mathbf{1''} \oplus \mathbf{1}   $ & $0, -1 $ & $1, 3, 5 $ & $1, 3 $ \\ \cline{2-2}\cline{4-10}$ $ & $\mathfrak{D}_{2}^{(3)}-\mathfrak{N}_{1}^{(3)}$ & $$ & $9$ & $ \mathbf{2''} \oplus \mathbf{1'}   $ & $\mathbf{1''} \oplus \mathbf{1'} \oplus \mathbf{1'}   $ & $\mathbf{1} \oplus \mathbf{1'}   $ & $1, 0 $ & $0, 2, 4 $ & $0, 2 $ \\ \hline$\mathcal{C}_{3}^{(1)}-\mathcal{W}_{1}^{(1)}$ & $\mathfrak{D}_{3}^{(1)}-\mathfrak{N}_{1}^{(1)}$ & $16$ & $9$ & $ \mathbf{2} \oplus \mathbf{1''}   $ & $\mathbf{1'} \oplus \mathbf{1} \oplus \mathbf{1''}   $ & $\mathbf{1''} \oplus \mathbf{1'}   $ & $1, 2 $ & $2, 2, 4 $ & $2, 2 $ \\ \hline$\mathcal{C}_{3}^{(1)}-\mathcal{W}_{3}^{(1)}$ & $\mathfrak{D}_{3}^{(1)}-\mathfrak{N}_{1}^{(2)}$ & $12$ & $9$ & $ \mathbf{2} \oplus \mathbf{1''}   $ & $\mathbf{1''} \oplus \mathbf{1'} \oplus \mathbf{1}   $ & $\mathbf{1'} \oplus \mathbf{1''}   $ & $1, 2 $ & $0, 2, 2 $ & $2, 2 $ \\ \hline\multirow{2}{*}{$\mathcal{C}_{4}^{(1)}-\mathcal{W}_{1}^{(1)}$} & $\mathfrak{D}_{1}^{(1)}-\mathfrak{N}_{1}^{(1)}$ & \multirow{2}{*}{$15$} & $9$ & $ \mathbf{2} \oplus \mathbf{1}   $ & $\mathbf{1} \oplus \mathbf{1''} \oplus \mathbf{1'}   $ & $\mathbf{1''} \oplus \mathbf{1}   $ & $2, 1 $ & $-1, 1, 3 $ & $1, 3 $ \\ \cline{2-2}\cline{4-10}$ $ & $\mathfrak{D}_{1}^{(1)}-\mathfrak{N}_{2}^{(1)}$ & $$ & $8$ & $ \mathbf{2''} \oplus \mathbf{1'}   $ & $\mathbf{1''} \oplus \mathbf{1} \oplus \mathbf{1'}   $ & $\mathbf{1''} \oplus \mathbf{1'}   $ & $5, 6 $ & $-2, -2, 0 $ & $0, 0 $ \\ \hline$\mathcal{C}_{4}^{(3)}-\mathcal{W}_{1}^{(1)}$ & $\mathfrak{D}_{1}^{(1)}-\mathfrak{N}_{1}^{(1)}$ & $14$ & $9$ & $ \mathbf{2''} \oplus \mathbf{1'}   $ & $\mathbf{2} \oplus \mathbf{1'}   $ & $\mathbf{1''} \oplus \mathbf{1'}   $ & $3, 4 $ & $-3, 0 $ & $2, 2 $ \\ \hline

 \end{longtable}
\end{small}
\end{center}

\begin{center}
\begin{small}
\begin{landscape}
\setlength\LTcapwidth{\textwidth}
\setlength\LTleft{0.7in}            
\setlength\LTright{0pt}           
 \begin{longtable}{|c|c|c|c|c|c|c|c|c|c|c|c|}
\caption{\label{tab:prediction_Majorana_SS_IO}The same as in Table~\ref{tab:prediction_Majorana_SS_NO} but for IO neutrino mass spectrum.} \\
\hline
\hline \multicolumn{12}{|c|}{Seesaw mechanism without gCP (IO)} \\ \hline
\multicolumn{2}{|c|}{\multirow{2}{*}{Combinations}}& \multicolumn{9}{c|}{Predictions for mixing parameters and neutrino masses at best fitting point} & \multirow{2}{*}{$\chi^{2}_{\text{min}}$} \\ \cline{3-11}
\multicolumn{2}{|c|}{} & $\sin^2\theta_{12}$ &$\sin^2\theta_{13}$ &$\sin^2\theta_{23}$&$\delta_{CP}/\pi$ & $\phi/\pi$ & $m_1$/meV & $m_2$/meV & $m_3$/meV & $m_{\beta\beta}$/meV &  \\ \hline
\endfirsthead

\multicolumn{12}{c}%
   {{\bfseries \tablename\ \thetable{} -- continued from previous page}} \\
\hline
\multicolumn{2}{|c|}{\multirow{2}{*}{Combinations}}& \multicolumn{9}{c|}{Predictions for mixing parameters and neutrino masses at best fitting point} & \multirow{2}{*}{$\chi^{2}_{\text{min}}$} \\ \cline{3-11}
\multicolumn{2}{|c|}{} & $\sin^2\theta_{12}$ &$\sin^2\theta_{13}$ &$\sin^2\theta_{23}$&$\delta_{CP}/\pi$ & $\phi/\pi$ & $m_1$/meV & $m_2$/meV & $m_3$/meV & $m_{\beta\beta}$/meV &  \\ \hline
\endhead

\hline \multicolumn{12}{|r|}{{continues on next page}} \\ \hline
\endfoot

\hline \hline
\endlastfoot

\multirow{2}{*}{$\mathcal{C}_{1}^{(1)}-\mathcal{W}_{3}^{(2)}$} & $\mathfrak{D}_{2}^{(3)}-\mathfrak{N}_{0}^{(1)}$ & $0.304$ & $0.02241$ & $0.572$ & $1.557$ & $0.343$ & $49.150$ & $49.899$ & $0$ & $42.529$ & $0.023$ \\ \cline{2-12}$ $ & $\mathfrak{D}_{4}^{(1)}-\mathfrak{N}_{0}^{(1)}$ & $0.304$ & $0.02241$ & $0.572$ & $1.557$ & $0.014$ & $49.150$ & $49.899$ & $0$ & $48.262$ & $0.023$ \\ \hline\multirow{2}{*}{$\mathcal{C}_{2}^{(2)}-\mathcal{W}_{3}^{(2)}$} & $\mathfrak{D}_{2}^{(3)}-\mathfrak{N}_{0}^{(1)}$ & $0.304$ & $0.02241$ & $0.572$ & $1.557$ & $0.343$ & $49.150$ & $49.899$ & $0$ & $42.529$ & $0.023$ \\ \cline{2-12}$ $ & $\mathfrak{D}_{4}^{(1)}-\mathfrak{N}_{0}^{(1)}$ & $0.304$ & $0.02241$ & $0.572$ & $1.557$ & $0.014$ & $49.150$ & $49.899$ & $0$ & $48.262$ & $0.023$ \\ \hline\multirow{2}{*}{$\mathcal{C}_{2}^{(3)}-\mathcal{W}_{3}^{(2)}$} & $\mathfrak{D}_{2}^{(3)}-\mathfrak{N}_{0}^{(1)}$ & $0.304$ & $0.02241$ & $0.570$ & $1.546$ & $1.589$ & $49.151$ & $49.900$ & $0$ & $40.168$ & $0.000$ \\ \cline{2-12}$ $ & $\mathfrak{D}_{4}^{(1)}-\mathfrak{N}_{0}^{(1)}$ & $0.304$ & $0.02241$ & $0.570$ & $1.543$ & $1.023$ & $49.151$ & $49.900$ & $0$ & $18.682$ & $0.000$ \\ \hline\multirow{2}{*}{$\mathcal{C}_{3}^{(1)}-\mathcal{W}_{3}^{(2)}$} & $\mathfrak{D}_{2}^{(3)}-\mathfrak{N}_{0}^{(1)}$ & $0.304$ & $0.02241$ & $0.570$ & $1.545$ & $0.654$ & $49.154$ & $49.903$ & $0$ & $29.609$ & $0.000$ \\ \cline{2-12}$ $ & $\mathfrak{D}_{4}^{(1)}-\mathfrak{N}_{0}^{(1)}$ & $0.304$ & $0.02241$ & $0.570$ & $1.546$ & $1.134$ & $49.151$ & $49.900$ & $0$ & $20.796$ & $0.000$ \\  \hline$\mathcal{C}_{1}^{(1)}-\mathcal{W}_{1}^{(1)}$ & $\mathfrak{D}_{1}^{(1)}-\mathfrak{N}_{2}^{(1)}$ & $0.303$ & $0.02242$ & $0.575$ & $1.586$ & $0.178$ & $49.150$ & $49.899$ & $0$ & $46.687$ & $0.206$ \\ \hline$\mathcal{C}_{1}^{(1)}-\mathcal{W}_{3}^{(2)}$ & $\mathfrak{D}_{2}^{(3)}-\mathfrak{N}_{2}^{(1)}$ & $0.304$ & $0.02241$ & $0.571$ & $1.552$ & $0.012$ & $49.151$ & $49.900$ & $0$ & $48.264$ & $0.007$ \\ \hline$\mathcal{C}_{2}^{(2)}-\mathcal{W}_{3}^{(2)}$ & $\mathfrak{D}_{2}^{(3)}-\mathfrak{N}_{2}^{(1)}$ & $0.304$ & $0.02241$ & $0.571$ & $1.552$ & $0.012$ & $49.151$ & $49.900$ & $0$ & $48.264$ & $0.007$ \\ \hline$\mathcal{C}_{2}^{(3)}-\mathcal{W}_{1}^{(1)}$ & $\mathfrak{D}_{1}^{(1)}-\mathfrak{N}_{2}^{(1)}$ & $0.304$ & $0.02241$ & $0.570$ & $1.542$ & $1.087$ & $49.151$ & $49.900$ & $0$ & $19.568$ & $0.000$ \\ \hline$\mathcal{C}_{2}^{(3)}-\mathcal{W}_{3}^{(1)}$ & $\mathfrak{D}_{2}^{(2)}-\mathfrak{N}_{2}^{(1)}$ & $0.303$ & $0.02242$ & $0.575$ & $1.583$ & $1.544$ & $49.152$ & $49.902$ & $0$ & $38.436$ & $0.222$ \\ \hline$\mathcal{C}_{3}^{(1)}-\mathcal{W}_{1}^{(1)}$ & $\mathfrak{D}_{1}^{(1)}-\mathfrak{N}_{2}^{(1)}$ & $0.304$ & $0.02241$ & $0.570$ & $1.542$ & $0.580$ & $49.151$ & $49.900$ & $0$ & $33.029$ & $0.000$ \\ \hline$\mathcal{C}_{3}^{(1)}-\mathcal{W}_{3}^{(1)}$ & $\mathfrak{D}_{2}^{(2)}-\mathfrak{N}_{2}^{(1)}$ & $0.303$ & $0.02242$ & $0.576$ & $1.584$ & $1.213$ & $49.152$ & $49.902$ & $0$ & $23.730$ & $0.235$ \\ \hline$\mathcal{C}_{4}^{(1)}-\mathcal{W}_{1}^{(1)}$ & $\mathfrak{D}_{1}^{(1)}-\mathfrak{N}_{2}^{(1)}$ & $0.303$ & $0.02242$ & $0.575$ & $1.586$ & $1.853$ & $49.151$ & $49.900$ & $0$ & $47.187$ & $0.220$ \\

\hline \multicolumn{12}{|c|}{Seesaw mechanism with gCP (IO)} \\ \hline
\multicolumn{2}{|c|}{\multirow{2}{*}{Combinations}}& \multicolumn{9}{c|}{Predictions for mixing parameters and neutrino masses at best fitting point} & \multirow{2}{*}{$\chi^{2}_{\text{min}}$} \\ \cline{3-11}
\multicolumn{2}{|c|}{} & $\sin^2\theta_{12}$ &$\sin^2\theta_{13}$ &$\sin^2\theta_{23}$&$\delta_{CP}/\pi$ & $\phi/\pi$ & $m_1$/meV & $m_2$/meV & $m_3$/meV & $m_{\beta\beta}$/meV &  \\ \hline\multirow{2}{*}{$\mathcal{C}_{1}^{(1)}-\mathcal{W}_{1}^{(1)}$} & $\mathfrak{D}_{1}^{(1)}-\mathfrak{N}_{1}^{(1)}$ & $0.304$ & $0.02241$ & $0.569$ & $1.540$ & $1.081$ & $49.151$ & $49.900$ & $0$ & $19.454$ & $0.001$ \\ \cline{2-12}$ $ & $\mathfrak{D}_{1}^{(1)}-\mathfrak{N}_{2}^{(1)}$ & $0.303$ & $0.02242$ & $0.575$ & $1.586$ & $1.745$ & $49.150$ & $49.899$ & $0$ & $45.044$ & $0.206$ \\ \hline$\mathcal{C}_{1}^{(1)}-\mathcal{W}_{3}^{(1)}$ & $\mathfrak{D}_{2}^{(2)}-\mathfrak{N}_{2}^{(1)}$ & $0.303$ & $0.02242$ & $0.576$ & $1.584$ & $0.632$ & $49.152$ & $49.902$ & $0$ & $30.693$ & $0.234$ \\ \hline\multirow{4}{*}{$\mathcal{C}_{1}^{(1)}-\mathcal{W}_{3}^{(2)}$} & $\mathfrak{D}_{2}^{(3)}-\mathfrak{N}_{1}^{(1)}$ & $0.304$ & $0.02241$ & $0.570$ & $1.547$ & $1.045$ & $49.151$ & $49.900$ & $0$ & $18.883$ & $0.001$ \\ \cline{2-12}$ $ & $\mathfrak{D}_{2}^{(3)}-\mathfrak{N}_{1}^{(2)}$ & $0.300$ & $0.02249$ & $0.589$ & $1.740$ & $0.031$ & $49.156$ & $49.905$ & $0$ & $48.222$ & $4.107$ \\ \cline{2-12}$ $ & $\mathfrak{D}_{2}^{(3)}-\mathfrak{N}_{1}^{(3)}$ & $0.304$ & $0.02241$ & $0.569$ & $1.414$ & $1.165$ & $49.150$ & $49.900$ & $0$ & $21.824$ & $0.609$ \\ \cline{2-12}$ $ & $\mathfrak{D}_{2}^{(3)}-\mathfrak{N}_{2}^{(1)}$ & $0.304$ & $0.02241$ & $0.571$ & $1.552$ & $0.009$ & $49.151$ & $49.900$ & $0$ & $48.267$ & $0.007$ \\ \hline$\mathcal{C}_{1}^{(1)}-\mathcal{W}_{1}^{(1)}$ & $\mathfrak{D}_{3}^{(1)}-\mathfrak{N}_{1}^{(1)}$ & $0.305$ & $0.02241$ & $0.570$ & $1.234$ & $1.067$ & $49.151$ & $49.900$ & $0$ & $19.080$ & $3.486$ \\ \hline$\mathcal{C}_{1}^{(1)}-\mathcal{W}_{3}^{(1)}$ & $\mathfrak{D}_{3}^{(1)}-\mathfrak{N}_{1}^{(2)}$ & $0.305$ & $0.02241$ & $0.570$ & $1.233$ & $0.921$ & $49.151$ & $49.900$ & $0$ & $19.303$ & $3.503$ \\ \hline\multirow{2}{*}{$\mathcal{C}_{2}^{(2)}-\mathcal{W}_{1}^{(1)}$} & $\mathfrak{D}_{1}^{(1)}-\mathfrak{N}_{1}^{(1)}$ & $0.304$ & $0.02241$ & $0.569$ & $1.540$ & $0.512$ & $49.151$ & $49.900$ & $0$ & $36.081$ & $0.001$ \\ \cline{2-12}$ $ & $\mathfrak{D}_{1}^{(1)}-\mathfrak{N}_{2}^{(1)}$ & $0.304$ & $0.02241$ & $0.571$ & $1.549$ & $1.149$ & $49.151$ & $49.900$ & $0$ & $21.302$ & $0.002$ \\ \hline\multirow{4}{*}{$\mathcal{C}_{2}^{(2)}-\mathcal{W}_{3}^{(2)}$} & $\mathfrak{D}_{2}^{(3)}-\mathfrak{N}_{1}^{(1)}$ & $0.304$ & $0.02241$ & $0.569$ & $1.536$ & $0.659$ & $49.152$ & $49.901$ & $0$ & $29.383$ & $0.005$ \\ \cline{2-12}$ $ & $\mathfrak{D}_{2}^{(3)}-\mathfrak{N}_{1}^{(2)}$ & $0.307$ & $0.02238$ & $0.539$ & $1.911$ & $0.846$ & $49.204$ & $49.952$ & $0$ & $21.219$ & $11.085$ \\ \cline{2-12}$ $ & $\mathfrak{D}_{2}^{(3)}-\mathfrak{N}_{1}^{(3)}$ & $0.304$ & $0.02241$ & $0.569$ & $1.415$ & $0.803$ & $49.151$ & $49.900$ & $0$ & $23.025$ & $0.608$ \\ \cline{2-12}$ $ & $\mathfrak{D}_{2}^{(3)}-\mathfrak{N}_{2}^{(1)}$ & $0.304$ & $0.02241$ & $0.571$ & $1.552$ & $1.009$ & $49.151$ & $49.900$ & $0$ & $18.641$ & $0.007$ \\ \hline$\mathcal{C}_{2}^{(2)}-\mathcal{W}_{1}^{(1)}$ & $\mathfrak{D}_{3}^{(1)}-\mathfrak{N}_{1}^{(1)}$ & $0.303$ & $0.02246$ & $0.506$ & $1.252$ & $1.791$ & $49.140$ & $49.890$ & $0$ & $46.075$ & $11.536$ \\ \hline$\mathcal{C}_{2}^{(2)}-\mathcal{W}_{3}^{(1)}$ & $\mathfrak{D}_{3}^{(1)}-\mathfrak{N}_{1}^{(2)}$ & $0.303$ & $0.02246$ & $0.506$ & $1.252$ & $0.912$ & $49.141$ & $49.890$ & $0$ & $19.692$ & $11.536$ \\ \hline\multirow{2}{*}{$\mathcal{C}_{2}^{(3)}-\mathcal{W}_{1}^{(1)}$} & $\mathfrak{D}_{1}^{(1)}-\mathfrak{N}_{1}^{(1)}$ & $0.307$ & $0.02243$ & $0.548$ & $1.335$ & $0.798$ & $49.149$ & $49.898$ & $0$ & $23.029$ & $2.675$ \\ \cline{2-12}$ $ & $\mathfrak{D}_{1}^{(1)}-\mathfrak{N}_{2}^{(1)}$ & $0.304$ & $0.02241$ & $0.570$ & $1.546$ & $0.800$ & $49.151$ & $49.900$ & $0$ & $23.142$ & $0.000$ \\ \hline$\mathcal{C}_{2}^{(3)}-\mathcal{W}_{3}^{(1)}$ & $\mathfrak{D}_{2}^{(2)}-\mathfrak{N}_{2}^{(1)}$ & $0.303$ & $0.02242$ & $0.575$ & $1.583$ & $0.129$ & $49.152$ & $49.902$ & $0$ & $47.438$ & $0.222$ \\ \hline\multirow{4}{*}{$\mathcal{C}_{2}^{(3)}-\mathcal{W}_{3}^{(2)}$} & $\mathfrak{D}_{2}^{(3)}-\mathfrak{N}_{1}^{(1)}$ & $0.304$ & $0.02241$ & $0.570$ & $1.545$ & $1.395$ & $49.150$ & $49.899$ & $0$ & $31.908$ & $0.000$ \\ \cline{2-12}$ $ & $\mathfrak{D}_{2}^{(3)}-\mathfrak{N}_{1}^{(2)}$ & $0.310$ & $0.02243$ & $0.521$ & $1.363$ & $0.986$ & $49.148$ & $49.898$ & $0$ & $18.010$ & $6.357$ \\ \cline{2-12}$ $ & $\mathfrak{D}_{2}^{(3)}-\mathfrak{N}_{1}^{(3)}$ & $0.304$ & $0.02241$ & $0.570$ & $1.536$ & $0.003$ & $49.151$ & $49.900$ & $0$ & $48.271$ & $0.002$ \\ \cline{2-12}$ $ & $\mathfrak{D}_{2}^{(3)}-\mathfrak{N}_{2}^{(1)}$ & $0.304$ & $0.02242$ & $0.571$ & $1.540$ & $1.008$ & $49.142$ & $49.891$ & $0$ & $18.627$ & $0.003$ \\ \hline$\mathcal{C}_{2}^{(3)}-\mathcal{W}_{1}^{(1)}$ & $\mathfrak{D}_{3}^{(1)}-\mathfrak{N}_{1}^{(1)}$ & $0.304$ & $0.02241$ & $0.572$ & $1.556$ & $1.066$ & $49.150$ & $49.899$ & $0$ & $19.203$ & $0.018$ \\ \hline$\mathcal{C}_{2}^{(3)}-\mathcal{W}_{3}^{(1)}$ & $\mathfrak{D}_{3}^{(1)}-\mathfrak{N}_{1}^{(2)}$ & $0.304$ & $0.02241$ & $0.572$ & $1.556$ & $1.042$ & $49.150$ & $49.899$ & $0$ & $18.869$ & $0.018$ \\ \hline$\mathcal{C}_{3}^{(1)}-\mathcal{W}_{1}^{(1)}$ & $\mathfrak{D}_{1}^{(1)}-\mathfrak{N}_{2}^{(1)}$ & $0.304$ & $0.02241$ & $0.570$ & $1.542$ & $1.099$ & $49.151$ & $49.900$ & $0$ & $19.855$ & $0.000$ \\ \hline$\mathcal{C}_{3}^{(1)}-\mathcal{W}_{3}^{(1)}$ & $\mathfrak{D}_{2}^{(2)}-\mathfrak{N}_{2}^{(1)}$ & $0.303$ & $0.02242$ & $0.576$ & $1.584$ & $0.046$ & $49.152$ & $49.902$ & $0$ & $48.164$ & $0.235$ \\ \hline\multirow{2}{*}{$\mathcal{C}_{3}^{(1)}-\mathcal{W}_{3}^{(2)}$} & $\mathfrak{D}_{2}^{(3)}-\mathfrak{N}_{1}^{(1)}$ & $0.304$ & $0.02241$ & $0.570$ & $1.545$ & $1.012$ & $49.151$ & $49.900$ & $0$ & $18.632$ & $0.000$ \\ \cline{2-12}$ $ & $\mathfrak{D}_{2}^{(3)}-\mathfrak{N}_{1}^{(3)}$ & $0.304$ & $0.02241$ & $0.570$ & $1.534$ & $1.006$ & $49.151$ & $49.900$ & $0$ & $18.616$ & $0.004$ \\ \hline$\mathcal{C}_{3}^{(1)}-\mathcal{W}_{1}^{(1)}$ & $\mathfrak{D}_{3}^{(1)}-\mathfrak{N}_{1}^{(1)}$ & $0.304$ & $0.02241$ & $0.572$ & $1.556$ & $1.553$ & $49.150$ & $49.899$ & $0$ & $38.765$ & $0.018$ \\ \hline$\mathcal{C}_{3}^{(1)}-\mathcal{W}_{3}^{(1)}$ & $\mathfrak{D}_{3}^{(1)}-\mathfrak{N}_{1}^{(2)}$ & $0.304$ & $0.02241$ & $0.572$ & $1.556$ & $1.130$ & $49.150$ & $49.899$ & $0$ & $20.717$ & $0.018$ \\ \hline\multirow{2}{*}{$\mathcal{C}_{4}^{(1)}-\mathcal{W}_{1}^{(1)}$} & $\mathfrak{D}_{1}^{(1)}-\mathfrak{N}_{1}^{(1)}$ & $0.303$ & $0.02242$ & $0.575$ & $1.583$ & $0.983$ & $49.152$ & $49.902$ & $0$ & $18.746$ & $0.219$ \\ \cline{2-12}$ $ & $\mathfrak{D}_{1}^{(1)}-\mathfrak{N}_{2}^{(1)}$ & $0.303$ & $0.02242$ & $0.575$ & $1.586$ & $0.231$ & $49.151$ & $49.900$ & $0$ & $45.611$ & $0.220$ \\ \hline$\mathcal{C}_{4}^{(3)}-\mathcal{W}_{1}^{(1)}$ & $\mathfrak{D}_{1}^{(1)}-\mathfrak{N}_{1}^{(1)}$ & $0.307$ & $0.02243$ & $0.547$ & $1.333$ & $1.994$ & $49.149$ & $49.898$ & $0$ & $48.269$ & $2.727$ \\

\end{longtable}
\end{landscape}
\end{small}
\end{center}

\end{appendix}



\providecommand{\href}[2]{#2}\begingroup\raggedright\endgroup

\end{document}